\documentclass[11pt,a4paper]{article}
\pdfoutput=1

\usepackage{jcappub}
\usepackage[T1]{fontenc}
\usepackage[utf8]{inputenc}
\usepackage{lmodern}
\usepackage{physics}
\usepackage{mathtools}
\usepackage{booktabs}
\usepackage{xcolor}
\usepackage{comment}
\usepackage{etoolbox}
\let\JCAPtableofcontents\tableofcontents
\renewcommand{\tableofcontents}{\begingroup\small\JCAPtableofcontents\endgroup}
\patchcmd{\maketitle}{\setcounter{footnote}{0}}{\clearpage\setcounter{footnote}{0}}{}{}
\graphicspath{{figures/}}
\hypersetup{
  pdftitle={Horizon kinematics and the fine structure induced by finite-energy Hawking radiation},
  pdfauthor={M. A. Aguirre Cortes; Moises Bravo-Gaete; Julio A. Mendez-Zavaleta; Efrain Rojas Marcial}
}

\newcommand{\e}{\mathrm{e}}

\newcommand{\Mcal}{\mathcal{M}}
\newcommand{\Qcal}{\mathcal{Q}}
\newcommand{\Scal}{\mathcal{S}}
\newcommand{\Ecal}{\mathcal{E}}

\newcommand{\Order}{\mathcal{O}}
\newcommand{\mgot}{\mathfrak{m}}

\title{Horizon kinematics and the fine structure induced by finite-energy Hawking radiation}

\author[a]{\mbox{M. A. Aguirre Cort\'es,}}
\author[b]{\mbox{Mois\'es Bravo-Gaete,}}
\author[a]{\mbox{Julio A. M\'endez-Zavaleta}}
\author[a]{\mbox{and Efra\'in Rojas Marcial}}

\affiliation[a]{Facultad de F\'isica, Universidad Veracruzana,\\
Paseo No. 112, Desarrollo Habitacional Nuevo Xalapa,\\
C.P. 91097, Xalapa-Enr\'iquez, M\'exico.}
\affiliation[b]{Departamento de Matem\'atica, F\'isica y Estad\'istica,\\
Facultad de Ciencias B\'asicas, Universidad Cat\'olica del Maule,\\
Casilla 617, Talca, Chile.}

\emailAdd{maria.aguirrec185@gmail.com}
\emailAdd{mbravo@ucm.cl}
\emailAdd{julmendez@uv.mx}
\emailAdd{efrojas@uv.mx}

\abstract{Hawking radiation can modify the horizon structure of a black hole
through the accompanying loss of mass and global charges. Investigating
this process through quantum tunneling, we develop a systematic
strategy to reconstruct horizon displacements under finite-energy
emission, even when their radii are not explicitly known.
Local inversion of the horizon condition allows us to follow simple
and degenerate roots along sequences of static geometries.
As a concrete realization, we consider a nonlinearly charged
extension of the Ba\~nados--Teitelboim--Zanelli (BTZ) family, whose
rich horizon structure offers an appealing setting to explore
this process.
Of particular interest is an initial configuration with a
double inner horizon coexisting with a simple event horizon.
We show that neutral emission necessarily splits this inner horizon
into two simple horizons, revealing an internal horizon fine structure.
For charged emission, the dependence of extracted charge on
emitted energy determines the separation law and allows channels
that preserve or later restore the degeneracy.
We construct such channels connecting the same initial and
final states through distinct internal histories and sharing the
same leading semiclassical tunneling factor.
As emission continues, the larger of the newly separated inner
horizons can coalesce with the event horizon, driving the sequence
towards an extremal black hole with vanishing temperature but
a finite tunneling probability.
Relating the duration of this approach to the radiation fluxes,
we find that zero temperature cannot be attained in finite boundary
time while a uniform adiabaticity bound holds.
We also establish a regime where the geometric expansion and the
canonical thermal approximation remain simultaneously controlled
near this endpoint.
Finally, applying this analysis to the sector reproducing a
quantum-backreacted BTZ geometry reveals how the leading semiclassical
corrections raise the temperature and enhance the neutral-probe
horizon emission factor relative to classical BTZ.}


\begin{document}
\maketitle
\flushbottom

\section{Introduction
\label{sec:introduction}}

Among all physics associated with black holes, Hawking radiation
offers one of the clearest links between gravitation and quantum
physics. In 1974, Hawking showed that black holes formed by
gravitational collapse emit radiation at a temperature determined
by their surface gravity \cite{Hawking:1974rv,Hawking:1975vcx}.
His approach treats quantum fields on a classical collapsing
spacetime, showing that a state with no incoming particles gives
rise to an outgoing flux at late times. The spectrum reaching
infinity is modulated by greybody factors, which account for
scattering by the exterior geometry. Beyond predicting radiation
from an object classically unable to emit it, this result gave
physical content to the analogy between black hole mechanics and
thermodynamics \cite{Bardeen:1973gs}. It also fixed the normalization
of Bekenstein's proposal that a black hole possesses an intrinsic
entropy directly proportional to its horizon area, an idea motivated
by the apparent loss of entropy when matter falls into a black hole
\cite{Bekenstein:1973ur,Bekenstein:1974ax}. The outgoing radiation,
in turn, carries energy away, so the geometry cannot remain unchanged
as the black hole loses mass and evaporates. An isolated Schwarzschild
black hole offers a simple illustration of this feedback; as its mass
decreases, its temperature rises and the emission accelerates. Even
so, the estimated evaporation times for stellar masses far exceed
the present age of the Universe \cite{Page:1976df}.

The low Hawking temperature of stellar-mass black holes also makes
their radiation exceedingly difficult to distinguish from the
astrophysical background \cite{Hawking:1975vcx}. While Hawking
radiation was originally a purely theoretical prediction, Unruh
opened a way to study its analogue in the laboratory by considering
sound in a transonic flow \cite{Unruh:1980cg}. In this setting,
sound perturbations propagate on an effective spacetime with a
sonic horizon, and where their quanta, the phonons, play the role of the
particles emitted in the gravitational case. The idea extends
beyond fluids. In a nonlinear optical fiber, for instance, a
light pulse can produce a moving perturbation of the refractive
index that acts as a horizon for probe light, an arrangement
used to observe stimulated analogue Hawking emission
\cite{Drori:2018ivu,Bernal:2026lme}. What makes these analogues
especially useful is that they bring the conversion of modes
near a horizon within experimental reach, allowing a key part
of the emission process to be studied under controlled conditions.

In the gravitational setting, Hawking radiation also raises a
question about what becomes of the initial quantum state. If a
black hole formed from a pure state evaporates completely, leaving
behind radiation in a mixed state, the evolution would not be
unitary \cite{Hawking:1976ra}. A major advance in describing the emission process came with a
series of seminal works
\cite{Kraus:1994fh,Kraus:1994by,Keski-Vakkuri:1996lbi,Parikh:1999mf}
that laid the foundations of the Kraus--Parikh--Wilczek (KPW)
approach, a semiclassical framework that describes Hawking
radiation as quantum tunneling across the event horizon. By
enforcing energy conservation, the approach accounts for the
energy carried away by the outgoing shell and the accompanying
shift of the horizon. It is this backreaction that gives rise
to corrections to the leading thermal tunneling factor. These
departures from thermality have been discussed in connection
with the information paradox \cite{Parikh:2004ih}, although the
emission probabilities alone do not establish whether the full
evolution is unitary.

This tunneling picture can be extended to anti-de Sitter spacetimes,
with an early application to Schwarzschild--AdS black holes
\cite{Hemming:2000as}. Closer to our problem is Medved's analysis
of neutral emission from the electrically charged
Ba\~nados--Teitelboim--Zanelli (BTZ) black hole, which also considered
the approach to extremality \cite{Medved:2001ca}. For rotating BTZ,
Wu and Jiang accounted for the angular momentum carried away by the
emitted shell \cite{Wu:2006pz}. These examples bring changes in the
black hole parameters into the calculation of the emission
probability. Here we turn the question around and ask what emission
does to the horizon structure itself. The nonlinearly charged BTZ
family of Ref.~\cite{Mendez-Zavaleta:2026rgg} is particularly well
suited to this question, since it admits configurations in which
simple and degenerate horizons coexist. The loss of mass and charge
can then affect not only the positions of the horizons but also their number and degeneracy.

A black hole can lose charge and still move closer to extremality,
as Hiscock and Weems found by following the coupled evolution of
mass and charge in sufficiently massive Reissner--Nordstr\"om
black holes early in their evaporation \cite{Hiscock:1990ex}.
In the BTZ family considered here, the charge carried away for
a given loss of energy selects a path through the stationary
configurations. Following that path lets us see how horizon
degeneracy changes along the way and whether the black hole
moves toward extremality. The rate of energy loss then sets
the pace of the evolution. As extremality draws near, the effect
of a single emission becomes especially important. Kraus and
Wilczek showed that self-interaction corrections can substantially
modify the spectrum in this regime \cite{Kraus:1994fj}. The
background may evolve slowly enough on the timescale of the
outgoing modes for an adiabatic description \cite{Barcelo:2010pj},
yet a single quantum can carry away an appreciable fraction of
the energy above extremality, making the canonical approximation
unreliable \cite{Preskill:1991tb}. More recently, Brown \emph{et al.}
found that quantum fluctuations of the near-extremal geometry can
substantially alter the emission spectrum of four-dimensional
charged black holes \cite{Brown:2024ajk}.

\emph{Organization of this work.} We begin in
Sec.~\ref{sec:horizon_kinematics} by studying
Hawking emission at finite energy within the KPW description and
developing a local inversion of the horizon-defining equation,
arriving at formal series expressions for the tunneling
probabilities and finite horizon displacements. This construction
is extended to degenerate horizons in
Sec.~\ref{sec:degenerate_unfolding}, where we examine their
splitting and possible subsequent coalescence as mass and charge
are extracted. We then turn to the nonlinearly charged BTZ family
in Sec.~\ref{sec:BTZ_laboratory}, following the horizon histories
selected by different emission profiles. The approach to outer
coalescence motivates the analysis of
Sec.~\ref{sec:BTZNLED_outer_endpoint}, which examines the limiting
tunneling exponent alongside the duration of the process and the
conditions for an adiabatic and thermal description.
A complementary application to the sector reproducing the
quantum-backreacted BTZ geometry is presented in
Sec.~\ref{sec:BTZNLED_semiclassical_emission}.
We close with a discussion of the results and further directions
in Sec.~\ref{sec:discussion}, leaving the detailed coefficient
formulas and complementary calculations to the appendices.

\section{KPW probability for finite-energy emission
\label{sec:horizon_kinematics}}

The KPW prescription describes a finite emission between members
of a stationary black hole family, with the mass lost by the black
hole carried by the outgoing shell. Consider a static family in
$D\geq3$ spacetime dimensions whose line element can be written
in Schwarzschild-like gauge,
\begin{equation}
\dd s^2
=
-f(r;\Mcal,\lambda^A)\dd t^2
+\frac{\dd r^2}{f(r;\Mcal,\lambda^A)}
+r^2\dd\Sigma_{D-2,k}^2 .
\label{eq:gen_static_metric}
\end{equation}
Here, $\Mcal$ denotes the black hole mass, while $\lambda^A$
($A=1,\ldots,n_p$) collectively denotes the integration constants,
couplings, and other parameters held fixed during neutral, radial
emission. The metric
$d\Sigma_{D-2,k}^2$ describes the transverse section of the horizon,
with normalized curvature $k$. For a noncompact section, extensive
thermodynamic quantities are understood as densities with respect to
its volume.

 At leading semiclassical order, the probability is given by
$\Gamma\simeq\exp(-2\operatorname{Im}I)$, where $I$ denotes the
semiclassical tunneling action and where we use $\hbar=1$ throughout. Its imaginary contribution arises
from the radial pole crossed by the outgoing trajectory. In the
gauge of Eq.~\eqref{eq:gen_static_metric}, the relevant pole occurs
at the event horizon, which marks the causal barrier
forbidding classical escape from the interior. The same function
therefore determines the event horizon, its kinematics as the mass
decreases due to emission, and the radial pole contributing to the
tunneling probability.

This class of spacetimes \eqref{eq:gen_static_metric} includes asymptotically flat, de Sitter (dS), and anti-de Sitter  (AdS)
black holes with admissible transverse sections. Since the
corresponding residue is determined locally near the horizon, the
asymptotic boundary conditions governing the subsequent propagation
of radiation do not enter its evaluation. The interpretation of a
sequence of such emissions as a complete evaporation process, however, requires
boundary conditions that allow the outgoing flux to escape the
system. This condition is particularly relevant for asymptotically
AdS geometries, where reflecting boundary conditions
would return the emitted radiation from infinity. Evaporation with absorbing boundary conditions at AdS infinity
has been analyzed, for instance, in Ref.~\cite{Page:2015rxa}.

Concretely, let $r=r_+$ denote the position of the initial event
horizon, which we first assume to be a simple outer zero of the
blackening function,
\begin{equation}
f(r_+;\Mcal,\lambda^A)=0,
\qquad
\partial_r f(r_+;\Mcal,\lambda^A)>0,
\label{eq:initial_simple_horizon}
\end{equation}
where for our notations $\partial_x f:=\partial f /\partial x$ throughout. In this case, the near-horizon zero is linear, and both the outgoing
trajectory and the horizon displacement admit the ordinary local
expansions developed below. Degenerate roots appear later as limiting configurations in which these expansions change character; they play a central role in what follows.

Throughout the emission, we follow the outer branch continuously
connected to this initial horizon. The stationary Killing field
$\partial_t$ is assumed to preserve its normalization across the family,
and the energy and temperature below use this normalization throughout
the sequence. Here, horizon motion refers to stationary Killing horizons
parametrized by the emitted energy.

\subsection{Neutral emission along the tunneling path
\label{sec:gen_neutral_channel}}

A Gullstrand--Painlev\'e-type time coordinate
\cite{PainleveGullstrand} makes the metric
\eqref{eq:gen_static_metric} regular near the horizon,
\begin{equation}
\dd\tau
=
\dd t
+\frac{\sqrt{1-f}}{f}\dd r.
\label{eq:gen_Painleve_time}
\end{equation}
In this chart, the outgoing radial null geodesic obeys
\begin{align}
\dot r&=1-\sqrt{1-f}
=\frac12\partial_r f(r_+)(r-r_+)
+\Order\!\left((r-r_+)^2\right),
\label{eq:gen_outgoing_null_geodesic}
\end{align}
where the dot denotes differentiation with respect to $\tau$ and
the expansion is about the initial horizon. The pole in $1/\dot r$
gives the usual Boltzmann factor when the geometry is kept fixed.

The KPW prescription follows the displacement of
this pole as energy is extracted from the black hole. Let $\Ecal>0$
denote the total emitted energy and $\Ecal'\in[0,\Ecal]$ its running
value. The instantaneous black hole mass and event horizon are
\begin{equation}
\Mcal'=\Mcal-\Ecal',
\qquad
R_+(\Ecal')=r_+(\Mcal-\Ecal',\lambda^A),
\label{eq:gen_running_mass}
\end{equation}
with $R_+(0)=r_+$ and satisfying the horizon condition
\begin{equation}
f\bigl[R_+(\Ecal');\Mcal-\Ecal',\lambda^A\bigr]=0.
\label{eq:gen_running_horizon}
\end{equation}
at each stage of the
emission process.
Throughout the text, we retain this convention for other
functions of the emitted energy and similar quantities, where a prime denotes an intermediate running value.

For this neutral contour calculation, we assume that the outer
branch shrinks as energy is emitted, equivalently
$\partial_{\Mcal}f<0$ at the running horizon.
Hamilton's equation replaces the radial momentum by the black hole
Hamiltonian ${\cal H}=\Mcal-\Ecal'$. The resulting imaginary part of
the action is
\begin{equation}
\operatorname{Im}I
=
\operatorname{Im}
\int_{r_{\rm in}}^{r_{\rm out}}
\int_{\Mcal}^{\Mcal-\Ecal}
\frac{d{\cal H}}
{\dot r(r;{\cal H},\lambda^A)}d r.
\label{eq:gen_KPW_action}
\end{equation}
The shell starts just inside the initial horizon and emerges
just outside the final one. The radial endpoints therefore approach
$r_{\rm in}=r_+$ and $r_{\rm out}=R_+(\Ecal)$, with
$r_{\rm out}<r_{\rm in}$ for the shrinking branch, despite the
outgoing character of the shell \cite{Parikh:1999mf}. At every
intermediate energy, the radial contour crosses $r=R_+(\Ecal')$.
The instantaneous Hawking temperature is
\begin{equation}
T_{\text{H}}(\Ecal')
=
\frac{1}{4\pi}
\partial_r f\bigl[R_+(\Ecal');\Mcal-\Ecal',\lambda^A\bigr].
\label{eq:gen_inst_T}
\end{equation}
At the running pole, this orientation and the positive-energy contour
prescription give
\[
\operatorname{Im}\int\frac{dr}{\dot r}
=-\frac{1}{2T_{\text H}(\Ecal')}.
\]
Combining this with $d{\cal H}=-d\Ecal'$ therefore gives
\begin{equation}
\operatorname{Im}I
=
\frac12\int_0^{\Ecal}
\frac{d\Ecal'}{T_{\text{H}}(\Ecal')}.
\label{eq:gen_KPW_T}
\end{equation}

At each $\Ecal'$, the black hole enclosed by the shell is identified
with a stationary member of the same family, to which an
instantaneous entropy and the stationary first law can be assigned.
The entropy $\Scal$ is the gravitational entropy entering that first
law and need not coincide with the area entropy in a general theory
\cite{Wald:1993nt,Iyer:1994ys}.

Let $\Qcal^a=\Qcal^a(\lambda^A)$, with $a=1,\ldots,n_q$, denote
the independent global charges determined by the parameters
$\lambda^A$, and let $\Phi_a$ be their conjugate potentials.
Keeping the couplings fixed, the first law at an intermediate stage
takes the form
\begin{equation}
d\Mcal'
=
T_{\text{H}}(\Ecal')d\Scal
+\sum_{a=1}^{n_q}\Phi_a(\Ecal')d\Qcal^a.
\label{eq:gen_stationary_1Law}
\end{equation}
 For the neutral radial channel, we assume that only the mass
varies among the independent solution parameters. Thus $\lambda^A$ remains fixed and $d\Qcal^a=0$, so
$d\Mcal'=T_{\text{H}}(\Ecal')d\Scal$ throughout the tunneling path. Since $d\Mcal'=-d\Ecal'$, the entropy difference is
\begin{align}
\Delta\Scal
={}&\Scal(\Mcal-\Ecal,\lambda^A)
-\Scal(\Mcal,\lambda^A)
={}-\int_0^{\Ecal}
\frac{d\Ecal'}{T_{\text{H}}(\Ecal')}.
\label{eq:gen_entropy_int}
\end{align}
Combining this expression with Eq.~\eqref{eq:gen_KPW_T} yields
the usual entropy form of the backreacting tunneling probability
\cite{Massar:1999wg,Sarkar:2007sx},
\begin{equation}
\operatorname{Im}I=-\frac12\Delta\Scal,
\qquad
\Gamma_{\rm KPW}\simeq\exp(\Delta\Scal).
\label{eq:gen_KPW_entropy}
\end{equation}
without assuming an
area law for $\Scal$, provided the stationary first law holds
along the tunneling path.

\subsection{Inversion of the horizon condition
\label{sec:gen_horizon_inversion}}

The finite-emission dependence of the KPW exponent can be recovered
from the horizon displacement once the entropy is parametrized along
the stationary family. This amounts to locally inverting the horizon equation.

At fixed $\lambda^A$, provided
$\partial_{\Mcal}f(r_+;\Mcal,\lambda^A)\neq0$, the implicit function
theorem allows the horizon equation to be solved locally in terms of a mass function
\begin{equation}
\Mcal=\mgot(r).
\label{eq:gen_local_mass}
\end{equation}
Thus $\Mcal=\mgot(r_+)$. Writing
$\Delta r_+(\Ecal):=R_+(\Ecal)-r_+$ for the finite displacement,
the final horizon position obeys
$\Mcal-\Ecal=\mgot(r_++\Delta r_+)$.
We reserve $\delta$ for infinitesimal variations and
perturbative corrections.
Defining
\begin{equation}
h(y):=\mgot(r_+)-\mgot(r_++y),
\label{eq:gen_h_definition}
\end{equation}
the displacement is determined by $\Ecal=h(\Delta r_+)$. Its local
inverse exists because
 \begin{equation}
\partial_y h(0)=-\partial_r\mgot(r_+)
=\frac{\partial_r f(r_+;\Mcal,\lambda^A)}
{\partial_{\Mcal}f(r_+;\Mcal,\lambda^A)}\neq0.
\label{eq:gen_inverse_slope}
\end{equation}
If $\mgot(r)$ is analytic near the initial simple horizon, the Lagrange--B\"urmann (LB) formula \cite{Gessel:2016Lagrange}
gives $\Delta r_+=\sum_{n\geq1}a_n\Ecal^n$, with
\begin{equation}
a_n=\frac{1}{n!}
\left[
\partial_y^{\,n-1}
\left(\frac{y}{h(y)}\right)^n
\right]_{y=0}.
\label{eq:gen_horizon_coefficients}
\end{equation}
Since $\partial_y h(0)\neq0$, the ratio $y/h(y)$ has a regular extension at
$y=0$.

Writing the entropy restricted to the stationary family as
$\Scal=\Scal(r,\lambda^A)$, the composed LB
formula, assuming this entropy is analytic near $r_+$, similarly gives
$\Delta\Scal=\sum_{n\geq1}b_n\Ecal^n$, where
\begin{equation}
b_n=\frac{1}{n!}
\left[
\partial_y^{\,n-1}
\left\{
\partial_y\Scal(r_++y,\lambda^A)
\left(\frac{y}{h(y)}\right)^n
\right\}
\right]_{y=0}.
\label{eq:gen_entropy_coefficients}
\end{equation}
These coefficients reconstruct the horizon displacement and
entropy exponent from the local mass--radius relation and gravitational
entropy, without requiring a closed-form horizon radius. The derivations
and explicit coefficient extractions are given in
Appendix~\ref{appx:Lagrange}. The inverse branch satisfying
$R_+(0)=r_+$ admits local continuation while the horizon remains
simple, although its Maclaurin series in $\Ecal$ about
$\Ecal=0$ need not converge throughout the emission interval. Independently of convergence, the construction
also holds for formal power series \cite{Sambale:2022fps}.

The construction separates the inversion from the specific form of
the entropy. In Einstein gravity with minimally coupled matter,
\begin{equation}
\Scal(r_+)
=\frac{\operatorname{Vol}(\Sigma_{D-2,k})}{4G_D}r_+^{D-2},
\label{eq:gen_Einstein_entropy}
\end{equation}
with the volume factor omitted for entropy densities. For $D=3$,
the entropy is linear in $r_+$ and the exponent is directly
proportional to the horizon displacement.

A particularly enlightening specialization of the above result occurs when the blackening
function is affine in the black hole mass with a constant coefficient,
\begin{equation}
f(r;\Mcal,\lambda^A)
=c\big[\mgot(r)-\Mcal\big],
\qquad c>0.
\label{eq:gen_affine_mass}
\end{equation}
The coefficient $c$ may depend on the fixed parameters but is
independent of $r$ and $\Mcal$. The geometric temperature and
the first law at fixed $\lambda^A$  then require
$d\Scal/dr_+=4\pi/c$. This is a consistency condition on the
chosen family and its gravitational entropy, including a Wald entropy.
It therefore applies directly to the three-dimensional Einstein
family considered in Sec.~\ref{sec:BTZ_laboratory}.
Writing $f_0(r):=f(r;\Mcal,\lambda^A)$ for the blackening
function of the initial state, neutral emission produces the constant shift
\begin{equation}
f(r;\Mcal-\Ecal,\lambda^A)=f_0(r)+c\Ecal.
\label{eq:gen_vertical_flow}
\end{equation}
The first inverse terms are
\begin{align}
\Delta r_+
={}&-\frac{c\Ecal}{\partial_r f_0(r_+)}
-\frac{c^2\partial_r^2f_0(r_+)}
{2[\partial_r f_0(r_+)]^3}\Ecal^2
+\Order(\Ecal^3).
\label{eq:gen_simple_horizon_shift}
\end{align}
Here $h(y)=-f_0(r_++y)/c$, so every higher order follows from application of
Eq.~\eqref{eq:gen_horizon_coefficients}.

\subsection{Thermal limit and nonthermal corrections
\label{sec:gen_thermal_corrections}}

The entropy dependence in Eq.~\eqref{eq:gen_KPW_entropy} also
admits a statistical-mechanical interpretation. For an isolated
composite system in equilibrium, the relative weight of a specified
radiation state is determined by the number of compatible states
available to the emitting body, giving the factor
$\exp(\Delta\Scal)$ \cite{Ryskin:2014wua}. The usual Boltzmann
factor follows upon linearizing the entropy change in the transferred
energy, while the quadratic correction accounts for the change in
the body's temperature due to its finite heat capacity.

Accordingly, expanding the entropy difference about the initial
configuration results in
\begin{equation}
\Delta\Scal
=-\frac{\Ecal}{T_{\text{H}}}
-\frac{\Ecal^2}{2C_{\Qcal}T_{\text{H}}^2}
+\Order(\Ecal^3),
\label{eq:gen_thermal_limit}
\end{equation}
where $T_{\text{H}}:=T_{\text{H}}(0)$ and
\begin{equation}
C_{\Qcal}
:=\left(\frac{\partial\Mcal}{\partial T_{\text{H}}}\right)_{\Qcal^a},
\qquad\forall\,\Qcal^a,
\label{eq:gen_heat_capacity}
\end{equation}
are evaluated at the initial state. Any remaining parameters in
$\lambda^A$ are also held fixed in the derivative defining
$C_{\Qcal}$. The first nonthermal correction therefore suppresses
emission relative to the initial temperature Boltzmann factor for
$C_{\Qcal}>0$ and enhances it for $C_{\Qcal}<0$, coinciding with the results in \cite{Ryskin:2014wua}. For a finite neutral emission, the suppression extends beyond this
expansion if $C_{\Qcal}>0$ throughout the simple outer branch.
The running temperature then decreases as energy is extracted, and
Eq.~\eqref{eq:gen_entropy_int} gives
$\Delta\Scal\leq-\Ecal/T_{\text{H}}$.

\subsection{Charged emission
\label{sec:gen_charged_channel}}

For a shell carrying one conserved $U(1)$ charge, the blackening
function depends explicitly on both the mass and electric charge,
\begin{equation*}
f=f(r;\Mcal,\Qcal,\lambda^A).
\end{equation*}
The remaining parameters $\lambda^A$, including charges associated
with other global symmetries and the couplings of the theory, are
held fixed. Let $\Qcal_{\rm em}$ be the signed charge transported by the
emitted shell. Emission of charge with the same sign as $\Qcal$
reduces the magnitude of the black hole charge, whereas oppositely
charged matter increases it. We first consider charge extraction within a fixed sign sector.
Opposite-sign emission is discussed in
Sec.~\ref{subsec:BTZNLED_outer_coalescence_exponent}.

An \emph{emission channel} connects $(\Mcal,\Qcal)$ to
$(\Mcal-\Ecal,\Qcal-\Qcal_{\rm em})$ through a quasistationary
path ${\cal C}$ in the space of black hole states. When the extracted
energy is strictly monotonic, this path can be parametrized by
$\Ecal'$ as
\begin{equation}
{\cal C}:\ [0,\Ecal]
\longrightarrow\mathbb{R}^2,\quad
\Ecal'
\longmapsto
\big(\Mcal-\Ecal',\Qcal-\Qcal_{\rm em}(\Ecal')\big).
\label{eq:gen_charged_running_state}
\end{equation}
Here, $\Qcal_{\rm em}(\Ecal')$ is a continuously differentiable
charge profile satisfying $\Qcal_{\rm em}(0)=0$ and
$\Qcal_{\rm em}(\Ecal)=\Qcal_{\rm em}$. The coordinates on
${\cal C}$ are denoted by $(\Mcal',\Qcal')$, and its image is
restricted to the admissible stationary branch at fixed $\lambda^A$.
At the endpoint, $\Mcal'(\Ecal)=\Mcal-\Ecal$ and
$\Qcal'(\Ecal)=\Qcal-\Qcal_{\rm em}(\Ecal)$, with
$\Mcal$ and $\Qcal$ referring to the initial state of the emission.
The running horizon now satisfies
\begin{equation}
f\bigl[R_+(\Ecal');\Mcal',\Qcal',\lambda^A\bigr]=0,
\qquad R_+(0)=r_+.
\label{eq:gen_charged_running_horizon}
\end{equation}
Conservation fixes the endpoints. It does not, by itself, determine
the charge profile $\Qcal_{\rm em}(\Ecal')$.

The gauge potential associated with this charge is
$A=A_\mu\dd x^\mu$. In the static, purely electric
sector considered here, it can be written as $A=A_t(r)\dd t$.
The potential $\Phi(r):=-A_t(r)$ and the conserved energy use the
same reference convention as the black hole Hamiltonian. With this
convention, the instantaneous horizon value $\Phi_+:=\Phi(R_+)$ is
the thermodynamic potential conjugate to $\Qcal$ along the radiation process. Both the potential
and the horizon position vary along ${\cal C}$.

For minimally coupled charged matter, the near-horizon radial phase
depends on the energy after the gauge work has been subtracted.
We adopt the quasistationary charged KPW prescription, which
subtracts the electrostatic work from the mass variation entering the
radial action \cite{Zhang:2005xt,Israel:2010gk}. Taking the residue
at the instantaneous outer horizon gives
\begin{equation}
\operatorname{Im}I
=-\frac12\int_{\cal C}
\frac{d\Mcal'-\Phi_+d\Qcal'}{T_{\text{H}}}.
\label{eq:gen_charged_KPW_residue}
\end{equation}
All quantities in this integral are evaluated at that horizon.
Other zeros of $f$, if present, are not swept by the contour for
emission across the exterior horizon. The charged horizon residue
and its integration along ${\cal C}$ are derived in
Appendix~\ref{appx:charge_transfer}, using a regular gauge at the
future horizon and the quasistationary KPW prescription.

By the stationary first law
$d\Mcal'=T_{\text{H}}d\Scal+\Phi_+d\Qcal'$ at fixed $\lambda^A$,
the one-form in Eq.~\eqref{eq:gen_charged_KPW_residue} is $d\Scal$.
Thus Eq.~\eqref{eq:gen_KPW_entropy} holds with
\begin{align}
\Delta\Scal
={}&\Scal(\Mcal-\Ecal,\Qcal-\Qcal_{\rm em},\lambda^A)
 -\Scal(\Mcal,\Qcal,\lambda^A),
\label{eq:gen_charged_KPW_entropy}
\end{align}
The leading horizon exponent depends only on the endpoints
for paths on the same admissible stationary branch.

In the parametrization of Eq.~\eqref{eq:gen_charged_running_state},
the running-horizon charged integral \eqref{eq:gen_charged_KPW_residue}  becomes
\begin{equation}
\operatorname{Im}I
=\frac12\int_0^{\Ecal}
\frac{1-\Phi_+(\Ecal')\,
\left( d\Qcal_{\rm em}(\Ecal')/d\Ecal'\right)}
{T_{\text{H}}(\Ecal')}d\Ecal'.
\label{eq:gen_charged_energy_integral}
\end{equation}
We restrict attention to ordinary positive horizon-energy channels,
whose extracted increments obey
$d\Ecal'-\Phi_+d\Qcal_{\rm em}>0$. They give a positive
imaginary action and a decreasing black hole entropy. For bosonic
modes, the opposite sign is associated with the superradiant sector,
where the exponential enters an emission-to-absorption ratio rather
than a standalone emission probability, see \cite{Vanzo:2011wq}.

The local mass function now becomes $\Mcal=\mgot(r,\Qcal)$,
with the remaining $\lambda^A$ suppressed. The final horizon
satisfies
\begin{equation}
\Mcal-\Ecal
=\mgot(r_++\Delta r_+,\Qcal-\Qcal_{\rm em}).
\label{eq:gen_charged_final_mass}
\end{equation}
 The radial displacement follows by inverting the mass--radius relation
at the final charge $\Qcal-\Qcal_{\rm em}$. This gives a family of inverses
parametrized by the charge transfer, whose evaluation along the chosen profile $\Qcal_{\rm em}(\Ecal)$ determines the evolving horizon. To retain a function vanishing at $y=0$,
define, analogously,
\begin{align}
h(y;\Qcal_{\rm em})
:={}&\mgot(r_+,\Qcal-\Qcal_{\rm em})
-\mgot(r_++y,\Qcal-\Qcal_{\rm em}).
\label{eq:gen_charged_h_definition}
\end{align}
The final horizon equation is then equivalent to
\begin{equation}
h(\Delta r_+;\Qcal_{\rm em})
=\Ecal+\mgot(r_+,\Qcal-\Qcal_{\rm em})-\Mcal.
\label{eq:gen_charged_inverse_problem}
\end{equation}
This choice of reference mass makes the neutral inversion formula applicable without assuming an explicit relation between emitted
energy and charge.
For a jointly analytic mass function with
$\partial_r\mgot\neq0$, the inverse coefficients depend analytically on the charge transfer $\Qcal_{\rm em}$
near $\Qcal_{\rm em}=0$. The resulting horizon and
entropy coefficients are worked out in Appendix~\ref{appx:Lagrange}.
If $\Scal(r,\Qcal)$ has explicit charge dependence at fixed radius,
the entropy change also contains that variation. Of course, such a contribution vanishes for the standard Einstein area entropy.

At first order, differentiation of the horizon condition gives
\begin{equation}
\Delta r_+
=\frac{(\partial_{\Mcal}f)\,\Ecal+
(\partial_{\Qcal}f)\,\Qcal_{\rm em}}{\partial_r f}
+\cdots,
\label{eq:gen_charged_linear_shift}
\end{equation}
with derivatives at the initial horizon and omitted terms of total
degree at least two. When $\Scal=\Scal(r)$ and  $\partial_r\Scal(r)>0$, the first law gives
$T_{\text H}(\partial_r\Scal)(R_+)\,dR_+
=-(d\Ecal'-\Phi_+d\Qcal_{\rm em})<0$
for an ordinary channel. Thus the outer horizon shrinks due to the emission, although the charge contribution may oppose the mass contribution.
With explicit charge dependence in the entropy, its decrease alone
does not determine the sign of the radial displacement.

A joint Taylor expansion about the initial state expresses
the quadratic correction in terms of the entropy Hessian
\cite{Aman:2003ug},
\begin{align}
\Delta\Scal
={}&-\frac{\Ecal-\Phi_+\Qcal_{\rm em}}{T_{\text H}}
+\frac12\Big[
(\partial_{\Mcal}^2\Scal)\,\Ecal^2
+2(\partial_{\Mcal}\partial_{\Qcal}\Scal)\,
\Ecal\Qcal_{\rm em}
\nonumber\\
&
+(\partial_{\Qcal}^2\Scal)\,\Qcal_{\rm em}^2
\Big]+\cdots .
\label{eq:gen_charged_entropy_Hessian}
\end{align}
 All derivatives, $T_{\text H}$, and $\Phi_+$ here refer to the
 initial state. The omitted terms have total degree at least three.
The linear term is the thermal exponent with its chemical potential
arising from the electromagnetic sector.

For the case of a linear charge profile
$\Qcal_{\rm em}(\Ecal)=\eta\Ecal$, substitution in
Eq.~\eqref{eq:gen_charged_entropy_Hessian} gives the correction
along the emission direction. As $\eta\rightarrow0$, this reduces
to Eq.~\eqref{eq:gen_thermal_limit}, since
$\partial_{\Mcal}^2\Scal=-1/(C_{\Qcal}T_{\text H}^2)$.
The mixed and charge derivatives also enter for charged emission,
so $C_{\Qcal}$ alone does not determine the sign.

\begin{samepage}
A smooth nonlinear profile with the same initial slope
$\left.d\Qcal_{\rm em}/d\Ecal\right|_0=\eta$
adds the following quadratic contribution to the
linear-profile expansion:
$$
\frac{\Phi_+}{2T_{\text H}}
(\partial_\Ecal^2\Qcal_{\rm em})(0)\Ecal^2.
$$
\par
\end{samepage} This contribution
is already included when the thermal exponent of the initial state
$-(\Ecal-\Phi_+\Qcal_{\rm em}(\Ecal))/T_{\text H}$ retains the
full charge profile.

The neutral and charged prescriptions above apply while the outer
horizon remains simple. If this branch approaches a degenerate
configuration, the entropy difference may still have a finite limit.
The change in the local inverse at degeneracy, and its consequences
for the other horizon branches, are examined next.

\section{Horizon splitting and coalescence under emission
\label{sec:degenerate_unfolding}}

\begin{figure}[t]
\centering
\includegraphics[width=\textwidth]{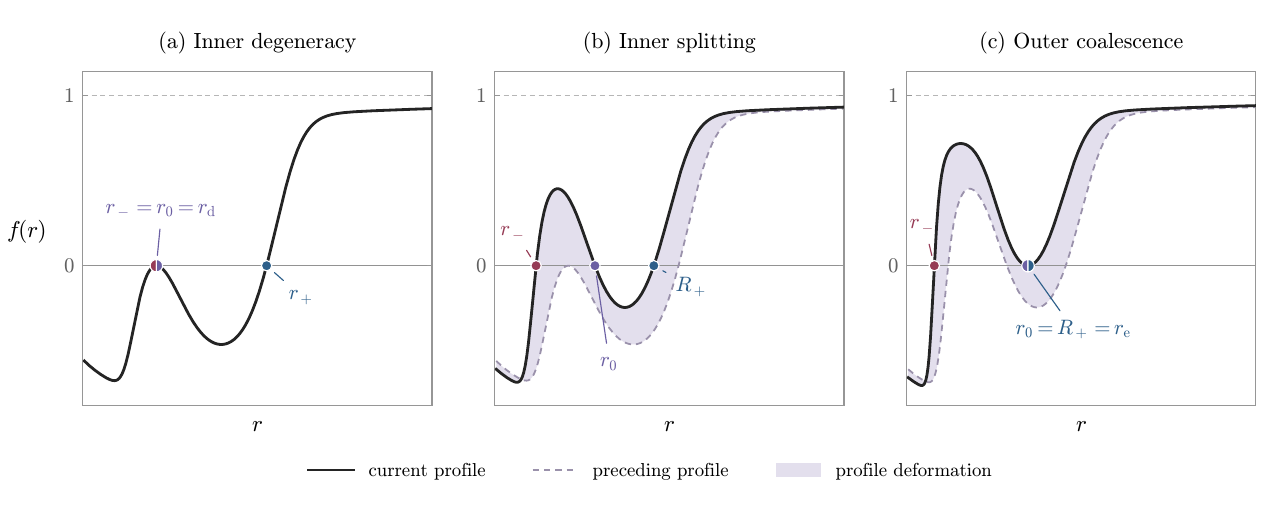}
\caption{Schematic horizon motion and rearrangement along an illustrative charged
emission channel. (a) An internal double zero at $r_{\rm d}$ coexists
with a simple event horizon at $r_+$.
(b) The inner pair separates immediately after emission, giving three simple roots
$r_-<r_0<R_+$.
(c) The middle and event horizons merge at $r_{\rm e}$, while
$r_-$ remains simple. Red, violet, and blue markers follow the
innermost, middle, and event horizons, respectively, with two-color
markers indicating double zeros.
In panels (b) and (c), the dashed curve is the preceding profile
and the shaded region lies between the two curves.
The event-horizon slope is positive in panels (a,b) and vanishes
at the merger in panel (c). The asymptotic form shown is
illustrative; the horizon sequence depends on the stationary family,
initial state, and chosen emission channel.}
\label{fig:general_horizon_deformation}
\end{figure}

At a degenerate horizon, the linear term in the radial
displacement vanishes, so the ordinary inverse of
Sec.~\ref{sec:horizon_kinematics} must be replaced by a Puiseux-type
expansion. We use it to determine the nearby real roots, when they
exist, and how their separation depends on the emission channel.
A sequence of quasistationary emissions can change both the
positions and the number of horizons. More precisely, an initially degenerate inner
horizon may split into two simple zeros, one of which can later
merge with the event horizon. Figure~\ref{fig:general_horizon_deformation}
illustrates one such sequence, allowing the profile of $f(r)$ to
deform as energy and charge are lost.

\subsection{Neutral splitting under a uniform mass shift
\label{sec:degenerate_neutral_unfolding}}

A good starting point is the affine mass dependence of
Eq.~\eqref{eq:gen_affine_mass}, for which neutral emission produces
the uniform shift
$f(r;\Mcal-\Ecal,\lambda^A)=f_0(r)+c\,\Ecal$, with
all parameters $\lambda^A$ held fixed. Let $r=r_{\rm d}$ be a double
zero of the initial blackening function,
\begin{equation}
f_0(r_{\rm d})
=\partial_r f_0(r_{\rm d})=0,
\qquad
\partial_r^2f_0(r_{\rm d})\neq0.
\label{eq:degenerate_horizon}
\end{equation}
The local analysis applies to any such zero. The causal character
of the associated horizons depends on the global geometry. In
particular, an internal double horizon may coexist with a simple
event horizon through which the tunneling process takes place.

Writing the displaced root at the final stage as
$r_{{\rm d},f}:=r_{\rm d}+\Delta r_{\rm d}$, its horizon condition
becomes
\begin{equation}
0=c\Ecal
+\frac{\partial_r^2f_0(r_{\rm d})}{2}(\Delta r_{\rm d})^2
+\frac{\partial_r^3f_0(r_{\rm d})}{6}(\Delta r_{\rm d})^3
+\Order((\Delta r_{\rm d})^4).
\label{eq:degenerate_local_equation}
\end{equation}
The absence of a linear term prevents a Maclaurin expansion
of the inverse about $\Ecal=0$. The leading quadratic equation has
discriminant $-2c\Ecal\,\partial_r^2f_0(r_{\rm d})$. For
sufficiently small $\Ecal>0$, a positive discriminant gives
two distinct real roots, whereas a negative one gives none. Thus, under a decrease of the
physical mass due to matter tunneling out of the black hole, a
double zero at a local maximum of $f_0$ will unfold into two real horizons,
whereas a double zero at a local minimum has no nearby real
continuation in the same direction. A vanishing second derivative
requires the higher-multiplicity analysis developed in
Appendix~\ref{appx:Puiseux}.

In the splitting sector,
\begin{equation}
\partial_r^2f_0(r_{\rm d})<0.
\label{eq:degenerate_splitting_sector}
\end{equation}
Assuming that $f_0$ is analytic near $r_{\rm d}$, each branch admits
a locally convergent Puiseux expansion
\cite{Puiseux:1850,Sambale:2022fps} of the form
\begin{equation}
\Delta r_{\rm d}^{(\epsilon)}(\Ecal)
=\sum_{n=1}^{\infty}a_n^{(\epsilon)}\Ecal^{n/2},
\qquad\epsilon=\pm1.
\label{eq:degenerate_puiseux_series}
\end{equation}
The half-integer powers follow from the multiplicity of the initial
zero. Applying the LB construction to the ordinary
inverse in the signed variable $\epsilon\sqrt{\Ecal}$ gives the
coefficients, with $a_n^{(-)}=(-1)^n a_n^{(+)}$. The integer powers
of $\Ecal$ therefore produce a common displacement, while the
half-integer powers separate the branches. The coefficient formulas
and their extension to arbitrary multiplicity are developed in
Appendix~\ref{appx:Puiseux}.

The first few terms are
\begin{align}
\Delta r_{\rm d}^{(\epsilon)}(\Ecal)
={}&\epsilon
\sqrt{-\frac{2c\Ecal}{\partial_r^2f_0(r_{\rm d})}}
+\frac{c\,\partial_r^3f_0(r_{\rm d})}
{3[\partial_r^2f_0(r_{\rm d})]^2}\Ecal
+\Order(\Ecal^{3/2}).
\label{eq:degenerate_first_terms}
\end{align}
The two signs identify simultaneous zeros of the same final
geometry, continuous in radius but nonanalytic in emitted energy.
The common term linear in $\Ecal$ cancels from their separation,
\begin{equation}
r_{{\rm d},f}^{(+)}-r_{{\rm d},f}^{(-)}
=2\sqrt{-\frac{2c\Ecal}{\partial_r^2f_0(r_{\rm d})}}
+\Order(\Ecal^{3/2}).
\label{eq:degenerate_root_separation}
\end{equation}
At these roots, the radial slopes of the final blackening function
behave as
\begin{align}
&\partial_r f\left(r_{{\rm d},f}^{(\epsilon)};
\Mcal-\Ecal,\lambda^A\right)
=-\epsilon\sqrt{-2c\Ecal\,\partial_r^2f_0(r_{\rm d})}
+\Order(\Ecal).
\label{eq:degenerate_unfolded_slopes}
\end{align}
Therefore, immediately after a sufficiently small emission of
energy, the originally degenerate horizon splits into two
simple horizons, with positive slope at the smaller radius and
negative slope at the larger one, while their surface gravities have
the same leading magnitude, proportional to $\sqrt{\Ecal}$.

\subsection{Charged emission and the critical mass curve
\label{sec:degenerate_charged_unfolding}}

Charge variation changes not only the level of $f$ but also its shape. Along the chosen emission path
${\cal C}$ in Eq.~\eqref{eq:gen_charged_running_state}, the mass and charge vary together, with the cumulative emitted charge
$\Qcal_{\rm em}(\Ecal')$ specified by the emission channel. To follow a double zero under this variation, suppose that
$f(r;\Mcal,\Qcal,\lambda^A)$ is jointly analytic near the initial
degenerate horizon and that $\partial_{\Mcal}f\neq0$ there.
The local mass function $\Mcal=\mgot(r,\Qcal)$ introduced
around Eq.~\eqref{eq:gen_charged_final_mass} is then well defined.
At the double zero, $\partial_r\mgot=0$ and
$\partial_r^2\mgot=-\partial_r^2f/\partial_{\Mcal}f\neq0$.
The remaining $\lambda^A$ are held fixed and suppressed below.

For each running value of the charge $\Qcal'$, we first determine the mass and
radius at which a double zero continuously connected to the initial
one occurs. We denote this radius by $r_{\rm d}(\Qcal')$ and the
corresponding critical mass by $\Mcal_{\rm d}(\Qcal')$, defined by
\begin{align}
\Mcal_{\rm d}(\Qcal')
&:=\mgot\bigl[r_{\rm d}(\Qcal'),\Qcal'\bigr],
\nonumber\\
\partial_r\mgot\bigl[r_{\rm d}(\Qcal'),\Qcal'\bigr]&=0.
\label{eq:degenerate_charged_critical_curve}
\end{align}
Since $\partial_r^2\mgot\neq0$ at the initial double zero, these
functions are analytic near $\Qcal$, with
$r_{\rm d}(\Qcal)=r_{\rm d}$ and
$\Mcal_{\rm d}(\Qcal)=\Mcal$.
The emission path generally leaves this critical curve.
At the emission endpoint, where
$\Qcal'(\Ecal)=\Qcal-\Qcal_{\rm em}$, define the signed mass
difference
\begin{equation}
\Delta\Mcal_{\rm d}
:=\Mcal_{\rm d}(\Qcal-\Qcal_{\rm em})-(\Mcal-\Ecal).
\label{eq:degenerate_charged_mass_offset}
\end{equation}
It vanishes on the critical curve and is positive when the actual
mass lies below it.
Writing
$r=r_{\rm d}[\Qcal'(\Ecal)]+\Delta r_{\rm d}$, with
$\Delta r_{\rm d}$ now measured from the critical radius at the
final charge, the horizon condition
$\Mcal-\Ecal=\mgot[r,\Qcal'(\Ecal)]$ becomes
\begin{equation}
0=\Delta\Mcal_{\rm d}
+\frac{\partial_r^2\mgot}{2}(\Delta r_{\rm d})^2
+\frac{\partial_r^3\mgot}{6}(\Delta r_{\rm d})^3
+\Order((\Delta r_{\rm d})^4).
\label{eq:degenerate_charged_local_equation}
\end{equation}
All radial derivatives here and in the following local expansion
are evaluated at
\[
\bigl[r_{\rm d}[\Qcal'(\Ecal)],\Qcal'(\Ecal)\bigr].
\]
The leading quadratic discriminant is therefore
\begin{equation}
-2\left(\partial_r^2\mgot\right)\,\Delta\Mcal_{\rm d}.
\label{eq:degenerate_charged_discriminant}
\end{equation}
For sufficiently small nonzero $\Delta\Mcal_{\rm d}$, a positive
discriminant gives two nearby real roots, whereas a negative one
gives none. At $\Delta\Mcal_{\rm d}=0$, the double zero persists
at $r_{\rm d}[\Qcal'(\Ecal)]$.

On the real-root side, the first terms are
\begin{align}
r_{{\rm d},f}^{(\epsilon)}
={}&r_{\rm d}[\Qcal'(\Ecal)]
+\epsilon\sqrt{-\frac{2\Delta\Mcal_{\rm d}}{\partial_r^2\mgot}}
+\frac{\partial_r^3\mgot}{3(\partial_r^2\mgot)^2}
\Delta\Mcal_{\rm d}
+\Order\!\left(|\Delta\Mcal_{\rm d}|^{3/2}\right),
\label{eq:degenerate_charged_first_terms}
\end{align}
with $\epsilon=\pm1$. Charge variation thus moves the critical
radius about which the two branches separate and changes their inverse coefficients. Their separation begins at
$2\sqrt{-2\Delta\Mcal_{\rm d}/\partial_r^2\mgot}$.
At a zero, $\partial_r f=-(\partial_{\Mcal}f)\,\partial_r\mgot$, hence
the two slopes have opposite signs and the same leading
magnitude, proportional to $|\Delta\Mcal_{\rm d}|^{1/2}$. For the
affine family with $c>0$ and $\partial_r^2\mgot<0$, the smaller radius
branch has positive slope, just as in the neutral unfolding.
For the internal pair in
Fig.~\ref{fig:general_horizon_deformation}(b), the signs
$\epsilon=-1$ and $\epsilon=+1$ identify $r_-$ and $r_0$,
respectively.

At each fixed charge, the inversion therefore follows the
neutral construction, with the displacement measured from
$r_{\rm d}[\Qcal'(\Ecal)]$. Its coefficients and the derivatives
of the critical curve are collected in Appendix~\ref{appx:Puiseux}.

To express this unfolding along ${\cal C}$, suppose that its charge
profile is analytic near $\Ecal'=0$.
The mass difference along the path has the expansion
\begin{equation}
\Delta\Mcal_{\rm d}(\Ecal)
=\left(1-\eta\frac{d\Mcal_{\rm d}}{d\Qcal}\right)\Ecal
+\Order(\Ecal^2),
\label{eq:degenerate_charged_path_offset}
\end{equation}
Here $\eta=(d\Qcal_{\rm em}/d\Ecal)|_0$, and the
critical derivatives are evaluated at the initial state. Three
local cases can be distinguished:
\begin{enumerate}
\renewcommand{\theenumi}{\roman{enumi}}
\renewcommand{\labelenumi}{(\theenumi)}
\item\label{case:critical_transverse_split}
 If $(1-\eta\,d\Mcal_{\rm d}/d\Qcal)\partial_r^2\mgot<0$,
the path enters the side with two real roots. Their separation
is proportional to $\Ecal^{1/2}$.

\item\label{case:critical_tangent}
If $1-\eta\,d\Mcal_{\rm d}/d\Qcal=0$, the path is tangent
to the critical curve. The first nonzero higher-order mass
difference decides whether a real pair is present. If it occurs
at order $\Ecal^p$ and has the sign required by
Eq.~\eqref{eq:degenerate_charged_discriminant}, the separation
is proportional to $\Ecal^{p/2}$; a quadratic departure gives
a separation of order $\Ecal$.

\item\label{case:critical_transverse_absent}
If $(1-\eta\,d\Mcal_{\rm d}/d\Qcal)\partial_r^2\mgot>0$,
the nearby real pair is absent immediately after emission.
\end{enumerate}
Initial tangency alone does not ensure that the double
horizon persists. The degeneracy is preserved only if
$\Mcal'=\Mcal_{\rm d}(\Qcal')$ throughout the path.
The higher coefficients are given in Appendix~\ref{appx:Puiseux}.

 When the charge derivative of the mass function satisfies
$\partial_{\Qcal}\mgot(r,\Qcal)=\Phi(r)$ in the potential
convention of Sec.~\ref{sec:gen_charged_channel}, the critical
slope equals the electric potential at the degenerate horizon,
\begin{equation}
\frac{d\Mcal_{\rm d}}{d\Qcal}=\Phi_{\rm d},
\qquad
\Phi_{\rm d}:=\Phi\bigl[r_{\rm d}(\Qcal)\bigr].
\label{eq:degenerate_critical_potential}
\end{equation}
Then the initial departure from the critical curve is 
governed by $1-\eta\Phi_{\rm d}$, whereas ordinary emission
through the simple event horizon requires $1-\eta\Phi_+>0$.
Since the two potentials are evaluated at different radii, this
condition alone does not determine whether the internal pair
splits or disappears.

\subsection{From splitting to outer coalescence
\label{sec:degenerate_outer_coalescence}}

 After the internal double horizon splits, the larger of
the two resulting roots can merge with the event horizon, leaving
the innermost horizon simple, as illustrated in
Fig.~\ref{fig:general_horizon_deformation}(b)-(c). The splitting and outer merger occur
at different extrema of the stationary mass--radius relation.

For the neutral affine family, suppose that $\mgot(r)$ has
an internal local maximum at $r_{\rm d}$ and an adjacent outer
local minimum at $r_{\rm e}>r_{\rm d}$. Denote the two
critical masses by $\Mcal_{\rm i}$ and $\Mcal_{\rm e}$, and their
positive difference by the critical energy $\Ecal_c$
\begin{align}
\Mcal_{\rm i}:=\mgot(r_{\rm d}),
\qquad
\Mcal_{\rm e}:=\mgot(r_{\rm e}), \qquad
\Ecal_c&=\Mcal_{\rm i}-\Mcal_{\rm e}>0.
\label{eq:degenerate_critical_levels}
\end{align}
Taking $\Mcal=\Mcal_{\rm i}$ as the initial mass, assume
that the double internal horizon coexists with a separate simple
event horizon at $r_+>r_{\rm e}$, and that these are the only positive
zeros. Neutral emission lowers the mass to
$\Mcal'=\Mcal_{\rm i}-\Ecal'$, initially splitting the internal
double zero into two simple horizons. Provided these three branches
persist without other horizon transitions up to the critical
energy, the larger of the internal pair
approaches the event horizon until they merge at $\Ecal'=\Ecal_c$,
while the innermost horizon remains simple.

Let $f_c(r):=f_0(r)+c\,\Ecal_c$ denote the limiting blackening
function. The outer coalescence occurs at
\begin{equation}
f_c(r_{\rm e})=\partial_r f_c(r_{\rm e})=0,
\qquad
\partial_r^2 f_c(r_{\rm e})>0.
\label{eq:degenerate_outer_endpoint}
\end{equation}
For $\Ecal<\Ecal_c$, the horizon equation is
$f_c(r)-c(\Ecal_c-\Ecal)=0$. The outer member of the pair obeys
\begin{equation}
R_+(\Ecal)
=r_{\rm e}
+\sqrt{\frac{2c}{\partial_r^2 f_c(r_{\rm e})}}
(\Ecal_c-\Ecal)^{1/2}
+\Order(\Ecal_c-\Ecal).
\label{eq:degenerate_outer_radius}
\end{equation}
The other member has the opposite leading radial displacement.
Their separation vanishes as $(\Ecal_c-\Ecal)^{1/2}$, and the
instantaneous Hawking temperature behaves as
\begin{equation}
T_{\text{H}}(\Ecal)
=\frac{\sqrt{2c\,\partial_r^2 f_c(r_{\rm e})}}{4\pi}
(\Ecal_c-\Ecal)^{1/2}
+\Order(\Ecal_c-\Ecal).
\label{eq:degenerate_outer_temperature}
\end{equation}
The inverse temperature is integrable at this endpoint. Using again
$d\Scal/d r_+=4\pi/c$ for the affine family gives
\begin{align}
\Delta\Scal(\Ecal)
={}&\Delta\Scal(\Ecal_c)
 +4\pi\sqrt{\frac{2}{c\,\partial_r^2 f_c(r_{\rm e})}}
(\Ecal_c-\Ecal)^{1/2}
+\Order(\Ecal_c-\Ecal).
\label{eq:degenerate_outer_entropy}
\end{align}
Here $\Delta\Scal(\Ecal_c)$ denotes the limit from simple outer
horizons. The next radial correction is given in
Appendix~\ref{appx:Puiseux_outer_endpoint}.

For charged emission, the outer critical level becomes a curve
$\Mcal_{\rm e}(\Qcal')=\mgot\bigl[r_{\rm e}(\Qcal'),\Qcal'\bigr]$,
defined by the same stationary point construction, now at the mass
minimum. In the affine charged family with $c>0$, the outer pair
exists on the side $\Mcal'>\Mcal_{\rm e}(\Qcal')$. Along
${\cal C}$, the mass difference appropriate to this branch is
\begin{equation}
\Delta\Mcal_{\rm e}(\Ecal')
:=\Mcal_{\rm e}\bigl[\Qcal-\Qcal_{\rm em}(\Ecal')\bigr]
-\Mcal+\Ecal'.
\label{eq:degenerate_charged_outer_offset}
\end{equation}
It is negative before coalescence and vanishes when the path reaches
the outer critical curve. Applying
Eq.~\eqref{eq:degenerate_charged_first_terms} at the minimum gives
a horizon separation of order $|\Delta\Mcal_{\rm e}|^{1/2}$.
The temperature has the same leading order at a regular double
outer zero. The relation between this mass difference and the remaining
emitted energy depends on the channel: a transverse approach gives
the square-root dependence found above, whereas a tangent approach
can change that power.

The charged limiting action is obtained from the full residue in
Eq.~\eqref{eq:gen_charged_energy_integral}, retaining the gauge electric work
factor $1-\Phi_+\left( d\Qcal_{\rm em}/d\Ecal'\right)$. For either channel,
a finite limiting entropy gives a finite leading tunneling exponent.
This limit is taken through simple outer horizons. Radiation from
an initially degenerate geometry requires a separate treatment
\cite{Angheben:2005rm}.

The nonlinearly charged BTZ family examined in
Sec.~\ref{sec:BTZ_laboratory} provides an explicit realization of these critical levels and the associated horizon rearrangement.

\section{Finite-energy emission from a nonlinearly charged BTZ black hole
\label{sec:BTZ_laboratory}}

The exact solution presented in Ref.~\cite{Mendez-Zavaleta:2026rgg}
is an extension of the BTZ geometry \cite{Banados:1992wn} through nonlinear
electrodynamics.
Finite-energy tunneling and the near-extremal regime were
studied for the standard charged BTZ black hole in
Ref.~\cite{Medved:2001ca}. Here the nonlinear matter permits a
degenerate inner horizon to coexist with a simple event horizon,
providing a setting for the distinct transitions described above.
This broader classical family contains, as a particular
member, the quantum-backreacted BTZ geometry of New Massive
Gravity (NMG) obtained to leading order in backreaction
in Ref.~\cite{Chernicoff:2024dll},
whose horizon emission is examined
in Sec.~\ref{sec:BTZNLED_semiclassical_emission}.
An explicit blackening function allows the horizons to be
followed along neutral and charged emission paths, including
the intervals between encounters with the critical curves.

\subsection{Tunneling probability from a two-horizon configuration}

The solution arises in three-dimensional Einstein gravity with a
negative cosmological constant $\Lambda=-1/\ell^2$, coupled to
nonlinear electrodynamics. We describe the supporting matter in
Plebanski's $\mathcal{H}(\mathcal{P})$ formulation
\cite{Plebanski1968}, with a gauge potential $A_\mu$ and an
independent 2-form with components $P^{\mu\nu}$. The structural function of the electrodynamics
depends on $\mathcal{P}=\frac14P_{\mu\nu}P^{\mu\nu}$, and the
field strength is $F_{\mu\nu}=2\partial_{[\mu}A_{\nu]}$.
The supporting structural function, first-order action, and field
equations are collected in Appendix~\ref{appx:BTZNLED_Plebanski}.
This description provides the matter realization throughout $r>0$.
The resulting spacetime is described by
the static metric
\begin{equation}
\dd s^2
=
-f(r)\dd t^2
+
\frac{\dd r^2}{f(r)}
+
r^2\dd\varphi^2,
\qquad
\varphi\sim\varphi+2\pi,
\label{eq:static_3D_metric}
\end{equation}
with
\begin{align}
f(r)
={}&
\frac{r^2}{\ell^2}
-
M
-
\frac{2\sqrt{2}\kappa q r}{\gamma}
\ln\left(
\frac{\beta\gamma r}{\sqrt{2}q}
\right),
\label{eq:BTZNLED_blackening}
\\
A
={}&
\frac{\sqrt{2}r}{\gamma}
\left[
\ln\left(
\frac{\beta\gamma r}{\sqrt{2}q}
\right)-1
\right]\dd t.
\label{eq:BTZNLED_gauge_potential}
\end{align}
The electric contribution in
Eq.~\eqref{eq:BTZNLED_blackening} contains an additional radial
factor multiplying the logarithm compared with the Maxwell-charged
BTZ solution \cite{Martinez:1999qi}.
Here, $\kappa=8\pi G_3$, $M$ and $q$ are integration constants,
whereas $\beta$ and $\gamma$ are couplings of the nonlinear
electromagnetic theory. The real configurations considered below require
\begin{equation}
\frac{\beta\gamma}{q}>0.
\label{eq:BTZNLED_reality_condition}
\end{equation}
The multihorizon configurations used here have
$q/\gamma>0$ and $\beta>0$,
where the mass--radius relation can develop both a local maximum
and a local minimum. Horizon-adapted representations and their
extremal limits are derived in Appendix~\ref{appx:BTZ_NLED}.

We shall focus on the critical two-horizon configuration identified in
Appendix~\ref{appx:BTZ_NLED}, in which a degenerate inner Killing
horizon at $r=r_-$ coexists with a simple event horizon at $r=r_+$,
with
\begin{equation*}
0<r_-<r_+.
\end{equation*}
At $r=r_-$, the initial blackening function has a local maximum that
touches zero, whereas $r=r_+$ is a simple outer zero with positive
slope. The corresponding horizon-adapted representation is given by
Eq.~\eqref{eq:f_2hor}, supplemented by the degeneracy condition
$\partial_r f_0(r_-)=0$. During emission we vary $M$ and, for charged shells, $q$ in the
original blackening function, keeping $(\kappa,\beta,\gamma,\ell)$
fixed.

 We begin with neutral emission from this state and then
allow the charge to vary. Throughout the analysis, $r_+$ denotes
the initial event-horizon radius.

The physical mass, electric charge, and entropy are
\begin{equation}
\Mcal
=
\frac{\pi M}{\kappa},
\qquad
\Qcal
=
2\pi q,
\qquad
\Scal
=
\frac{4\pi^2}{\kappa}r_+.
\label{eq:BTZNLED_physical_quantities}
\end{equation}
The Hawking temperature and electric potential in the gauge
\eqref{eq:BTZNLED_gauge_potential} are
\begin{align}
T_{\text H}
={}&
\frac{\partial_r f(r_+)}{4\pi}
={}
\frac{r_+}{2\pi\ell^2}
-
\frac{\sqrt{2}\kappa q}{2\pi\gamma}
\left[
\ln\left(
\frac{\beta\gamma r_+}{\sqrt{2}q}
\right)+1
\right],
\label{eq:BTZNLED_temperature}
\\
\Phi_e
={}&
-A_t(r_+)
={}
\frac{\sqrt{2}r_+}{\gamma}
\left[
1-
\ln\left(
\frac{\beta\gamma r_+}{\sqrt{2}q}
\right)
\right].
\label{eq:BTZNLED_electric_potential}
\end{align}
The instantaneous value of $\Phi_e$ is the horizon potential
$\Phi_+$ used in Sec.~\ref{sec:gen_charged_channel}.

 The thermodynamic description provided in Appendix~\ref{appx:termo}
follows the standard approach with cosmological constant and nonlinear electromagnetic
couplings fixed. The Hamiltonian variation then identifies $\Mcal$
as the internal energy exchanged with the shell. This is the
fixed-coupling restriction of the extended phase-space treatment in
Ref.~\cite{Mendez-Zavaleta:2026rgg}, where the mass has an enthalpy
interpretation. The first law reads
\begin{equation}
\dd\Mcal
=
T_{\text H}\dd\Scal
+
\Phi_e\dd\Qcal,
\label{eq:BTZNLED_first_law}
\end{equation}
together with the corresponding Smarr relation
\begin{equation}
\Mcal
=
\frac12
\left(
T_{\text H}\Scal
+
\Phi_e\Qcal
\right).
\label{eq:BTZNLED_Smarr}
\end{equation}

The horizon equation takes the mass--radius form introduced in
Secs.~\ref{sec:gen_horizon_inversion} and
\ref{sec:gen_charged_channel}, with the charge argument written as
$q=\Qcal/(2\pi)$ and the identification
\begin{equation}
\mgot(r,q)
:=
\frac{\pi}{\kappa}
\left[
\frac{r^2}{\ell^2}
-
\frac{2\sqrt{2}\kappa q r}{\gamma}
\ln\left(
\frac{\beta\gamma r}{\sqrt{2}q}
\right)
\right].
\label{eq:BTZNLED_mass_radius}
\end{equation}
Equivalently, the blackening function takes the affine form
\begin{equation}
f(r;\Mcal,q)
=
\frac{\kappa}{\pi}
\left[
\mgot(r,q)-\Mcal
\right].
\label{eq:BTZNLED_affine_mass}
\end{equation}

For the emission of a neutral shell, $(q,\beta,\gamma,\ell)$ remain
fixed and the mass parameters change according to
\begin{equation}
\Mcal
\longmapsto
\Mcal-\Ecal,
\qquad
M
\longmapsto
M-\frac{\kappa}{\pi}\Ecal.
\label{eq:BTZNLED_mass_loss}
\end{equation}
The final horizon equation therefore becomes
\begin{equation}
f(r;\Mcal-\Ecal,q)
=
f_0(r)
+
\frac{\kappa}{\pi}\Ecal
=
0.
\label{eq:BTZNLED_final_horizon_equation}
\end{equation}

For a charged shell, the mass changes as in
Eq.~\eqref{eq:BTZNLED_mass_loss}, while the integration constant $q$
varies along the channel ${\cal C}$ of
Sec.~\ref{sec:gen_charged_channel}. Its running value and the
corresponding event horizon satisfy
\begin{equation}
\begin{aligned}
q(\Ecal')
=q-\frac{\Qcal_{\rm em}(\Ecal')}{2\pi}, \quad
\mgot\bigl[R_+(\Ecal'),q(\Ecal')\bigr]=\Mcal-\Ecal',
\quad R_+(0)=r_+.
\end{aligned}
\label{eq:BTZNLED_charged_running_horizon}
\end{equation}
The charge profile is specified by the emission channel, with the
running states restricted by Eq.~\eqref{eq:BTZNLED_reality_condition}.
Charge variation therefore changes the radial profile of $f$ in
addition to the mass shift in
Eq.~\eqref{eq:BTZNLED_final_horizon_equation}.

Since the entropy is linear in the event horizon radius and has no
explicit charge dependence, its displacement determines the tunneling
exponent for either channel while the instantaneous event horizon
remains simple,
\begin{equation}
\ln\Gamma_{\rm KPW}\simeq\Delta\Scal
=\frac{4\pi^2}{\kappa}\,\Delta r_+(\Ecal).
\label{eq:BTZNLED_probability_radius}
\end{equation}
For neutral emission, $q(\Ecal')=q$ and the local series about the
initial simple event horizon gives
\begin{equation}
\ln\Gamma_{\rm KPW}
\simeq
\frac{4\pi^2}{\kappa}
\sum_{n=1}^{\infty}a_n\Ecal^n.
\label{eq:gen_KPW_series}
\end{equation}
Within the convergence domain of the inverse, this series
reconstructs the finite-emission exponent. For the initial two-horizon
configuration, its coefficients acquire the compact form
\begin{equation}
a_n
=
\left(
\frac{\kappa\ell^2}{\pi}
\right)^n
r_+^{\,1-2n}
\mathcal{A}_n\left(
\frac{r_+}{r_-}
\right),
\label{eq:BTZNLED_two_horizon_coefficients}
\end{equation}
where the dimensionless function $\mathcal{A}_n$ encodes the
degeneracy condition at the internal root. An explicit closed
expression for $\mathcal{A}_n$, together with the first coefficients
and their derivation from the LB formula, is presented
in Appendix~\ref{appx:BTZNLED_coefficients}.
For charged emission, the corresponding inverse is obtained from the
coefficients in Appendix~\ref{appx:Lagrange}, using
Eq.~\eqref{eq:BTZNLED_mass_radius} at the final charge. The prescribed
profile $\Qcal_{\rm em}(\Ecal)$ then gives the displacement along
${\cal C}$.

The dependence on $r_+/r_-$ in
Eq.~\eqref{eq:BTZNLED_two_horizon_coefficients} comes from the
initial degeneracy condition, which constrains the geometry at the
event horizon. The internal roots are not crossed by the outgoing
contour. Their rearrangement nevertheless follows from the same
change of mass and charge, as we now describe.

\subsection{From inner horizon fine-structure to extremality
\label{subsec:BTZNLED_fine-structure}}

For fixed $(q,\beta,\gamma,\ell)$, every positive horizon
solves the level set equation $\mgot(r,q)=\Mcal$, corresponding to
an intersection of the mass--radius curve with the horizontal line
at mass $\Mcal$. This picture places the initial inner degeneracy
and the later outer merger at the two stationary points of the same
curve. Its first derivatives are
\begin{align}
\partial_r\mgot
={}&
\frac{\pi}{\kappa}
\left\{
\frac{2r}{\ell^2}
-
\frac{2\sqrt{2}\kappa q}{\gamma}
\left[
\ln\left(
\frac{\beta\gamma r}{\sqrt{2}q}
\right)+1
\right]
\right\},
\nonumber\\
\partial_r^2\mgot
={}&
\frac{\pi}{\kappa}
\left(
\frac{2}{\ell^2}
-
\frac{2\sqrt{2}\kappa q}{\gamma r}
\right),
\nonumber\\
\partial_r^3\mgot
={}&
\frac{2\sqrt{2}\pi q}{\gamma r^2}.
\label{eq:BTZNLED_mass_derivatives}
\end{align}

When $q/\gamma<0$, the curve is strictly convex and admits only one
positive intersection for $\Mcal>0$. A configuration with several
positive horizons instead requires
\begin{equation*}
\frac{q}{\gamma}>0.
\end{equation*}
Together with Eq.~\eqref{eq:BTZNLED_reality_condition}, this condition
also implies $\beta>0$.

\begin{figure}[tbp]
\centering
\includegraphics[width=0.76\linewidth]{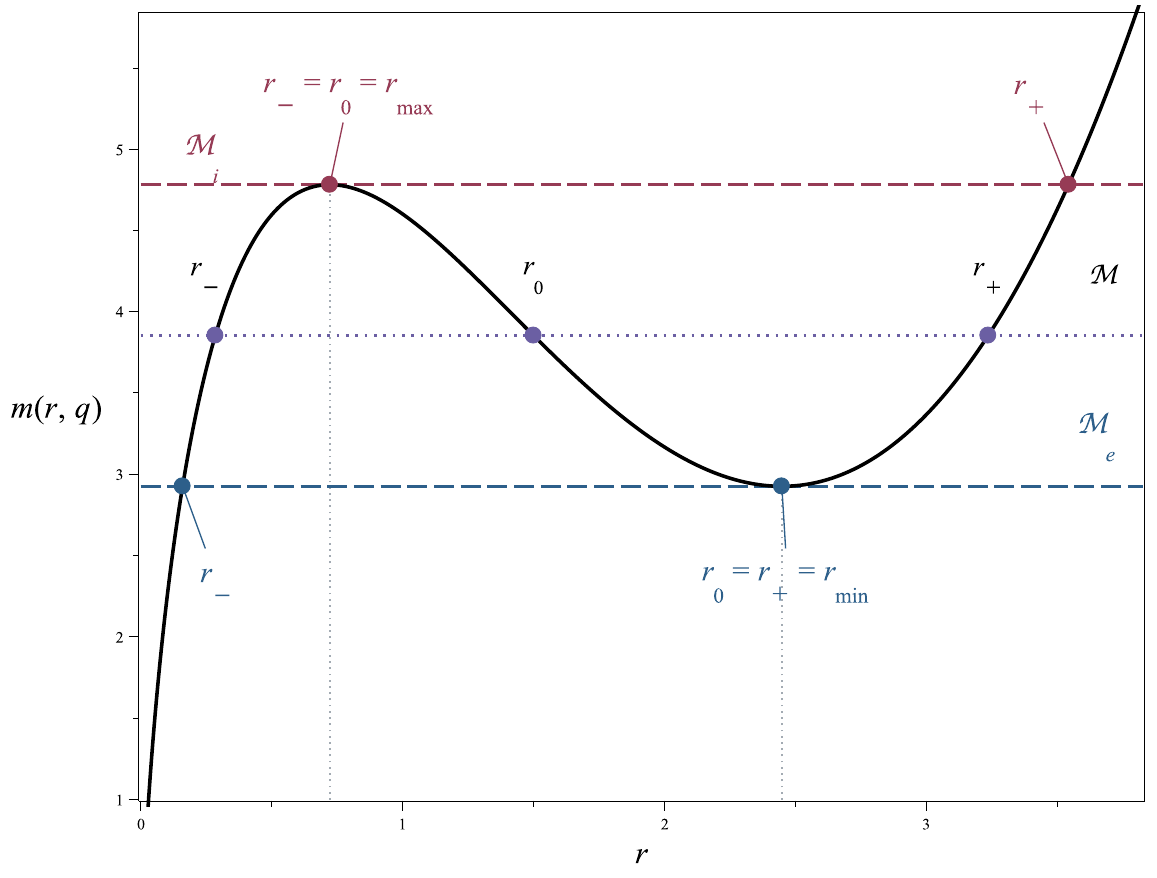}
\caption{Structure of the local mass function for sampling constant values
$\kappa=\ell=\gamma=q=1$ and $\beta=6/5$.
At the upper critical mass $\Mcal_{\rm i}$, the two inner
horizons coincide at $r_-=r_0=r_{\rm max}$, while a separate simple
event horizon remains. Intermediate mass levels intersect the curve
at three simple horizons. At the lower critical mass $\Mcal_{\rm e}$,
the middle and event horizons merge at $r_0=r_+=r_{\rm min}$,
while the innermost root survives and remains simple.
The plotted $r_+$ denotes $R_+(\Ecal)$ along the neutral
sequence, with $r_{\rm min}=r_{\rm e}$ in the notation of
Fig.~\ref{fig:general_horizon_deformation}. Neutral emission connects these
two critical levels at fixed charge, with
$\Ecal_{\rm c}=\Mcal_{\rm i}-\Mcal_{\rm e}\simeq1.856$.}
\label{fig:BTZNLED_mass_radius_portrait}
\end{figure}

For $\kappa\beta\ell^2>1$, the stationary condition
$\partial_r\mgot=0$ admits two positive solutions. These are
expressed through the two real branches of Lambert's $W$ function,
defined implicitly by $W(x) \exp\bigl[W(x)\bigr]=x$. On the interval
$-1/\e<x<0$, they correspond to the principal branch $W_0(x)$ and the
lower branch $W_{-1}(x)$ \cite{veberivc2012lambert}. Since their common
argument appears repeatedly, we introduce the shorthand
\begin{equation}
w_k
:=
W_k\left(
-\frac{1}{\e\kappa\beta\ell^2}
\right),
\qquad
k=0,-1.
\label{eq:BTZNLED_Lambert_values}
\end{equation}
The two stationary radii can then be written as
\begin{align}
r_{\rm max}
&=
-\frac{\sqrt{2}\kappa q\ell^2}{\gamma}\,w_0,
\label{eq:BTZNLED_rmax}
\\
r_{\rm min}
&=
-\frac{\sqrt{2}\kappa q\ell^2}{\gamma}\,w_{-1}.
\label{eq:BTZNLED_rmin}
\end{align}
The ordering of the two real Lambert branches,
\begin{equation*}
-1<w_0<0,
\qquad
w_{-1}<-1,
\end{equation*}
implies the corresponding ordering of the radii
\begin{equation*}
0
<
r_{\rm max}
<
\frac{\sqrt{2}\kappa q\ell^2}{\gamma}
<
r_{\rm min}.
\end{equation*}
These are the local maximum and minimum of $\mgot(r,q)$,
respectively, as shown in Fig.~\ref{fig:BTZNLED_mass_radius_portrait}.
For the neutral sequence of Sec.~\ref{sec:degenerate_outer_coalescence},
they give $r_{\rm d}=r_{\rm max}$ and $r_{\rm e}=r_{\rm min}$.
The critical masses defined in Eq.~\eqref{eq:degenerate_critical_levels}
therefore evaluate to
\begin{align}
\Mcal_{\rm i}
&=
\mgot(r_{\rm max},q)
=
-\frac{2\pi\kappa q^2\ell^2}{\gamma^2}
w_0\big(w_0+2\big),
\label{eq:BTZNLED_Mi}
\\
\Mcal_{\rm e}
&=
\mgot(r_{\rm min},q)
=
-\frac{2\pi\kappa q^2\ell^2}{\gamma^2}
w_{-1}\big(w_{-1}+2\big).
\label{eq:BTZNLED_Me}
\end{align}
These values are intrinsic thresholds of the black hole family. At
fixed $q$, the number and multiplicity of the positive roots change
when the mass crosses either of them. The subscripts indicate the
causal character of the corresponding degeneracy:
$\Mcal_{\rm i}$ marks the merger of the two inner roots, whereas
$\Mcal_{\rm e}$ marks the coalescence of the middle and event
horizons.

The upper critical mass is positive for every
$\kappa\beta\ell^2>1$, while the lower one is positive only within
\begin{equation}
1
<
\kappa\beta\ell^2
<
\frac{\e}{2}.
\label{eq:BTZNLED_positive_mass_window}
\end{equation}
 For the multihorizon emission paths considered below, we
restrict to this interval, for which
\begin{equation*}
0<\Mcal_{\rm e}<\Mcal_{\rm i}.
\end{equation*}
This is the sector in which the outer coalescence has positive mass
and leaves a separate positive innermost horizon. Since
$\mgot(0^+,q)=0$ and $\mgot(r,q)\rightarrow+\infty$ at large
radius, the three monotonic portions separated by the maximum and
minimum intersect every mass level between
$\Mcal_{\rm e}$ and $\Mcal_{\rm i}$ exactly once.

Starting from the upper critical configuration,
$\Mcal=\Mcal_{\rm i}$, neutral emission at fixed $q$ lowers
the mass to $\Mcal'(\Ecal)=\Mcal_{\rm i}-\Ecal$. It reaches
the lower critical level at the neutral critical energy of
Eq.~\eqref{eq:degenerate_critical_levels},
\begin{equation}
\Ecal_{\rm c}=\Mcal_{\rm i}-\Mcal_{\rm e}.
\label{eq:BTZNLED_critical_energy}
\end{equation}
The intersections in Fig.~\ref{fig:BTZNLED_mass_radius_portrait}
show the resulting sequence from inner splitting to outer
coalescence.
Above the initial mass level $\Mcal_{\rm i}$, the mass curve admits only one positive intersection, corresponding to a simple event
horizon. The two inner branches have the common initial limit $r_-(0)=r_0(0)=r_{\rm max}$, while $r_-(\Ecal)$ denotes the
smaller branch after splitting. The initial configuration satisfies
\begin{align}
f(r_{\rm max}) = \partial_r f(r_{\rm max}) =0, \qquad
\partial_r^2 f(r_{\rm max})< 0.
\label{eq:BTZNLED_upper_boundary}
\end{align}
A separate simple event horizon $r_+>r_{\rm min}$ remains, with finite
positive temperature. The initial double zero consequently lies in the
splitting sector \eqref{eq:degenerate_splitting_sector}.

Any positive neutral emission lowers the mass below $\Mcal_{\rm i}$
and immediately unfolds the double zero into two simple roots.
Throughout $0<\Ecal<\Ecal_{\rm c}$, the geometry therefore
possesses three distinct positive horizons,
\begin{equation*}
0<r_-(\Ecal)<r_0(\Ecal)<R_+(\Ecal).
\end{equation*}
Writing $r_a=r_a(\Ecal)$ for $a\in\{-,0,+\}$, with
$r_+(\Ecal)=R_+(\Ecal)$,
the metric function at mass $\Mcal-\Ecal$ and charge $q$
is negative for $0<r<r_-$ and changes sign across each
of the three simple horizons.
The outermost root $R_+(\Ecal)$ corresponds to the event horizon, while the intermediate root $r_0(\Ecal)$ defines the
Cauchy horizon associated with the exterior region.
The smallest root $r_-(\Ecal)$ corresponds to an additional inner Killing horizon. At fixed charge, their motion under emission
obeys
\begin{equation}
\frac{d r_a}{d\Ecal}
=
-\frac{\kappa}{\pi\,\partial_r f(r_a)}.
\label{eq:BTZNLED_neutral_horizon_motion}
\end{equation}
The positive slopes at the innermost and event horizons make both
contract, while the negative slope at the middle horizon makes it
expand as energy is radiated away. The middle and event horizons consequently approach one
another until they reach the minimum of the mass--radius curve.

At the critical energy, the middle and event horizons merge
at the local minimum,
\begin{equation*}
r_0(\Ecal_{\rm c})=R_+(\Ecal_{\rm c})=r_{\rm min},
\end{equation*}
with
\begin{align}
f(r_{\rm min})
=
\partial_r f(r_{\rm min})
=
0,\qquad
\partial_r^2 f(r_{\rm min})
>
0.
\label{eq:BTZNLED_lower_boundary}
\end{align}
The resulting event horizon is degenerate and has zero Hawking temperature,
while the innermost simple root survives. If the family is formally
continued to positive masses below $\Mcal_{\rm e}$, the outer pair is absent and the
continuation of the smallest root becomes the sole event horizon.

The same structure can be represented globally in the charge--mass
state space. At fixed $(\ell,\beta,\gamma)$, the local mass function can be checked to be homogeneous of second degree according to
\begin{equation}
\mgot(\lambda r,\lambda q)
=
\lambda^2\mgot(r,q),
\qquad
\lambda>0.
\label{eq:BTZNLED_homogeneity}
\end{equation}
In particular, for $q'=\Qcal'/(2\pi)$, any positive horizon satisfies
\begin{equation}
r_a(\Mcal',q')
=
\frac{q'}{q}\,
r_a\left(\Mcal'\frac{q^2}{q'^2},q\right),
\qquad
\frac{q'}q>0.
\label{eq:BTZNLED_horizon_scaling}
\end{equation}
Charge variation thus changes the radial scale, while the ratio
$\Mcal'/\Qcal'^2$ determines the positions of the scaled horizons.
This scaling also relates the tunneling exponents of
corresponding channels. After scaling the initial state as
$(\Mcal,\Qcal)\mapsto(\lambda^2\Mcal,\lambda\Qcal)$
and choosing the corresponding charge profile to satisfy
$\Qcal_{{\rm em},\lambda}(\lambda^2\Ecal)
=\lambda\Qcal_{\rm em}(\Ecal)$,
Eq.~\eqref{eq:BTZNLED_probability_radius} gives
\begin{equation}
\Delta\Scal_\lambda(\lambda^2\Ecal)
=\lambda\Delta\Scal(\Ecal).
\label{eq:BTZNLED_exponent_scaling}
\end{equation}
Hence $\Delta\Scal/\Scal(0)$ has the same dependence on
$\Ecal/\Mcal$ for these charge-scaled states, while their absolute
exponents, and therefore their tunneling factors, differ.
Since $\Qcal=2\pi q$,  Eqs.~\eqref{eq:BTZNLED_Mi} and
\eqref{eq:BTZNLED_Me} extend to the two critical curves
\begin{align}
\Mcal_{\rm i}(\Qcal)
&=
-\frac{\kappa\ell^2\Qcal^2}{2\pi\gamma^2}
w_0\big(w_0+2\big),
\label{eq:BTZNLED_upper_critical_curve}
\\
\Mcal_{\rm e}(\Qcal)
&=
-\frac{\kappa\ell^2\Qcal^2}{2\pi\gamma^2}
w_{-1}\big(w_{-1}+2\big).
\label{eq:BTZNLED_lower_critical_curve}
\end{align}
In the sign sector $\Qcal/\gamma>0$, they delimit the strip
\begin{equation}
\Mcal_{\rm e}(\Qcal)
<
\Mcal
<
\Mcal_{\rm i}(\Qcal),
\label{eq:BTZNLED_phase_strip}
\end{equation}
occupied by the three-horizon geometries. Its upper boundary contains
a degenerate inner horizon together with a simple event horizon,
whereas its lower boundary contains a simple innermost root and a
degenerate event horizon.

The two-boundary structure itself collapses at the limiting coupling
$\kappa\beta\ell^2=1$. In this particular choice, both negative branches of Lambert's function coincide
$w_0=w_{-1}=-1$, and
\begin{align}
r_{\rm max}
=
r_{\rm min}
=
r_{\rm t}
&:=
\frac{\sqrt{2}\kappa q\ell^2}{\gamma},
\nonumber\\
\Mcal_{\rm i}
=
\Mcal_{\rm e}
=
\Mcal_{\rm t}
&:=
\frac{2\pi\kappa q^2\ell^2}{\gamma^2}.
\label{eq:BTZNLED_triple_data}
\end{align}
The local maximum and minimum have then merged into a stationary
inflection point, with $f(r_{\rm t})=\partial_r f(r_{\rm t})
=\partial_r^2 f(r_{\rm t})=0$ and $\partial_r^3 f(r_{\rm t})>0$,
producing a horizon of multiplicity three. Its local
response belongs to the higher-multiplicity hierarchy developed in
Appendix~\ref{appx:Puiseux}. At fixed couplings in
Eq.~\eqref{eq:BTZNLED_positive_mass_window}, varying the charge scales
both radii by the same positive factor and cannot bring
them together at nonzero charge. The triple zero is therefore a
coupling-space limit, distinct from the emission paths considered
here.

Each charge profile defines a \emph{horizon history}, namely
the sequence of stationary horizon configurations as $\Ecal$
increases along ${\cal C}$. To follow it, we evaluate
the critical masses at the running charge.
With $\Mcal=\Mcal_{\rm i}(\Qcal)$ as the initial mass, their signed
mass differences are explicitly
\begin{equation}
\Delta\Mcal_l(\Ecal)
=\Mcal_l(\Qcal)
\left[1-\frac{\Qcal_{\rm em}(\Ecal)}{\Qcal}\right]^2
-\Mcal_{\rm i}(\Qcal)+\Ecal,
\qquad l\in\{{\rm i},{\rm e}\}.
\label{eq:BTZNLED_charged_mass_differences}
\end{equation}
Here the charge keeps the sign of the chosen sector,
$\Qcal'(\Ecal)=\Qcal-\Qcal_{\rm em}(\Ecal)$ with
$\Qcal'(\Ecal)/\gamma>0$.
The three-horizon geometry occurs precisely when
$\Delta\Mcal_{\rm i}>0$ and $\Delta\Mcal_{\rm e}<0$.
The first inequality places the state below the inner critical curve, resolving the inner double zero. The second keeps it above the outer critical curve, retaining the middle and event horizons.
At either equality the corresponding pair coincides.
Their local separation is given by the charged inverse of
Sec.~\ref{sec:degenerate_charged_unfolding}, evaluated about the
critical radius at the running charge.

Differentiating the horizon condition along any differentiable
charge profile, with $\partial_q\mgot=2\pi\Phi$, gives
\begin{equation}
\frac{d r_a}{d\Ecal}
=-\frac{1-\Phi(r_a)\,d\Qcal_{\rm em}/d\Ecal}
{ \partial_r\mgot\bigl[r_a,q(\Ecal)\bigr]},
\label{eq:BTZNLED_charged_horizon_motion}
\end{equation}
with $\Phi(r)=-A_t(r)$ evaluated at the running charge and the formula
applied to simple roots.

For linear charge extraction, take
$\Qcal_{\rm em}(\Ecal)=\eta\Ecal$, with constant $\eta$.
The inner mass difference is then exactly
\begin{equation}
\Delta\Mcal_{\rm i}(\Ecal)
=
\left(
1-\frac{2\eta\Mcal_{\rm i}}{\Qcal}
\right)\Ecal
+
\frac{\eta^2\Mcal_{\rm i}}{\Qcal^2}\Ecal^2,
\label{eq:BTZNLED_linear_channel_inner_difference}
\end{equation}
 Since the inner critical point is a maximum,
$1-2\eta\Mcal_{\rm i}/\Qcal>0$ realizes
case~(\ref{case:critical_transverse_split}), whereas a negative
value realizes case~(\ref{case:critical_transverse_absent}). At
$\eta=\Qcal/(2\Mcal_{\rm i})$, the positive quadratic term
resolves the tangent case~(\ref{case:critical_tangent}) in favor
of a real pair, with separation proportional to $\Ecal$.

Each of these cases can occur under ordinary emission through the event horizon. At the tangent value, homogeneity gives
\[
1-\eta\Phi_+
=
\frac{r_+\partial_r\mgot(r_+,q)}{2\Mcal_{\rm i}}>0,
\]
at the initial state, so a neighborhood on either side of this value
also obeys the condition ensuring a positive horizon energy.
For constant $\eta$, an initially positive factor remains positive
along the simple outer horizon \footnote{For $\eta\neq0$, if
$1-\eta\Phi_+$ vanishes at a simple outer horizon, differentiating
the horizon equation gives $dR_+/d\Ecal=0$. Since
$\partial_q\Phi=\sqrt{2}R_+/[\gamma q(\Ecal)]$, its derivative there is
$d(1-\eta\Phi_+)/d\Ecal
=\eta^2R_+/[\sqrt{2}\pi\gamma q(\Ecal)]>0$.
The factor cannot cross from positive to negative while the outer
horizon is simple and the charge remains in the chosen sign sector.
For $\eta=0$, the factor is identically one.}.

For $1/2<\eta\Mcal_{\rm i}/\Qcal<1$, the inner pair initially
absent in case~(\ref{case:critical_transverse_absent}) reappears at the positive emitted energy value
\begin{equation}
\Ecal_{\rm s}
:=
\frac{\Qcal^2}{\eta^2\Mcal_{\rm i}}
\left(
\frac{2\eta\Mcal_{\rm i}}{\Qcal}-1
\right),
\label{eq:BTZNLED_delayed_splitting_energy}
\end{equation}
where the path meets the upper critical curve again. The charge and
mass are still nonzero there. A double inner zero is recovered at
$\Ecal_{\rm s}$ and splits into two simple horizons with separation
proportional to $(\Ecal-\Ecal_{\rm s})^{1/2}$, before the middle
horizon later meets the event horizon.
The charged history can therefore contain an interval with a single
event horizon between two encounters with the inner critical curve.
 For a constant-$\eta$ emission channel with
$0<\eta\Mcal_{\rm i}/\Qcal<1$, the outer merger is reached
at positive mass and nonzero charge of the initial sign, at the
unique positive zero of $\Delta\Mcal_{\rm e}$. If
$\eta\Mcal_{\rm i}/\Qcal\geq1$, that curve is not reached before the charge vanishes, where the nonzero-charge construction ends.

\begin{figure}[t]
\centering
\includegraphics[width=0.84\linewidth]{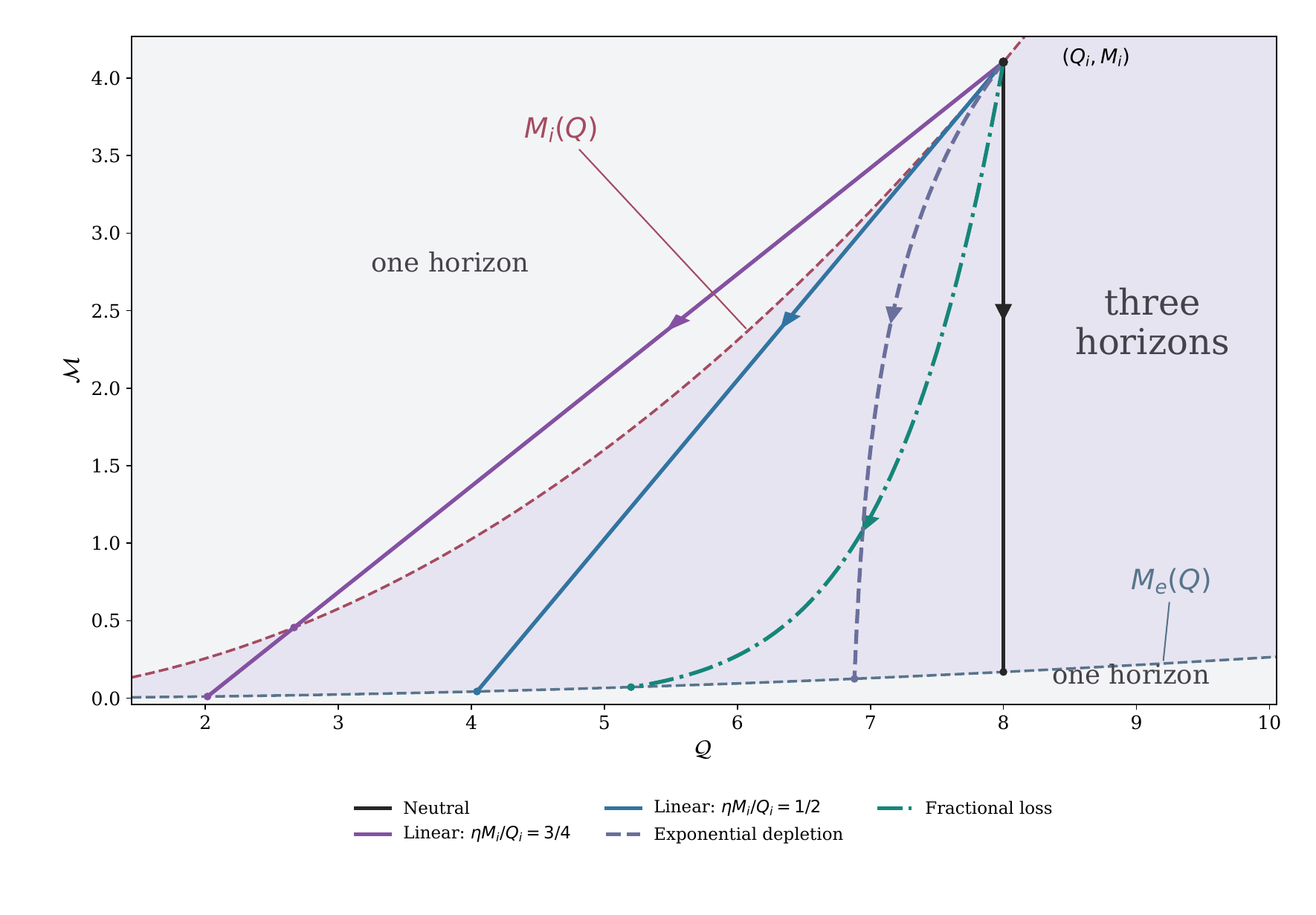}
\caption{Critical charge--mass structure and emission paths
of the logarithmically deformed BTZ family for
$\kappa=\ell=1$, $\beta=27/20$, and $q=\gamma=4/\pi$,
so the initial charge is $\Qcal_{\rm i}=\Qcal=8$.
The critical curves $\Mcal_{\rm i}(\Qcal)$ and
$\Mcal_{\rm e}(\Qcal)$ delimit the shaded three-horizon strip.
The pale regions contain a single event horizon. At nonzero charge,
the upper curve contains a double internal horizon and a simple
event horizon, whereas the lower curve contains a simple innermost
horizon and a double event horizon. All five paths start at
$(\Qcal,\Mcal_{\rm i})$ and are followed in the direction of
increasing emitted energy. The solid black path is neutral. The
solid blue and violet paths have $\Qcal_{\rm em}=\eta\Ecal$,
with $\eta\Mcal_{\rm i}/\Qcal=1/2$ and $3/4$, respectively:
the former enters tangentially, while the latter leaves the strip
and returns through the upper critical curve at
$\Ecal_{\rm s}>0$. The dashed indigo path is the exponential
profile \eqref{eq:BTZNLED_exponential_channel}, with $a=3$ and
$\Qcal_f/\Qcal=0.86$. The dash-dotted green path is the fractional
profile \eqref{eq:BTZNLED_fractional_channel}, with
$p\simeq0.1063$ and $\Qcal_f/\Qcal=0.65$.
Lower-boundary markers identify the merger of the middle and event
horizons. The additional upper-boundary marker on the $3/4$ path
identifies the delayed recovery of the inner double zero.}
\label{fig:BTZNLED_phase_diagram}
\end{figure}

The nonlinear paths in Fig.~\ref{fig:BTZNLED_phase_diagram}
allow the charge extracted per unit energy to vary during emission.
We take $\Qcal>0$ and retain
$\Mcal'=\Mcal_{\rm i}-\Ecal$ and
$\Qcal'(\Ecal)=\Qcal-\Qcal_{\rm em}(\Ecal)$,
where unprimed $\Mcal_{\rm i}$ and $\Mcal_{\rm e}$ denote
the critical masses evaluated at the initial charge. For each path,
$\Ecal_{\rm c}$ will denote its own outer coalescence energy. Concretely, we consider the following two profiles.

\noindent\textbf{ \emph{Exponential charge depletion.}}
Suppose that charge extraction becomes less efficient as
the remaining charge approaches a residual value
$0\leq\Qcal_{\rm res}<\Qcal$. A relaxation law analogous to Newton's
cooling law, here parametrized by emitted energy, is
\begin{equation*}
\frac{d\Qcal'}{d\Ecal}
=-\frac{a}{\Ecal_{\rm c}}
\big(\Qcal'-\Qcal_{\rm res}\big),
\qquad a>0.
\end{equation*}
Integrating with $\Qcal'(0)=\Qcal$ gives the exponential
charge-depletion profile
\begin{equation}
\Qcal_{\rm em}(\Ecal)
=\big(\Qcal-\Qcal_{\rm res}\big)
\left[1-e^{-a\Ecal/\Ecal_{\rm c}}\right].
\label{eq:BTZNLED_exponential_channel}
\end{equation}
Here $a$ is dimensionless and the charge extracted per unit energy
decreases exponentially from
$\eta_0=a(\Qcal-\Qcal_{\rm res})/\Ecal_{\rm c}$.

\noindent\textbf{\emph{Proportional fractional loss.}}
A different prescription takes the fractional charge loss to be
proportional to the fractional mass loss,
\begin{equation*}
\frac{d\Qcal'}{\Qcal'}
=p\,\frac{d\Mcal'}{\Mcal'},
\qquad p>0.
\end{equation*}
Integrating with $\Mcal'=\Mcal_{\rm i}-\Ecal$ and
$\Qcal'(0)=\Qcal$ gives
\begin{equation}
\Qcal_{\rm em}(\Ecal)
=\Qcal\left[
1-\left(1-\frac{\Ecal}{\Mcal_{\rm i}}\right)^p
\right],
\qquad 0\leq\Ecal<\Mcal_{\rm i}.
\label{eq:BTZNLED_fractional_channel}
\end{equation}
Here $\eta_0=p\Qcal/\Mcal_{\rm i}$, and
$d\Qcal_{\rm em}/d\Ecal
=\eta_0(1-\Ecal/\Mcal_{\rm i})^{p-1}$.
For $0<p<1/2$, charge extraction accelerates as the mass decreases,
while the path remains below the upper critical curve. Indeed,
Eq.~\eqref{eq:BTZNLED_charged_mass_differences} gives
\begin{equation*}
\Delta\Mcal_{\rm i}
=\Mcal_{\rm i}\left[
\left(1-\frac{\Ecal}{\Mcal_{\rm i}}\right)^{2p}
-\left(1-\frac{\Ecal}{\Mcal_{\rm i}}\right)
\right]>0
\end{equation*}
for $0<\Ecal<\Mcal_{\rm i}$. The profile is analytic for $\Ecal<\Mcal_{\rm i}$ for any
real $p>0$.

For either of these profiles, the endpoint can be selected by prescribing
the remaining charge fraction $s_f:=\Qcal_f/\Qcal$, where
$\Qcal_f=\Qcal'(\Ecal_{\rm c})$. The lower critical curve then
fixes
\begin{equation}
\Ecal_{\rm c}
=\Mcal_{\rm i}-\Mcal_{\rm e}s_f^2.
\label{eq:BTZNLED_profile_endpoint_calibration}
\end{equation}
The endpoint condition determines the residual charge in
Eq.~\eqref{eq:BTZNLED_exponential_channel} and the exponent in
Eq.~\eqref{eq:BTZNLED_fractional_channel}, respectively:
\begin{equation}
\begin{aligned}
\frac{\Qcal_{\rm res}}{\Qcal}
&=\frac{s_f-e^{-a}}{1-e^{-a}},
\\
p&=\frac{\ln s_f}
{\ln\!\left[(\Mcal_{\rm e}/\Mcal_{\rm i})s_f^2\right]}.
\end{aligned}
\label{eq:BTZNLED_nonlinear_endpoint_parameters}
\end{equation}
A nonnegative residual requires $s_f\geq e^{-a}$. For
$0<s_f<1$ and $0<\Mcal_{\rm e}<\Mcal_{\rm i}$, the fractional
prescription automatically gives $0<p<1/2$.

\begin{figure}[p]
\centering
\includegraphics[width=0.82\linewidth]{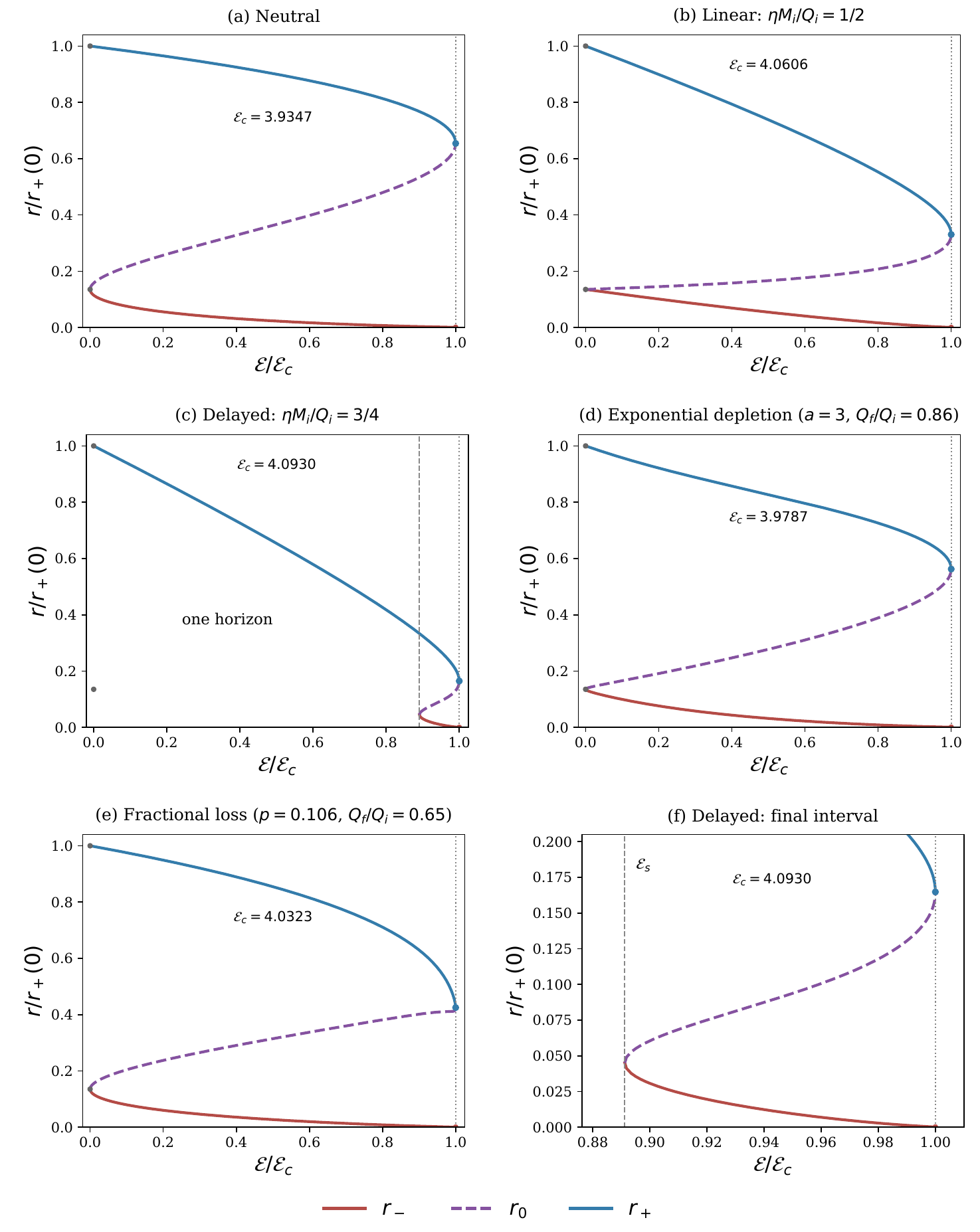}
\caption{Running horizons along the five paths in
Fig.~\ref{fig:BTZNLED_phase_diagram}, with the same parameters and
initial inner-degenerate state. Red, dashed violet, and blue denote
the innermost, middle, and event horizons, respectively. Radii are
normalized by the common initial event-horizon radius $r_+(0)$,
and the emitted energy by the coalescence energy
$\Ecal_{\rm c}$ of each path.
(a) Neutral emission resolves the internal double zero immediately,
with separation proportional to $\Ecal^{1/2}$.
(b) The linear $\eta\Mcal_{\rm i}/\Qcal=1/2$ path enters
tangentially and yields a linear separation.
(c) The linear $3/4$ path initially retains only the simple event
horizon. The isolated internal point at $\Ecal=0$ records the
initial double zero, recovered later at
$\Ecal_{\rm s}\simeq3.6477$ before splitting as
$(\Ecal-\Ecal_{\rm s})^{1/2}$.
(d) Exponential depletion and (e) fractional loss both produce
immediate square-root splitting, with different subsequent horizon
histories. The first five panels cover the full emission interval
and have the same radial range. Their coalescence energies are
$\Ecal_{\rm c}\simeq3.9347,\,4.0606,\,4.0930,\,3.9787,\,4.0323$,
respectively. In each case the middle and event horizons merge
while a simple innermost horizon survives.
(f) Zoom to the final interval of panel (c), with a different
radial range. Dashed and dotted guides mark delayed inner splitting
and outer coalescence, respectively.}
\label{fig:BTZNLED_charged_horizon_scenarios}
\end{figure}

For the nonlinear examples in
Figs.~\ref{fig:BTZNLED_phase_diagram} and
\ref{fig:BTZNLED_charged_horizon_scenarios}, we choose the parameters
so that both initial slopes lie below the tangent value,
$\eta_0\Mcal_{\rm i}/\Qcal<1/2$.
The inner pair therefore splits immediately with square-root
separation in both cases, even though the charge extracted per
unit energy decreases in one profile and increases in the other.
The subsequent horizon positions depend on the full charge
profile. In particular, the remaining charge at coalescence fixes
the merger radius through $r_{\rm min}\propto\Qcal'$ and the
total emitted energy through
Eq.~\eqref{eq:BTZNLED_profile_endpoint_calibration}. The fractional-loss example reaches the lower critical curve
with less remaining charge than the exponential one. Its merger
therefore occurs at a smaller radius and a larger emitted
energy, as seen in
Fig.~\ref{fig:BTZNLED_charged_horizon_scenarios}(d) and \ref{fig:BTZNLED_charged_horizon_scenarios}(e).
For the parameters shown, both paths remain inside the three-horizon
strip until this merger and satisfy
$1-\Phi_+d\Qcal_{\rm em}/d\Ecal>0$ throughout the simple outer
branch.
Both profiles prescribe charge loss as a function of energy.
Their realization by a particular radiating field depends on its
energy and charge fluxes.

The horizon histories in
Fig.~\ref{fig:BTZNLED_charged_horizon_scenarios} follow from the full
horizon condition, evaluated using
Eq.~\eqref{eq:BTZNLED_horizon_scaling}.

At the internal transitions, the outer denominator in
Eq.~\eqref{eq:BTZNLED_charged_horizon_motion} stays positive and
nonzero. For the analytic profiles considered here, $R_+$ and
$\Delta\Scal$ are therefore analytic through these transitions,
including the delayed splitting at $\Ecal_{\rm s}$. The simple-root description of the outer horizon ends
at the later coalescence.
The internal rearrangement therefore produces no nonanalytic
feature in the leading horizon emission exponent. Smooth
differences between charge profiles remain possible, and the full radiation spectrum also depends on propagation outside the horizon.

\subsection{Further emission channels
\label{subsec:BTZNLED_further_channels}}

\begin{figure}[tbp]
\centering
\includegraphics[width=0.48\linewidth]{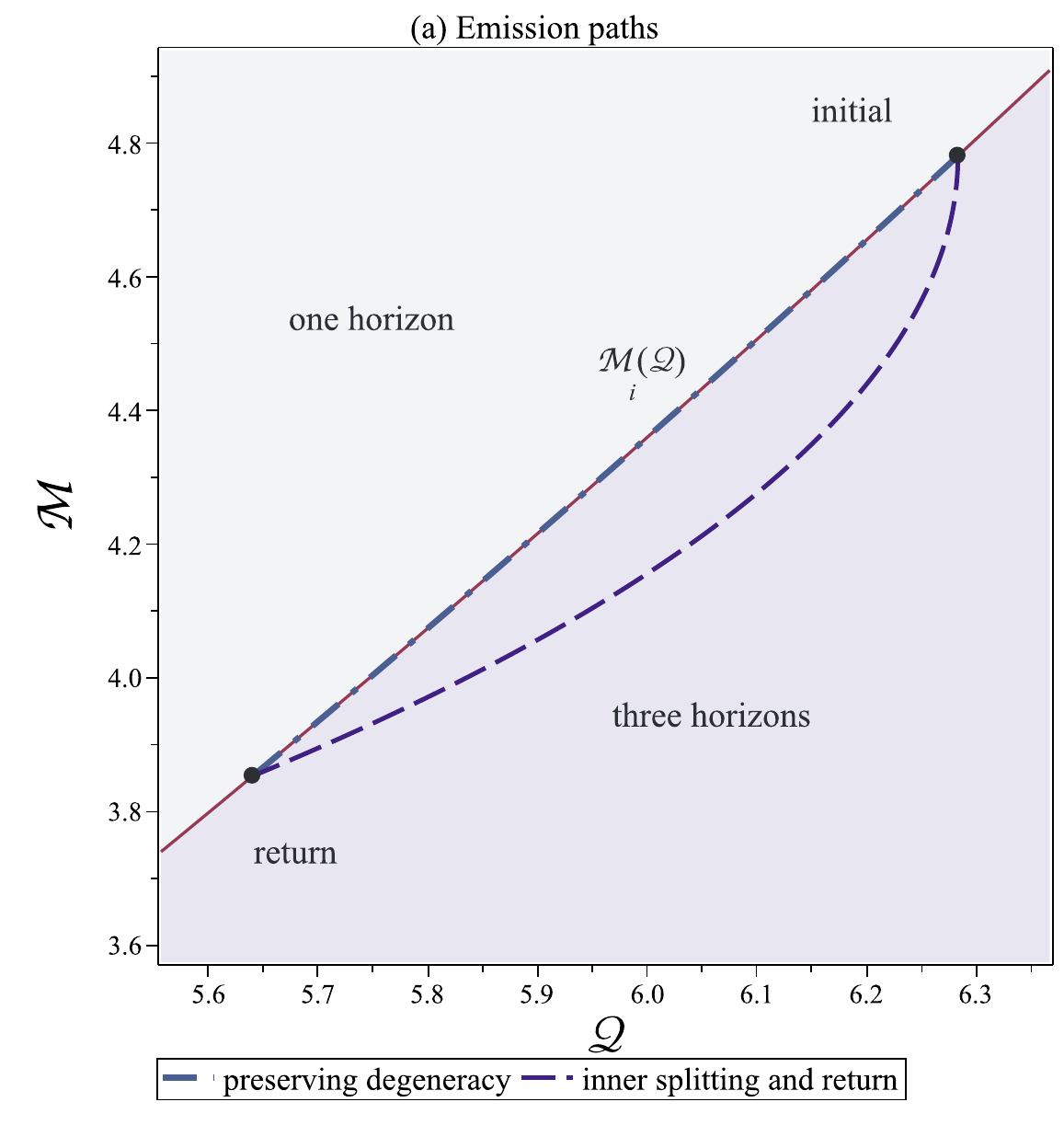}\hfill
\includegraphics[width=0.48\linewidth]{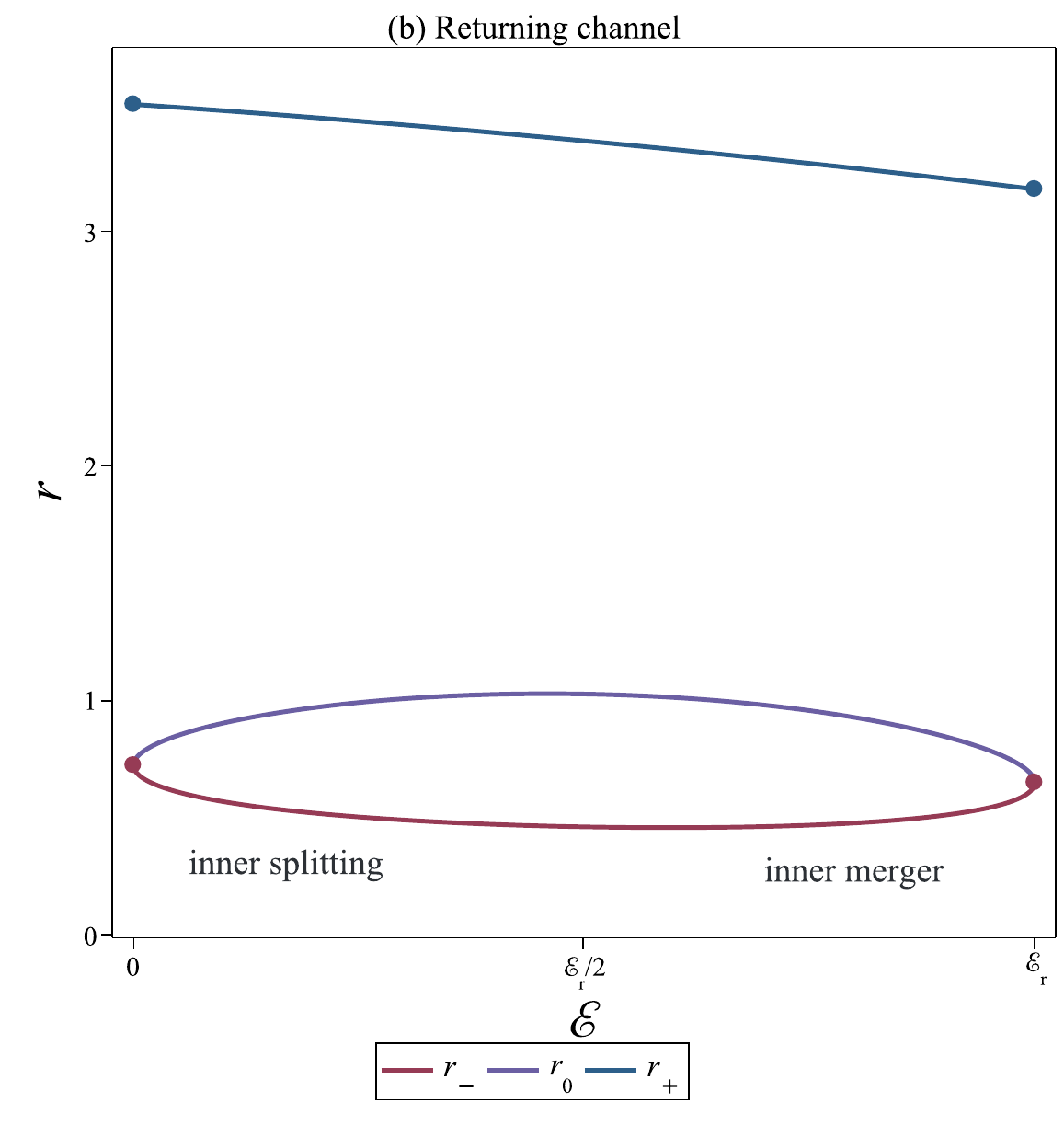}
\caption{Further charged emission channels for
$\kappa=\ell=\gamma=q=1$, $\beta=6/5$, and
$\Ecal_{\rm r}=(\Mcal_{\rm i}-\Mcal_{\rm e})/2$.
(a) The dash-dotted profile
\eqref{eq:BTZNLED_inner_critical_emission_profile} strictly follows the
$\Mcal_{\rm i}(\Qcal)$ curve, whereas the dashed profile
\eqref{eq:BTZNLED_returning_inner_profile} enters the shaded
three-horizon region and returns to this curve. Both paths are shown up to $\Ecal_{\rm r}$ and have the same
initial and final states.
 (b) Full horizon radii along the returning channel. Red, violet,
and blue denote the innermost, middle, and event horizons.
Both endpoints contain an internal double horizon and a simple
event horizon.}
\label{fig:BTZNLED_further_channels}
\end{figure}

The interplay between charge and energy emission can also
produce profiles that preserve the inner degeneracy throughout
the whole process. The limiting choice $p=1/2$ in
Eq.~\eqref{eq:BTZNLED_fractional_channel} gives such a channel,
\begin{equation}
\Qcal_{\rm em}(\Ecal)
=
\Qcal\left[
1-\sqrt{1-\frac{\Ecal}{\Mcal_{\rm i}}}
\right],
\qquad 0\leq\Ecal<\Mcal_{\rm i},
\label{eq:BTZNLED_inner_critical_emission_profile}
\end{equation}
for which $\Mcal-\Ecal= \Mcal_{\rm i}\bigl[\Qcal'(\Ecal)\bigr]$
at every stage.
All horizon radii scale by
$\sqrt{1-\Ecal/\Mcal_{\rm i}}$, so the double inner horizon and
the simple event horizon persist without an intervening
three-horizon geometry. Its outer horizon-energy factor is the positive constant
\[
\frac{r_+\partial_r\mgot(r_+,q)}{2\Mcal_{\rm i}}.
\]
The limit $\Qcal'\rightarrow0$ lies outside the nonzero-charge
sector used here.
\begin{samepage}
Another possibility is that a curved channel instead unfolds the inner double horizon at the beginning of the process
and later brings the pair together again at a lower point on the same
critical curve. An explicit profile of such a channel is
\begin{equation}
\Qcal_{\rm em}(\Ecal)
=\Qcal\left[
1-\sqrt{1-\frac{\Ecal^2}{\Mcal_{\rm i}\Ecal_{\rm r}}}
\right],
\label{eq:BTZNLED_returning_inner_profile}
\end{equation}
\par
\end{samepage}
where $\Ecal_{\rm r}$ is the return energy, chosen within
$0<\Ecal_{\rm r}<\Mcal_{\rm i}-\Mcal_{\rm e}$, and
$0\leq\Ecal\leq\Ecal_{\rm r}$.
Charge extraction remains monotonic, with
$(\partial_\Ecal\Qcal_{\rm em})(0)=0$. It starts at order
$\Ecal^2$, so the leading response is the neutral square-root
splitting. The inner mass difference is exactly
\begin{equation}
\Delta\Mcal_{\rm i}(\Ecal)
=\Ecal
\left(1-\frac{\Ecal}{\Ecal_{\rm r}}\right).
\label{eq:BTZNLED_returning_inner_difference}
\end{equation}
Together with the bound on $\Ecal_{\rm r}$, this places the path
inside the three-horizon strip for $0<\Ecal<\Ecal_{\rm r}$.
 The two histories are compared in
Fig.~\ref{fig:BTZNLED_further_channels}. Along the returning path,
the inner pair separates as $\Ecal^{1/2}$ and merges again as
$(\Ecal_{\rm r}-\Ecal)^{1/2}$. The event horizon remains simple and at nonzero temperature
throughout this return to inner degeneracy. For the parameters shown, the positive
horizon energy condition
$1-\Phi_+d\Qcal_{\rm em}/d\Ecal>0$ also holds throughout.

At $\Ecal_{\rm r}$, the returning channel and the profile
preserving the inner degeneracy reach the same black hole state.
Equation~\eqref{eq:gen_charged_KPW_entropy} therefore gives their
common leading tunneling exponent,
\begin{equation}
\Delta\Scal(\Ecal_{\rm r})
=\Scal(0)\left[
\sqrt{1-\frac{\Ecal_{\rm r}}{\Mcal_{\rm i}}}-1
\right],
\label{eq:BTZNLED_inner_return_exponent}
\end{equation}
The same leading exponent thus accompanies two different internal horizon histories.

\section{ Approach to outer coalescence}
\label{sec:BTZNLED_outer_endpoint}

The histories presented in
Fig.~\ref{fig:BTZNLED_charged_horizon_scenarios} reach the lower
critical curve at points determined by their charge profiles.
There the event horizon merges with the middle root and its
temperature vanishes. The limiting tunneling exponent can still
be obtained from the stationary family, while the duration of
this approach requires the energy and charge fluxes. The
finite-energy exponent also determines when a single emission
can be treated thermally.

\subsection{ Limiting tunneling exponent
\label{subsec:BTZNLED_outer_coalescence_exponent}}

For an approach from the three-horizon sector
at nonzero charge, the coalescence energy satisfies
\begin{equation}
\Mcal-\Ecal_{\rm c}
=
\Mcal_{\rm e}
\bigl[\Qcal-\Qcal_{\rm em}(\Ecal_{\rm c})\bigr],
\label{eq:BTZNLED_charged_coalescence_condition}
\end{equation}
and therefore depends on the charge profile. Since the lower
critical mass scales as $\Qcal'^2$, removing charge of the same
sign as the black hole, within the chosen fixed sign sector, lowers
this mass and requires more emitted energy than the neutral value
in Eq.~\eqref{eq:BTZNLED_critical_energy}.
Emission of oppositely charged matter instead increases
$|\Qcal'|$ and raises the critical mass, allowing coalescence at a smaller emitted energy. For either sign, the channel must satisfy
$1-\Phi_+d\Qcal_{\rm em}/d\Ecal>0$ while the event horizon
remains simple.

 At fixed couplings, charge variation moves $r_{\rm min}$
without changing the radial curvature of $\mgot$ at this minimum.
On the three-horizon side $\Delta\Mcal_{\rm e}<0$, the event
horizon and its temperature consequently behave as
\begin{align}
R_+(\Ecal)
={}&
 r_{\rm min}\bigl[q(\Ecal)\bigr]
+
\ell\sqrt{\frac{\kappa w_{-1}}{\pi(1+w_{-1})}}\,
\sqrt{-\Delta\Mcal_{\rm e}}
 +\Order(|\Delta\Mcal_{\rm e}|),
\nonumber\\
T_{\text H}(\Ecal)
={}&
\frac{1}{2\pi\ell}
\sqrt{\frac{\kappa(1+w_{-1})}{\pi w_{-1}}}\,
\sqrt{-\Delta\Mcal_{\rm e}}+\Order(|\Delta\Mcal_{\rm e}|).
\label{eq:BTZNLED_outer_coalescence_asymptotics}
\end{align}
The middle horizon approaches the same running minimum with
the opposite leading displacement. For neutral emission,
$-\Delta\Mcal_{\rm e}=\Ecal_{\rm c}-\Ecal$, recovering the
square-root merger and cooling of
Sec.~\ref{sec:degenerate_outer_coalescence}.
For charged emission, it is the mass difference from the
running critical curve that enters this square-root law. If
$-\Delta\Mcal_{\rm e}$ is proportional to
$(\Ecal_{\rm c}-\Ecal)^n$, the horizon separation and temperature
vanish as $(\Ecal_{\rm c}-\Ecal)^{n/2}$. For a path analytic at the endpoint, $n\geq1$ is the first nonzero integer order.
The scaling analysis below also allows a leading power $n>0$
when the approach is not analytic.

\begin{figure}[tbp]
\centering
\includegraphics[width=0.76\linewidth]{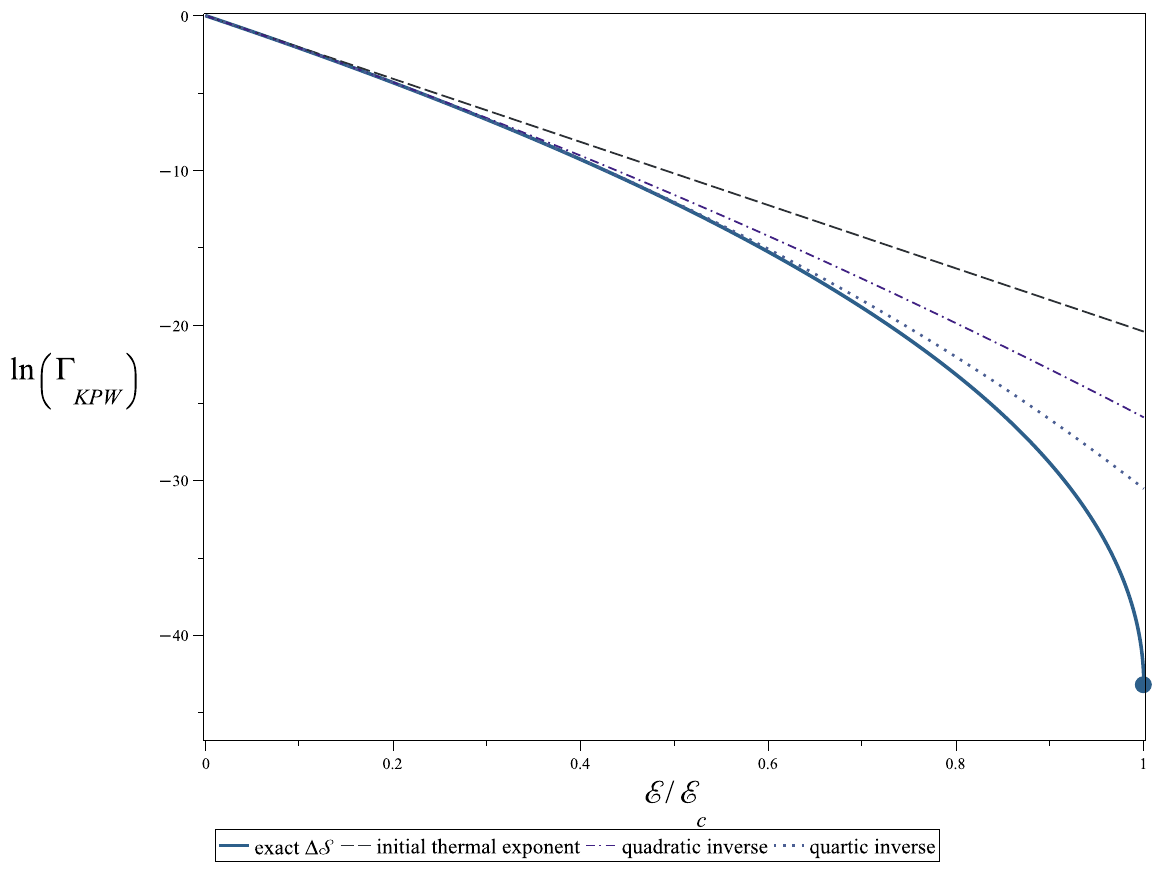}
\caption{Finite-energy entropy exponent for neutral emission,
for $\kappa=\ell=\gamma=q=1$ and $\beta=6/5$, as in Fig.~\ref{fig:BTZNLED_mass_radius_portrait}.
The solid curve is the full $\Delta\Scal$, evaluated on the outer
root giving $\ln\Gamma_{\rm KPW}$ at leading semiclassical order.
The dashed line is the thermal exponent $-\Ecal/T_{\text H}(0)$.
The dash-dotted and dotted curves truncate the inverse expansion in
Appendix~\ref{appx:BTZNLED_coefficients} at second and fourth order,
respectively. The horizontal axis is normalized by the neutral
$\Ecal_{\rm c}=\Mcal_{\rm i}-\Mcal_{\rm e}$.
The exponent obeys the global bound
\eqref{eq:BTZNLED_global_nonthermal_bound} and approaches
$\Delta\Scal(\Ecal_{\rm c})\simeq-43.209$ with a divergent negative
slope. The dot denotes this limit from simple outer horizons.
At the endpoint, the quadratic and quartic exponents have relative
errors of approximately $40\%$ and $29\%$, respectively.
These finite-order expansions resolve the initial emission,
but lose accuracy as the outer root approaches its double-zero limit.}
\label{fig:BTZNLED_entropy_exponent}
\end{figure}

Throughout the neutral trajectory, $R_+\geq r_{\rm min}$ and
$\partial_r^2f(R_+)>0$. Together with
Eq.~\eqref{eq:BTZNLED_neutral_horizon_motion}, this implies
$dT_{\text H}/d\Ecal<0$ along the simple outer branch and hence
\begin{equation}
\frac{d^2\Delta\Scal}{d\Ecal^2}
=\frac{1}{T_{\text H}^2}\frac{dT_{\text H}}{d\Ecal}<0.
\label{eq:BTZNLED_neutral_entropy_concavity}
\end{equation}
The finite-emission exponent lies strictly below its initial thermal
tangent throughout the trajectory:
\begin{equation}
\Delta\Scal(\Ecal)<-\frac{\Ecal}{T_{\text H}(0)},
\qquad 0<\Ecal\leq\Ecal_{\rm c}.
\label{eq:BTZNLED_global_nonthermal_bound}
\end{equation}
The endpoint is understood as a limit through simple outer horizons.
Figure~\ref{fig:BTZNLED_entropy_exponent} compares the full entropy
difference with the initial thermal exponent and the quadratic and
quartic inverse truncations. The higher-order truncation improves
the local approximation, but the square-root endpoint cannot be
reproduced by a finite polynomial in $\Ecal$. The entropy difference
remains finite while its derivative
$d\Delta\Scal/d\Ecal=-1/T_{\text H}$ diverges.

The limiting exponent is finite for both neutral and charged
paths approaching the lower critical curve at nonzero final charge
$\Qcal_f=\Qcal'(\Ecal_{\rm c})$,
\begin{equation}
\lim_{\Ecal\rightarrow\Ecal_{\rm c}^{-}}
\ln\Gamma_{\rm KPW}
\simeq
\frac{4\pi^2}{\kappa}
\left[
r_{\rm min}\left(\frac{\Qcal_f}{2\pi}\right)-r_+
\right].
\label{eq:BTZNLED_limiting_probability}
\end{equation}
For neutral emission, $\Qcal_f=\Qcal$.
For a charged approach, the full electric work factor in
Eq.~\eqref{eq:gen_charged_energy_integral} must be retained, as
shown in Appendix~\ref{appx:Puiseux_outer_endpoint}. For a tangent approach, it is this full integrand, rather than
$1/T_{\text H}$ alone, that determines the limiting action.
The resulting positive-mass, zero-temperature configuration contains
the double event horizon and a surviving simple innermost horizon.
For $0<\Mcal'<\Mcal_{\rm e}(\Qcal')$, the surviving
innermost root is the only positive horizon. Thus the lower critical
mass terminates the tracked outer branch, even though smaller-mass
black holes remain in the stationary family. A transition to this
smaller horizon would require a dynamical description.

\subsection{Time dependence and adiabaticity
\label{subsec:BTZNLED_endpoint_time}}

The horizon histories above are parametrized by emitted
energy. The KPW integral compares states along such a path, but
does not determine the time separating successive emissions.
To discuss this duration, we now use the stationary family as an
instantaneous description of an averaged evaporation process.
The cumulative energy becomes $\Ecal(t)$, where $t$ is the time at
the AdS boundary with the Killing normalization used for the mass
and temperature. The evolution of the parameters is supplied by energy and charge
balance.

Let $L_{\rm rad}=-d\Mcal'/dt>0$ denote the net luminosity,
measured as the black hole energy lost per unit boundary time.
Similarly, $J_{\rm rad}=-d\Qcal'/dt$ is the rate of loss of signed
electric charge. Conservation gives
\begin{align*}
\Mcal'(t)&=\Mcal-\Ecal(t),
&\Qcal'(t)&=\Qcal-\Qcal_{\rm em}\bigl[\Ecal(t)\bigr].
\end{align*}
In particular, an evolution following the prescribed path must
satisfy
\begin{equation}
\frac{d\Ecal}{dt}=L_{\rm rad},
\qquad
\frac{J_{\rm rad}}{L_{\rm rad}}
=\frac{d\Qcal_{\rm em}}{d\Ecal}.
\label{eq:BTZNLED_flux_path_compatibility}
\end{equation}
The charge profile fixes this ratio, while the luminosity sets
the rate at which the path is traversed. Its determination requires
the radiation spectrum and its transmission through the exterior,
together with boundary conditions allowing a net loss of energy,
as discussed in Sec.~\ref{sec:horizon_kinematics}.
The stationary first law also gives the instantaneous entropy rate,
\begin{equation*}
\frac{d\Scal}{dt}
=-\frac{L_{\rm rad}-\Phi_+J_{\rm rad}}{T_{\text H}}.
\end{equation*}
The positive horizon-energy condition used above therefore
becomes $L_{\rm rad}-\Phi_+J_{\rm rad}>0$.

We measure the approach to the selected outer coalescence by
the remaining energy
\begin{equation}
\Delta\Ecal(t):=\Ecal_{\rm c}-\Ecal(t)>0,
\qquad
\frac{d\Delta\Ecal}{dt}=-L_{\rm rad}.
\label{eq:BTZNLED_remaining_energy}
\end{equation}
For neutral emission this is the mass above the lower critical
value. For charged emission the critical mass itself varies, and
the mass difference governing the double zero instead has the form
\begin{equation*}
-\Delta\Mcal_{\rm e}\sim K(\Delta\Ecal)^n,
\qquad K>0,\quad n>0.
\end{equation*}
Here $K$ and $n$ are fixed by the chosen path. The analytic
case is derived in Appendix~\ref{appx:Puiseux_outer_endpoint}.
For a more general power-law approach, we assume that the leading
expansion can also be differentiated along the path. The two gaps
coincide for neutral emission, with $K=1$ and $n=1$.
Equation~\eqref{eq:BTZNLED_outer_coalescence_asymptotics} then
determines the temperature near the endpoint,
\begin{align}
T_{\text H}&\sim A_T(\Delta\Ecal)^{n/2},
\nonumber\\
A_T&:=\frac{1}{2\pi\ell}
\sqrt{\frac{\kappa K(1+w_{-1})}{\pi w_{-1}}}>0.
\label{eq:BTZNLED_time_temperature}
\end{align}

Choose $t_0$ sufficiently near coalescence that these
expansions apply, and write
$\Delta\Ecal_0:=\Delta\Ecal(t_0)>0$. Expressing the luminosity
along the path as $L_{\rm rad}(\Delta\Ecal)$, integration of
Eq.~\eqref{eq:BTZNLED_remaining_energy} gives
\begin{equation}
t-t_0
=\int_{\Delta\Ecal(t)}^{\Delta\Ecal_0}
\frac{d\xi}{L_{\rm rad}(\xi)}.
\label{eq:BTZNLED_endpoint_time}
\end{equation}
The duration of the final approach is the limit as the lower
integration bound tends to zero. For a luminosity whose leading
behavior is $L_{\rm rad}\sim L_0(\Delta\Ecal)^\nu$, with
$L_0>0$, retaining this leading term yields
\begin{equation}
\Delta\Ecal(t)\simeq
\begin{cases}
\left[(\Delta\Ecal_0)^{1-\nu}
-(1-\nu)L_0(t-t_0)\right]^{\frac{1}{1-\nu}},&\nu\ne1,\\[3pt]
\Delta\Ecal_0 e^{-L_0(t-t_0)},&\nu=1.
\end{cases}
\label{eq:BTZNLED_gap_time_power_law}
\end{equation}
The endpoint is reached in finite time only for $\nu<1$,
with $t_{\rm c}-t_0\simeq
(\Delta\Ecal_0)^{1-\nu}/[(1-\nu)L_0]$. Otherwise the approach
is exponential for $\nu=1$ and algebraic for $\nu>1$.
The convergence criterion follows directly from
Eq.~\eqref{eq:BTZNLED_endpoint_time}, independently of the
subleading terms in the luminosity.

To approximate the radiation temperature by the stationary
$T_{\text H}$, the inverse surface gravity timescale
$(2\pi T_{\text H})^{-1}$ must be short compared with the
temperature-variation timescale
$T_{\text H}/|dT_{\text H}/dt|$. Their ratio gives the
adiabaticity condition
\begin{equation}
\epsilon_{\rm ad}:=
\frac{|dT_{\text H}/dt|}{2\pi T_{\text H}^{\,2}}\ll1.
\label{eq:BTZNLED_endpoint_adiabaticity}
\end{equation}
This implements the slow variation criterion for a Hawking-like
spectrum \cite{Barcelo:2010pj}, assuming that the instantaneous
surface gravity approximates the peeling rate of the outgoing
rays. Using the differentiable endpoint expansion and
$d\Delta\Ecal/dt=-L_{\rm rad}$ gives
\begin{equation}
\epsilon_{\rm ad}
\sim\frac{nL_0}{4\pi A_T}
(\Delta\Ecal)^{\nu-1-n/2}.
\label{eq:BTZNLED_adiabatic_power_law}
\end{equation}
Consequently, $\epsilon_{\rm ad}$ decreases towards coalescence
when $\nu>1+n/2$. At $\nu=1+n/2$, it remains small if
$nL_0/(4\pi A_T)\ll1$. For a transverse approach the marginal
power is $\nu=3/2$, while quadratic contact raises it to $\nu=2$.
For $\nu<1+n/2$, the adiabaticity parameter grows without bound.
In particular, an infinite approach time does not suffice: the
interval $1\leq\nu<1+n/2$ has an infinite duration but still
loses adiabatic control.

The same calculation locates this loss of control. For a
fixed small tolerance $\epsilon_*>0$, a power law with
$\nu<1+n/2$ reaches $\epsilon_{\rm ad}=\epsilon_*$ at
\begin{equation}
\Delta\Ecal_{\rm ad}\simeq
\left(\frac{nL_0}{4\pi A_T\epsilon_*}\right)^{
\frac{1}{1+n/2-\nu}}.
\label{eq:BTZNLED_adiabatic_gap}
\end{equation}
This estimate applies if the evolution was initially adiabatic
and the resulting gap lies within the endpoint expansion.
Without specifying a luminosity law, a uniform bound
$\epsilon_{\rm ad}\leq\epsilon_*$ implies
$|d(1/T_{\text H})/dt|\leq2\pi\epsilon_*$. Integrating gives
\begin{equation}
T_{\text H}(t)\geq
\frac{T_{\text H}(t_0)}{
1+2\pi\epsilon_*T_{\text H}(t_0)(t-t_0)}.
\label{eq:BTZNLED_temperature_time_bound}
\end{equation}
A zero temperature is therefore excluded at finite boundary
time while this adiabatic bound remains satisfied. The
incompatibility between finite evaporation time and continued
adiabaticity was also found for regular four-dimensional black holes
with a Stefan--Boltzmann loss law \cite{Carballo-Rubio:2018pmi}.

For neutral emission into massless bulk modes, an ideal
blackbody estimate in $2+1$ dimensions takes
$L_{\rm rad}\propto(2\pi R_+)T_{\text H}^3$
\cite{Page:2015rxa}. Since $R_+\to r_{\rm min}>0$, this yields $\nu=3/2$, the marginal adiabatic case for $n=1$.
The actual net luminosity depends on greybody transmission and
returning fluxes, and this estimate requires the thermal regime
examined next.

\subsection{Finite-energy emission and the thermal regime
\label{subsec:BTZNLED_endpoint_thermal_regime}}

Slow variation in time does not ensure that one emission
produces only a small change in the black hole state. A canonical
thermal treatment also requires this latter condition
\cite{Preskill:1991tb}. We examine it on the neutral branch, where
$\Mcal'=\Mcal_{\rm e}+\Delta\Ecal$ and
$T_{\text H}=A_T\sqrt{\Delta\Ecal}+\Order(\Delta\Ecal)$,
with $K=1$ and $n=1$ in Eq.~\eqref{eq:BTZNLED_time_temperature}.
Writing $\Scal_{\rm e}$ for the limiting entropy, the first law and
the fixed-charge heat capacity give
\begin{align}
\Scal-\Scal_{\rm e}
&=\frac{2\sqrt{\Delta\Ecal}}{A_T}+\Order(\Delta\Ecal),
\nonumber\\
C_{\Qcal}
&=\frac{2\sqrt{\Delta\Ecal}}{A_T}+\Order(\Delta\Ecal).
\label{eq:BTZNLED_neutral_entropy_gap}
\end{align}

Consider one additional neutral emission of Killing energy
$0<\omega\leq\Delta\Ecal$, leaving a gap $\Delta\Ecal-\omega$
on the tracked outer branch.  The same $\omega$ denotes the angular frequency conjugate to the normalized time $t$.
The leading KPW exponent is
$\ln\Gamma_{\rm KPW}(\omega)\simeq\Delta\Scal_\omega$, where
\begin{align}
\Delta\Scal_\omega
&:=\Scal(\Mcal'-\omega,\Qcal)-\Scal(\Mcal',\Qcal)
\nonumber\\
&=\frac{2}{A_T}
\left[\sqrt{\Delta\Ecal-\omega}-\sqrt{\Delta\Ecal}\right]
+\Order(\omega).
\label{eq:BTZNLED_endpoint_step_exponent}
\end{align}
The remainder follows from the regular next term in the entropy
expansion.

For this neutral NLED branch, the same square-root expression
also gives a bound on the full entropy exponent. Equation~
\eqref{eq:BTZNLED_mass_derivatives} shows that
second and third derivatives of the mass function are positive for
$r\geq r_{\rm min}$. Since $\partial_r\mgot(r_{\rm min})=0$,
integration of
$\partial_r[(\partial_r\mgot)^2]
=2(\partial_r\mgot)(\partial_r^2\mgot)$ gives
\begin{equation*}
[\partial_r\mgot(r)]^2
<2[\mgot(r)-\Mcal_{\rm e}]\,\partial_r^2\mgot(r),
\qquad r>r_{\rm min}.
\end{equation*}
Together with $T_{\text H}=\kappa\partial_r\mgot/(4\pi^2)$,
this means that $T_{\text H}/\sqrt{\Delta\Ecal}$ increases with
the mass above coalescence. Integrating the fixed-charge first law
from $\Mcal'-\omega$ to $\Mcal'$ therefore yields
\begin{equation}
-\Delta\Scal_\omega
>
\frac{\omega}{T_{\text H}}
\frac{2}{1+\sqrt{1-\omega/\Delta\Ecal}},
\qquad 0<\omega\leq\Delta\Ecal.
\label{eq:BTZNLED_finite_energy_suppression_bound}
\end{equation}
Here $T_{\text H}$ is evaluated before the emission, and
$\omega=\Delta\Ecal$ is understood as a limit through simple
outer horizons. The bound is saturated at leading order in the
endpoint expansion. At that order, the ratio of the suppression
exponent to its thermal estimate increases from one to two as the
fraction $\omega/\Delta\Ecal$ increases from zero to one.

For $\omega\ll\Delta\Ecal$, the thermal expansion
\eqref{eq:gen_thermal_limit}, evaluated at the current state, has
linear term $-\omega/T_{\text H}$ and quadratic term
$-\omega^2/(2C_{\Qcal}T_{\text H}^{\,2})$.
The relative size of the quadratic correction is
\begin{equation}
\frac{\omega}{2C_{\Qcal}T_{\text H}}
=\frac{\omega}{4\Delta\Ecal}
\bigl[1+\Order(\sqrt{\Delta\Ecal})\bigr].
\label{eq:BTZNLED_endpoint_thermal_correction}
\end{equation}
For radiation with characteristic energies
$\omega\sim T_{\text H}$, the ratio
$\omega/\Delta\Ecal$ measures the fraction of the remaining
energy removed by one quantum. The thermal approximation therefore
requires
\begin{equation}
\frac{T_{\text H}}{\Delta\Ecal}\ll1,
\qquad\text{or equivalently}\qquad
\Delta\Ecal\gg A_T^2.
\label{eq:BTZNLED_endpoint_thermal_window}
\end{equation}
For the neutral solution the scale on the right is explicit,
\begin{equation}
A_T^2=\frac{\kappa(1+w_{-1})}{4\pi^3\ell^2w_{-1}}.
\label{eq:BTZNLED_neutral_thermal_scale}
\end{equation}
At fixed couplings, $w_{-1}$ is independent of charge by
Eq.~\eqref{eq:BTZNLED_Lambert_values}. The leading energy scale
$A_T^2$ is therefore the same at different charges, even though
the critical mass and entropy scale quadratically and linearly
with the charge magnitude, respectively. This follows from the
charge-independent radial curvature at the nondegenerate minimum.

Due to Eq.~\eqref{eq:BTZNLED_neutral_entropy_gap}, the thermal
window is also the regime
$C_{\Qcal}\sim\Scal-\Scal_{\rm e}\gg1$. The entropy excess,
rather than the possibly large limiting entropy, controls the
change caused by a typical quantum. To use the endpoint expansion
in this regime, the higher radial terms in the mass--radius
relation must remain small as well. Equation~
\eqref{eq:BTZNLED_mass_derivatives} gives the ratio of its cubic
to quadratic term as
$(\Scal-\Scal_{\rm e})/[3\Scal_{\rm e}|1+w_{-1}|]$,
where $\Scal-\Scal_{\rm e}=4\pi^2(R_+-r_{\rm min})/\kappa$
is exact. The two requirements consequently give
\begin{equation}
1\ll\Scal-\Scal_{\rm e}
\ll\Scal_{\rm e}|1+w_{-1}|.
\label{eq:BTZNLED_endpoint_entropy_hierarchy}
\end{equation}

This competition can also be expressed in energy. Define the
scale associated with the quadratic endpoint expansion by
\begin{align}
\Ecal_{\rm geom}
&:=\frac{A_T^2}{4}
\bigl[\Scal_{\rm e}|1+w_{-1}|\bigr]^2
 =\frac{2\pi\kappa q^2\ell^2}{\gamma^2}
|w_{-1}|\,|1+w_{-1}|^3.
\label{eq:BTZNLED_geometric_endpoint_scale}
\end{align}
It is a characteristic scale, not a sharp boundary. Within the
endpoint expansion, Eq.~\eqref{eq:BTZNLED_endpoint_entropy_hierarchy}
is equivalent parametrically to
\begin{equation}
A_T^2\ll\Delta\Ecal\ll\Ecal_{\rm geom}.
\label{eq:BTZNLED_endpoint_energy_hierarchy}
\end{equation}
This interval requires $\Scal_{\rm e}|1+w_{-1}|\gg1$.
Increasing the charge at fixed couplings raises its upper scale
as $q^2$, while leaving its lower scale unchanged.

The approach to triple degeneracy has the opposite effect.
Writing $d:=|1+w_{-1}|\in(0,1)$ in the positive-mass sector
\eqref{eq:BTZNLED_positive_mass_window}, the Lambert relation gives
$\kappa\beta\ell^2=e^d/(1+d)$, so
$d\sim\sqrt{2(\kappa\beta\ell^2-1)}$ as
$\kappa\beta\ell^2\to1^+$. The two scales then behave as
\begin{align}
A_T^2&\sim\frac{\kappa}{4\pi^3\ell^2}d,
\nonumber\\
\Ecal_{\rm geom}
&\sim\frac{2\pi\kappa q^2\ell^2}{\gamma^2}d^3.
\label{eq:BTZNLED_near_triple_scales}
\end{align}
At fixed $(q,\kappa,\ell,\gamma)$, the upper scale therefore
contracts as $d^3$,
faster than the thermal scale, which decreases as $d$. A large
limiting entropy alone does not ensure that the two approximations
overlap near triple degeneracy. The conclusion concerns the
quadratic double-root description. In the triple-zero regime the
cubic term must be retained.

A luminosity computed from a thermal spectrum must be reconsidered
when typical quanta carry a substantial fraction of the remaining
energy. Here $\Delta\Ecal$ bounds emissions that stay on the
tracked outer branch. Smaller masses still admit a different
event horizon. The adiabaticity condition and the finite-energy
condition thus test the variation in time and the change of state
during one emission, respectively.

The distinction can be seen in the neutral marginal case,
with $n=1$ and $\nu=3/2$. Taking
$L_{\rm rad}\simeq L_0(\Delta\Ecal)^{3/2}$ as an averaged
loss law, the leading solution reads
\begin{align}
\Delta\Ecal(t)
&\simeq\left[(\Delta\Ecal_0)^{-1/2}
+\frac{L_0}{2}(t-t_0)\right]^{-2},
\nonumber\\
T_{\text H}(t)
&\simeq\frac{A_T}{
(\Delta\Ecal_0)^{-1/2}+L_0(t-t_0)/2}.
\label{eq:BTZNLED_neutral_marginal_history}
\end{align}
At this order, the adiabaticity parameter is constant,
$\epsilon_{\rm ad}\simeq L_0/(4\pi A_T)$, and the
slow-variation condition requires this ratio to be small.
The horizon separation and entropy excess both decrease as
$t^{-1}$, so $\Scal_{\rm e}$ is approached only at infinite
time along this formal history.

For this marginal history to begin in a controlled thermal
endpoint regime, its initial gap must satisfy
Eq.~\eqref{eq:BTZNLED_endpoint_energy_hierarchy}.
As $\Delta\Ecal$ decreases, the endpoint expansion improves, but
the thermal approximation eventually fails even if
$\epsilon_{\rm ad}$ stays small. Continuing a luminosity derived
from a thermal spectrum beyond that range requires a treatment of
finite-energy emission. The entropy exponent, the time to
coalescence, and the validity of a thermal spectrum thus require
separate information. The stationary family fixes the first and
constrains the other two once a radiation law is supplied. The
complementary one-horizon sector considered next allows us to
examine the local temperature and the finite-energy exponent away
from these horizon mergers.

\section{ Horizon emission from the quantum-backreacted BTZ geometry
\label{sec:BTZNLED_semiclassical_emission}
\label{subsec:BTZNLED_semiclassical_emission}}

Finally, a complementary application of particular interest is provided by exploring the one-horizon sector under a suitable choice of parameters. This subfamily  incorporates the leading quantum-backreacted BTZ geometry
coupled to NMG. Accordingly,  we provide a comparison of the neutral probe horizon
factor before following finite-energy emission in the
Einstein--NLED realization. Corrections to the BTZ horizon and
temperature from the backreaction of conformally coupled scalars were studied in
Einstein gravity by Mart\'inez and Zanelli \cite{Martinez:1996uv},
where the blackening function acquires a $1/r$ term. In NMG, the
solution obtained in Ref.~\cite{Chernicoff:2024dll} at leading order
in the backreaction of $N\gg1$ conformal scalar fields has
\begin{align}
f_{\rm sc}(r;M)
&=\frac{r^2}{\ell^2}-M
+\alpha_{\rm sc}(M)r\ln\left(\frac{r}{\e\ell}\right),
\label{eq:BTZNLED_semiclassical_metric}
\\
\alpha_{\rm sc}(M)
&:=\frac{2N\ell_{\rm p}F_{\rm sc}(M)}{M\ell^2}>0.
\label{eq:BTZNLED_semiclassical_coefficient}
\end{align}
Here $M>0$ is the seed BTZ mass parameter and $\ell_{\rm p}=8\pi G_3$.
The positive, dimensionless function $F_{\rm sc}(M)$
enters the scalar expectation value $\langle T^\mu{}_{\nu}\rangle$
with transparent boundary conditions at infinity.
Matching Eq.~\eqref{eq:BTZNLED_blackening} at a given $M$ requires
\cite{Mendez-Zavaleta:2026rgg}
\begin{equation}
\frac{q}{\gamma}
=-\frac{\alpha_{\rm sc}(M)}{2\sqrt{2}\kappa},
\qquad
\beta=-\frac{\alpha_{\rm sc}(M)}{2\e\kappa\ell}.
\label{eq:BTZNLED_semiclassical_matching}
\end{equation}

For a neutral minimally coupled test scalar, the same
normalization of $\partial_t$ in the two realizations gives the
same leading WKB pole. It can be evaluated on the corrected mean
geometry without using a gravitational entropy formula.
Denote the corrected event horizon radius by $r_{\rm h}$,
and introduce
\begin{equation*}
r_{\rm B}:=\ell\sqrt{M},
\qquad
\varepsilon_{\rm q}(M)
:=\frac{\alpha_{\rm sc}(M)\ell}{2\sqrt{M}}
=\frac{N\ell_{\rm p}F_{\rm sc}(M)}{\ell M^{3/2}},
\end{equation*}
where $r_{\rm B}$ is the uncorrected BTZ radius. Expanding
the horizon condition about this simple root recovers the
displacement obtained in Ref.~\cite{Chernicoff:2024dll},
\begin{equation}
\frac{r_{\rm h}}{r_{\rm B}}
=1+\varepsilon_{\rm q}(1-\ln\sqrt{M})
+\Order(\varepsilon_{\rm q}^2).
\label{eq:BTZNLED_quantum_radius_shift}
\end{equation}
The remainder is an expansion at fixed $M$ and $\ell$.
Small relative corrections to the horizon radius and
temperature require
\begin{equation*}
\varepsilon_{\rm q}\left|1-\ln\sqrt{M}\right|\ll1,
\qquad
\varepsilon_{\rm q}\ll1.
\end{equation*}

The temperature must be evaluated at the displaced horizon.
The radial derivative of Eq.~\eqref{eq:BTZNLED_semiclassical_metric}
is $2r/\ell^2+\alpha_{\rm sc}\ln(r/\ell)$, so the first-order
change in its horizon value is
\begin{equation*}
4\pi\,\delta T_{\text H}
=\frac{2(r_{\rm h}-r_{\rm B})}{\ell^2}
+\alpha_{\rm sc}\ln\left(\frac{r_{\rm B}}{\ell}\right)
=\alpha_{\rm sc}.
\end{equation*}
The logarithms cancel between the displacement and the explicit
change in the radial slope.
Writing $T_{\rm sc}$ for the corrected Hawking temperature,
we obtain
\begin{equation}
\frac{T_{\rm sc}}{T_{\rm B}}
=1+\varepsilon_{\rm q}+\Order(\varepsilon_{\rm q}^2),
\qquad
T_{\rm B}:=\frac{\sqrt{M}}{2\pi\ell}.
\label{eq:BTZNLED_quantum_temperature_shift}
\end{equation}
The correction to the temperature is positive at fixed seed mass
and Killing-time normalization. For $M>\e^2$ the horizon correction
in Eq.~\eqref{eq:BTZNLED_quantum_radius_shift} is negative, so the
quantum-backreacted horizon is smaller than the BTZ horizon and
nevertheless hotter.

For a neutral excitation of Killing energy $\omega>0$,
treated here without its own backreaction, the leading horizon
emission-to-absorption ratio is
$\mathcal{R}_{\rm H}(\omega)\simeq\exp(-\omega/T_{\text H})$
\cite{Vanzo:2011wq}. Equations~\eqref{eq:BTZNLED_semiclassical_coefficient}
and \eqref{eq:BTZNLED_quantum_temperature_shift} then give
\begin{equation}
\ln\left(\frac{\mathcal{R}_{\rm H}^{\rm sc}(\omega)}
{\mathcal{R}_{\rm H}^{\rm BTZ}(\omega)}\right)
=\frac{2\pi N\ell_{\rm p}F_{\rm sc}(M)}{M^2}\,\omega
+\Order\!\left(\frac{\omega}{T_{\rm B}}\varepsilon_{\rm q}^2\right).
\label{eq:BTZNLED_quantum_horizon_factor}
\end{equation}
The correction enhances the horizon emission factor relative
to BTZ at the same $M$ and $\omega$ in both matched geometries.
This comparison retains the leading WKB action on the background
corrected to first order in $N\ell_{\rm p}$.

Emission at finite energy additionally probes the path followed
through the family. In the semiclassical NMG solution, the source
fixes $\alpha_{\rm sc}$ as a function of the seed mass.
By Eq.~\eqref{eq:BTZNLED_semiclassical_matching}, this variation
differs from neutral emission at fixed NLED couplings, since fixed
$\beta$ fixes $\alpha_{\rm sc}$. The finite-energy NMG calculation
also requires its own Hamiltonian energy and entropy, including
the quantum source \footnote{The
distinction is present even for the uncorrected BTZ metric. For the
NMG action used in Ref.~\cite{Chernicoff:2024dll}, the curvature
coupling is $m_{\rm g}^2\ell^2=-1/2$ at the special point.
The Wald entropy \cite{Wald:1993nt} is then
$\Scal_{\rm NMG}=(1-1/(2m_{\rm g}^2\ell^2))\Scal_{\rm E}
=2\Scal_{\rm E}$, where $\Scal_{\rm E}$ is the Einstein area
entropy of the same BTZ horizon.}. For the Einstein--NLED realization,
the first law at fixed couplings derived in Appendix~\ref{appx:termo}
allows an explicit finite-energy comparison.

Match the NLED solution at an initial seed mass $M_0$ and
write $\alpha_0:=\alpha_{\rm sc}(M_0)$. Neutral emission keeps $\alpha_0$ and the
logarithmic reference scale fixed. The final mass parameter is
\begin{equation*}
M_f=M_0-\frac{\kappa\Ecal}{\pi}>0,
\end{equation*}
and the final blackening function is
$r^2/\ell^2-M_f+\alpha_0r\ln[r/(\e\ell)]$.
Applying the horizon expansion at $M_0$ and $M_f$ and using
the Einstein entropy in Eq.~\eqref{eq:BTZNLED_probability_radius}
gives $\ln\Gamma_{\rm NLED}\simeq\Delta\Scal_{\rm NLED}$, with
\begin{align}
\Delta\Scal_{\rm NLED}
={}&\frac{4\pi^2\ell\sqrt{M_0}}{\kappa}
\left[\sqrt{1-\frac{\kappa\Ecal}{\pi M_0}}-1\right]
 -\frac{2\pi^2N\ell_{\rm p}F_{\rm sc}(M_0)}{\kappa M_0}
\ln\left(1-\frac{\kappa\Ecal}{\pi M_0}\right)
\nonumber\\
&+\Order\bigl[\varepsilon_{\rm q}(M_0)^2\bigr].
\label{eq:BTZNLED_matched_finite_energy_exponent}
\end{align}
This expression retains the full dependence on emitted energy at
first order in the initial background correction. The first term is
the finite-energy BTZ exponent. The second is positive for
$0<\Ecal<\pi M_0/\kappa$ and expresses the correction in the
quantum parameters of the initial matched geometry. Its small
$\Ecal$ limit reproduces $-\Ecal/T_{\rm sc}(M_0)$ to the same
perturbative order. The remainder is taken at fixed $M_0$ and
$\Ecal$.

Along the fixed-$\alpha_0$ continuation,
$\partial_r^2 f=2/\ell^2+\alpha_0/r>0$, so the temperature decreases
as the event horizon shrinks. The exact NLED exponent consequently
obeys $\Delta\Scal_{\rm NLED}<-\Ecal/T_{\text H}(0)$ for a
nonzero emission while this horizon remains simple. The initial
temperature $T_{\text H}(0)$ agrees with $T_{\rm sc}(M_0)$ at the
retained order.
The positive first-order background correction thus raises
the exponent relative to the BTZ emission result, while the
exact NLED result remains below its own initial thermal tangent.

The expansion requires small relative corrections to both
the temperature and the horizon radius,
\begin{equation*}
\frac{\alpha_0\ell}{2\sqrt{M'}}\ll1,
\qquad
\frac{\alpha_0\ell}{2\sqrt{M'}}
\left|1-\ln\sqrt{M'}\right|\ll1,
\end{equation*}
throughout $M_f\leq M'\leq M_0$.
The logarithmic correction in
Eq.~\eqref{eq:BTZNLED_matched_finite_energy_exponent} cannot be
continued to $M_f=0$ using this expansion.

Within its domain of validity, the comparison separates the
part of horizon emission fixed by the shared geometry from the
additional input needed for a finite change of state. The corrected
temperature determines the neutral-probe horizon factor in both
realizations. The finite-energy exponent computed here follows the
Einstein--NLED family at fixed couplings and its related first law. The matching
at the initial geometry does not make it the evaporation exponent
of the semiclassical NMG family.

\section{Discussion and concluding remarks
\label{sec:discussion}}

The KPW approach to Hawking radiation has found extensive application
in black hole physics by incorporating energy conservation into the
semiclassical description of emission. Revisiting the horizon
condition within this framework reveals how the accompanying loss
of mass and charge can affect the horizon locations, their number
and causal nature. For the analytic static spacetimes of the form
\eqref{eq:gen_static_metric}, wherever the local mass inversion
is well defined, the Lagrange--B\"urmann formula and its extension
through Puiseux series provide local expressions for the motion of
simple and degenerate roots without requiring an explicit solution
for the horizon radii. Furthermore, composing this inverse with
the appropriate gravitational entropy determines the leading KPW
exponent, provided the stationary first law holds along the emission
path. At the same time, the coefficient formulas developed in
Sec.~\ref{sec:horizon_kinematics} and
Appendices~\ref{appx:Lagrange} and \ref{appx:Puiseux} make
the corrections to the thermal exponent accessible to arbitrary
order.

A natural extension of this construction arises when a horizon
becomes degenerate, since an emission can then change the number of
nearby real roots. The corresponding critical curve consists of
the mass and charge values for which the double horizon exists.
The sign of the mass difference from this curve, together with the
radial curvature of the mass function, determines whether two
simple horizons emerge or the nearby real pair is absent.
The charge emission profile is therefore an essential part of
the description. Once the pair exists, its separation is proportional
to the square root of the magnitude of this mass difference,
giving a square-root dependence on the emitted energy for a
transverse departure and a linear dependence for a quadratic one.
The double horizon itself is preserved along paths that remain on the critical curve throughout.

We found a particularly well-suited case study in the recently
presented logarithmic deformation of the BTZ black hole
\cite{Mendez-Zavaleta:2026rgg}, which realizes these possibilities
within a single matter theory. Starting with a double inner horizon
and a separate simple event horizon, we show that neutral emission
necessarily splits the former into two simple horizons, revealing
an internal horizon fine structure. The behavior is richer for
charged channels, which allow for immediate splitting, delayed
recovery of an inner pair, or persistent inner degeneracy, while
the event horizon remains simple. Here the analyticity of
the charge emission profile has a useful consequence: the outer
radius and entropy exponent remain locally analytic through these
internal transitions, so the creation or loss of an inner pair
produces no nonanalytic marker in the leading horizon emission
factor. An instructive comparison is given in
Fig.~\ref{fig:BTZNLED_further_channels}, where we present one
channel that preserves the double inner horizon, while another
separates it into two simple horizons and later brings them
together again. Both connect the same initial and final states
and consequently have the same leading KPW exponent, despite
their different intermediate horizon configurations.

Through a formal continuation of the process, we find that the
larger of the newly separated inner horizons can coalesce with
the event horizon, driving the sequence towards an extremal black
hole. Its temperature then tends to zero but, surprisingly, the
entropy difference and hence the leading tunneling exponent have
a finite limit. For charged approaches tangent to the outer
critical curve, retaining the electric work in the residue is
essential to recover this behavior. Smaller positive masses below
the coalescence value still admit the continuation of the innermost
root as a simple event horizon. This critical mass therefore ends
the tracked outer branch, with a transition to the smaller horizon
requiring a dynamical description.

This approach towards extremality has an interesting counterpart
in Ref.~\cite{Madler:2026fve}.
The fully nonlinear evolutions of the four-dimensional
Einstein--Maxwell--charged-scalar system produce near-extremal
black holes from regular, horizon-free initial data, with this
character independently identified through radiation at future
null infinity. There, the near-extremal regime emerges during
formation, whereas in our case Hawking emission can drive an
already formed black hole towards a degenerate event horizon.
A common feature of the two settings is the role of the relative
loss of mass and charge. In their evolutions these fluxes are determined
dynamically, while the emission profiles considered here make
their influence on the horizon motion explicit.

The prospect of reaching extremality naturally raises the question
of how long this approach takes, since the entropy exponent does
not specify the interval between successive emissions. To that end,
the energy and charge balance examined in
Sec.~\ref{subsec:BTZNLED_endpoint_time} relates the charge emission
profile to the ratio of charge and energy fluxes, while the
luminosity determines how quickly the path is traversed.
A sufficiently rapid decrease of the luminosity makes the approach
asymptotic in time. Adiabaticity imposes a stronger restriction,
however, because the inverse surface gravity also diverges and the
geometry must evolve slowly on this increasing time scale.
More generally, the cooling bound in
Eq.~\eqref{eq:BTZNLED_temperature_time_bound} excludes reaching
zero temperature in finite boundary time while a uniform bound
on adiabaticity remains valid, independently of the particular
luminosity law.

Interestingly, even such a slow approach need not remain thermal.
A quantum with energy of order $T_{\text H}$ can eventually carry
away an appreciable fraction of the energy remaining before
coalescence, so that the temperature can change substantially
during a single emission. For neutral emission, this effect is
captured by the exact bound
\eqref{eq:BTZNLED_finite_energy_suppression_bound}, which quantifies
the suppression relative to the Boltzmann estimate without
truncating the entropy change at quadratic order. Near extremality,
the available thermal reservoir is measured by the entropy above
the limiting configuration, rather than by the total entropy
\cite{Preskill:1991tb}. Requiring this entropy excess to be large
while remaining within the quadratic radial expansion gives the
overlap interval in
Eq.~\eqref{eq:BTZNLED_endpoint_energy_hierarchy}.
This interval widens with increasing charge magnitude at fixed
couplings, but contracts towards triple degeneracy. In the latter
case, the geometric upper scale decreases faster than the thermal
lower scale. To examine the thermal regime near a triple horizon,
the mass--radius relation must therefore be expanded through cubic
order in the displacement from the critical radius.

Our final application concerns the one-horizon sector, which
reproduces the quantum-backreacted BTZ geometry of NMG
\cite{Chernicoff:2024dll}. It offers an opportunity to examine
Hawking emission from a black hole whose geometry already
incorporates a semiclassical effect. At the matched background and
with the same normalization of time, both realizations give the
same temperature and leading horizon emission factor for a neutral
minimally coupled probe. At fixed seed mass, the leading
semiclassical correction raises the temperature, even when the
corrected horizon is smaller than the BTZ horizon.
Following a finite emission in the Einstein--NLED realization at
fixed couplings gives, in addition, a positive logarithmic correction
to the BTZ entropy exponent. The resulting exponent recovers the
corrected temperature in the limit of small emitted energy.
This enhancement relative to BTZ coexists with suppression relative to the initial thermal
estimate of the corrected black hole, since the NLED configuration
cools as energy is extracted. These results apply within the
perturbative regime specified in
Sec.~\ref{sec:BTZNLED_semiclassical_emission}. Following the
family obtained through semiclassical backreaction in NMG \cite{Chernicoff:2024dll} would
instead change the quantum source with the mass and require the
energy and entropy of that theory for the finite-emission
calculation.

An interesting question is what happens if the initial state is
already an extremal black hole. In the family studied here, neutral
mass loss from the outer critical curve removes the double outer
horizon, leaving the previously innermost root as a simple event
horizon. A charged path along that curve, on the other hand,
preserves extremality within the stationary family as the mass and
charge magnitude decrease. To recover two distinct outer horizons, the
remaining mass must lie above the critical mass at the new charge.
A transverse departure into this region requires negative horizon
energy at leading order, outside the ordinary channels considered
here, while tangential departures must be decided at higher orders.
The role of this energy--charge balance recalls classical absorption
tests of cosmic censorship \cite{Wald:1974hkz,Sorce:2017dst}, with
the distinction that the smaller horizon survives here and the
issue is a transition between horizon configurations. Determining
a physical emission rate for any of these paths calls for a
calculation of tunneling directly from the degenerate state
\cite{Angheben:2005rm}.

One of the main remaining questions is which of these
stationary histories can be realized during evaporation. This
requires the radiation spectrum and an analysis of the exterior's
linear stability. Backscattering in the exterior, or reflection
at the AdS boundary, can return energy and, for charged radiation,
charge to the hole, changing the net fluxes that select the
emission path. Alongside this scattering problem, the coupled
gravitational and NLED perturbations would determine whether
growing modes can disrupt the slow evolution assumed here.
Charged-scalar studies of other three-dimensional AdS black holes
illustrate the dependence of superradiance and mode stability on
the matter model and boundary conditions
\cite{Gonzalez:2021vwp,Konewko:2023gbu}.

The fate of the internal horizon fine structure also depends
on how these perturbations evolve inside the black hole. The mass
inflation mechanism of Poisson and Israel \cite{Poisson:1989zz}
and the analysis of strong cosmic censorship for linearly charged
BTZ \cite{Singha:2022bvr} make this a necessary part of the
dynamical problem. Semiclassical effects can also destabilize an
inner horizon with vanishing surface gravity, as shown for
stationary regular black holes in Ref.~\cite{McMaken:2023uue}.
The role of time dependence is illustrated by the finite
renormalized stress tensor obtained for trapped regions of finite
lifetime in Ref.~\cite{Carballo-Rubio:2026gwg}, within the $s$-wave
Polyakov approximation. Bringing these effects into the present
family would help determine which features of its horizon fine
structure persist during evaporation.

Finally, a natural extension of this study is to account for
non-static spacetimes. Recent calculations of near-extremal
Kerr--Newman evaporation at fixed electric charge show how the
competition between $s$-wave and superradiant emission, with
quantum corrections included, can slow the late-time evolution
\cite{Luo:2026epp}. Closer to our setting, generalized
Chern--Simons interactions offer a way to support charged
rotating geometries by supplying force-free electromagnetic
currents without contributing directly to the stress tensor
\cite{Deshpande:2024vbn,Hale:2024zvu}. This approach accommodates
the original charged rotating BTZ solution and extends to
nonlinear electrodynamics, with further applications to
rotating AdS and Lifshitz black holes
\cite{Bravo-Gaete:2025lgs,Bravo-Gaete:2026bdq}. These constructions
suggest a concrete route to rotating counterparts of the family
studied here through an enlargement of its matter sector,
allowing angular-momentum extraction to join the loss of mass
and charge in selecting paths through the critical configurations.
A separate extension to genuinely time-dependent geometries can
build on the tunneling description of evolving trapping horizons
in spherical symmetry \cite{Vanzo:2011wq}. In both directions,
an appealing question is whether the radiation itself favors
histories that preserve horizon degeneracy or lead towards
extremality.

\acknowledgments

The authors are thankful to Eloy Ay\'on-Beato and Borja Diez for their
feedback on this work, and to Radouane Gannouji for useful
discussions on recent results complementary to those presented here. MAAC was partially supported by the \emph{Pro-Mejoras} program of Universidad Veracruzana.
MB is supported by \emph{Proyecto Interno UCM, Línea Regular UCM-IN-25202} and FONDECYT 1262452. ER aknowledges encouragement from PRODEP-M\'exico, CA-UV-320:
Algebra, Geometr\'ia y Gravitaci\'on.
JAMZ was partially funded by the \emph{Estancias Posdoctorales por México} program of SECIHTI. ER and JAMZ wer partially supported by the \emph{Sistema Nacional de Investigadoras e Investigadores} (SNII), also under SECIHTI.

\appendix

\section{Lagrange--B\"urmann formulae
\label{appx:Lagrange}}

The horizon and entropy coefficients in
Eqs.~\eqref{eq:gen_horizon_coefficients} and
\eqref{eq:gen_entropy_coefficients} follow from the same inversion
problem set by the horizon condition. We give its derivation in the notation of
Sec.~\ref{sec:horizon_kinematics}, keeping all parameters
$\lambda^A$ fixed. The analytic and formal versions of the
LB construction are extensively discussed in
Refs.~\cite{Gessel:2016Lagrange,Sambale:2022fps}.

\subsection{The inverse horizon map}

Consider the function $h$ defined in
Eq.~\eqref{eq:gen_h_definition}. Its Maclaurin expansion about
$y=0$ is
\begin{equation}
h(y)
=
-\sum_{j=1}^{\infty}
\frac{\partial_r^j\mgot(r_+)}{j!}y^j,
\qquad
\partial_y h(0)=-\partial_r\mgot(r_+)\ne0.
\label{eq:LB_mass_expansion}
\end{equation}
If $\mgot$ is analytic near $r_+$, the inverse function
theorem gives a unique analytic solution
$y=\Delta r_+(\Ecal)$ of $\Ecal=h(y)$ with
$\Delta r_+(0)=0$. The ratio $y/h(y)$ has a removable singularity
at the origin, with value $-1/[\partial_r\mgot(r_+)]$.

Let $g(y)$ be analytic in the same neighborhood, and denote by
$[\Ecal^n]$ the coefficient of $\Ecal^n$. For $n\geq1$, a change
of variable in the residue gives
\begin{align}
[\Ecal^n]g\bigl[\Delta r_+(\Ecal)\bigr]
&=
\operatorname*{Res}_{\Ecal=0}
\frac{g\bigl[\Delta r_+(\Ecal)\bigr]}{\Ecal^{n+1}}
 =
\operatorname*{Res}_{y=0}
\frac{g(y)\partial_y h(y)}{h(y)^{n+1}}
\nonumber\\
&=
\frac{1}{n}\operatorname*{Res}_{y=0}
\frac{\partial_y g(y)}{h(y)^n}.
\label{eq:LB_residue_identity}
\end{align}
The last equality follows because the residue of
$d[g(y)/h(y)^n]/d y$ vanishes. Extracting the coefficient of
$y^{-1}$ yields
\begin{equation}
g\bigl[\Delta r_+(\Ecal)\bigr]
=g(0)+\sum_{n=1}^{\infty}\frac{\Ecal^n}{n!}
\left[\partial_y^{\,n-1}
\left\{\partial_y g(y)\left(\frac{y}{h(y)}\right)^n\right\}
\right]_{y=0}.
\label{eq:LB_composed_inverse}
\end{equation}
Taking $g(y)=y$ reproduces
Eq.~\eqref{eq:gen_horizon_coefficients}. Taking instead
$g(y)=\Scal(r_++y,\lambda^A)$ gives
Eq.~\eqref{eq:gen_entropy_coefficients}, including the subtraction
of the initial entropy through the term $g(0)$.

For a computation at a prescribed order, only finitely many
radial derivatives of the mass function are needed. Direct coefficient
extraction gives
\begin{align}
a_n
={}&
\frac{(-1)^n}{n[\partial_r\mgot(r_+)]^n}
 \big[y^{n-1}\big]
\left[
1+
\sum_{j=1}^{n-1}
\frac{\partial_r^{j+1}\mgot(r_+)}
{(j+1)!\,\partial_r\mgot(r_+)}y^j
\right]^{-n},
\label{eq:LB_finite_extraction}
\end{align}
where the sum is understood to not contribute for $n=1$. In particular,
\begin{align}
a_1={}&-\tfrac{1}{\partial_r\mgot},\qquad
a_2=-\tfrac{\partial_r^2\mgot}{2(\partial_r\mgot)^3},
\label{eq:LB_first_horizon_coefficients} \qquad
a_3=\tfrac{
(\partial_r\mgot)(\partial_r^3\mgot)
-3(\partial_r^2\mgot)^2}
{6(\partial_r\mgot)^5},\\ \nonumber
a_4={}&\tfrac{
10(\partial_r\mgot)(\partial_r^2\mgot)(\partial_r^3\mgot)
-(\partial_r\mgot)^2\partial_r^4\mgot
-15(\partial_r^2\mgot)^3}
{24(\partial_r\mgot)^7}.
\nonumber
\end{align}
with all derivatives evaluated at $r_+$.

\subsection{Entropy coefficients and specialization for affine mass dependence}

Expanding the composed entropy to third order gives
\begin{align}
b_1
&=-\frac{\partial_r\Scal}{\partial_r\mgot},
\nonumber\\
b_2
&=\frac{(\partial_r\mgot)(\partial_r^2\Scal)
-(\partial_r^2\mgot)(\partial_r\Scal)}
{2(\partial_r\mgot)^3},
\nonumber\\
b_3
&=\frac{(\partial_r\mgot)(\partial_r^3\mgot)
-3(\partial_r^2\mgot)^2}
{6(\partial_r\mgot)^5}\,\partial_r\Scal
+\frac{(\partial_r^2\mgot)(\partial_r^2\Scal)}
{2(\partial_r\mgot)^4}
-\frac{\partial_r^3\Scal}{6(\partial_r\mgot)^3}.
\label{eq:LB_first_entropy_coefficients}
\end{align}
All radial derivatives here are evaluated at the initial $r_+$
along the stationary family at fixed $\lambda^A$.
Using the neutral first law,
$\partial_r\Scal=\partial_r\mgot/T_{\text H}$ and
$dT_{\text H}/dr_+=\partial_r\mgot/C_{\Qcal}$,
the first two coefficients recover the thermal expansion in
Eq.~\eqref{eq:gen_thermal_limit}.

There is a useful consistency condition between the mass dependence
of the metric and the entropy. Differentiating the horizon equation
along the stationary family yields
\begin{equation}
\partial_r\mgot(r_+)
=
-\frac{\partial_r f}{\partial_{\Mcal}f},
\qquad
\partial_r\Scal(r_+)
=
-\frac{4\pi}{\partial_{\Mcal}f},
\label{eq:LB_mass_entropy_consistency}
\end{equation}
where the partial derivatives are evaluated at
$(r_+,\mgot(r_+),\lambda^A)$ and the second equality uses the
fixed-parameter first law and $T_{\text H}=\partial_r f/(4\pi)$
at a simple horizon. Thus, for the radially constant affine coefficient
in Eq.~\eqref{eq:gen_affine_mass},
\begin{equation}
\partial_r\Scal(r_+)=\frac{4\pi}{c},
\qquad
b_n=\frac{4\pi}{c}a_n.
\label{eq:LB_affine_entropy}
\end{equation}
The entropy is consequently linear in the horizon radius along
each connected portion of this fixed-parameter family. For the
charged BTZ solution, $c=\kappa/\pi$, and
Eq.~\eqref{eq:LB_affine_entropy} exactly matches
Eq.~\eqref{eq:BTZNLED_probability_radius}.

In the same specialization,
$h(y)=-f_0(r_++y)/c$ and
$\partial_r^j\mgot(r_+)=\partial_r^j f_0(r_+)/c$ for $j\geq1$.
These substitutions in
Eq.~\eqref{eq:LB_first_horizon_coefficients} give
Eq.~\eqref{eq:gen_simple_horizon_shift}, while
Eq.~\eqref{eq:LB_finite_extraction} determines every higher order.

The residue identities also hold for formal power series. In
the analytic case, the convergence radii of the horizon and entropy
series are set by the nearest complex singularities of the chosen
inverse and of the composed entropy, respectively. A multiple zero
on another branch does not, by itself, obstruct the event-horizon
expansion.

\subsection{Charged horizon and entropy coefficients}

For charged emission, the remaining parameters $\lambda^A$ are
held fixed, while the final charge is the endpoint value
$\Qcal'(\Ecal)=\Qcal-\Qcal_{\rm em}$. The function
$h(y;\Qcal_{\rm em})$ in
Eq.~\eqref{eq:gen_charged_h_definition} satisfies
$h(0;\Qcal_{\rm em})=0$. Its inverse argument is now
\begin{equation}
\widetilde{\Ecal}
:=\Ecal+\mgot(r_+,\Qcal'(\Ecal))-\Mcal,
\qquad
\widetilde{\Ecal}=h(\Delta r_+;\Qcal_{\rm em}).
\label{eq:LB_charged_inverse_argument}
\end{equation}
This shift accounts for the change of the reference mass at fixed
radius. For sufficiently small $\Qcal_{\rm em}$, the initial simple zero ensures
$\partial_y h(0;\Qcal_{\rm em})=-\partial_r\mgot(r_+,\Qcal'(\Ecal))\neq0$.
Holding this endpoint charge fixed in the radial residue calculation gives
\begin{equation}
\Delta r_+
=\sum_{n=1}^{\infty}a_n(\Qcal_{\rm em})
\widetilde{\Ecal}^{\,n},
\label{eq:gen_charged_horizon_series}
\end{equation}
where the coefficients are given by
Eq.~\eqref{eq:gen_horizon_coefficients} with
$h(y)$ replaced by $h(y;\Qcal_{\rm em})$.
Equations~\eqref{eq:LB_finite_extraction} and
\eqref{eq:LB_first_horizon_coefficients} remain valid with all
radial derivatives evaluated at $(r_+,\Qcal'(\Ecal))$.

Writing the entropy on the stationary family as $\Scal(r,\Qcal)$
and taking $g(y)=\Scal(r_++y,\Qcal'(\Ecal))$ at fixed $\Qcal'(\Ecal)$, the
same composed inverse gives
\begin{align}
\Delta\Scal
={}&\Scal(r_+,\Qcal'(\Ecal))-\Scal(r_+,\Qcal)
 +\sum_{n=1}^{\infty}b_n(\Qcal_{\rm em})
\widetilde{\Ecal}^{\,n},
\label{eq:gen_charged_entropy_series}
\end{align}
with $b_n(\Qcal_{\rm em})$ given by
Eq.~\eqref{eq:gen_entropy_coefficients} at fixed $\Qcal'(\Ecal)$,
using the same replacement of $h$.
The fixed-radius entropy difference vanishes for the standard
Einstein area entropy, but need not do so for a general
gravitational entropy.

The charged first law, restricted to the stationary family, reads
\begin{equation}
\partial_r\Scal=\frac{\partial_r\mgot}{T_{\text H}},
\qquad
\partial_\Qcal\Scal
=\frac{\partial_\Qcal\mgot-\Phi_+}{T_{\text H}}.
\label{eq:LB_charged_radial_first_law}
\end{equation}
Consequently, the joint expansion of
Eq.~\eqref{eq:gen_charged_entropy_series} begins with
$-(\Ecal-\Phi_+\Qcal_{\rm em})/T_{\text H}$ and reproduces
the entropy Hessian in
Eq.~\eqref{eq:gen_charged_entropy_Hessian} at quadratic order.
For the affine family, Eq.~\eqref{eq:LB_affine_entropy}
also applies to the coefficients at each fixed $\Qcal'(\Ecal)$; the
fixed-radius charge contribution must still be retained when
present.

Along the chosen channel ${\cal C}$, one finally substitutes
$\Qcal_{\rm em}=\Qcal_{\rm em}(\Ecal)$ to obtain the horizon and
its entropy at the emission endpoint.

\section{Charged horizon residue and the emission path}
\label{appx:charge_transfer}

The radial contour calculation of
Sec.~\ref{sec:gen_neutral_channel} extends to charged emission once
the electrostatic work is included. We use a regular gauge and follow the outgoing pole along
the path ${\cal C}$ of Eq.~\eqref{eq:gen_charged_running_state}. The finite-emission
construction uses the quasistationary charged KPW prescription
\cite{Zhang:2005xt,Israel:2010gk,Vanzo:2011wq}.

At each fixed intermediate state, the time transformation in
Eq.~\eqref{eq:gen_Painleve_time} sends $A=A_t\dd t$ to a potential
with radial component $-A_t\sqrt{1-f}/f$. A gauge transformation
removes this component:
\begin{equation}
\begin{aligned}
A^{\rm reg}&=A+\dd\Lambda=A_t\dd\tau,
\\
\partial_r\Lambda&=\frac{A_t\sqrt{1-f}}{f},
\qquad I^{\rm reg}=I+e\Lambda.
\end{aligned}
\label{eq:charge_transfer_regular_gauge}
\end{equation}
The potential is regular at the future horizon, although $\Lambda$
may be logarithmically singular on the overlap with the original
exterior chart. The kinetic momentum $\partial_\mu I-eA_\mu$ is
unchanged. We use this representative below, with
$A_\tau=-\Phi$ and $A_r=0$, retaining the energy reference of
Sec.~\ref{sec:gen_charged_channel}.

For a radial excitation of charge $e$, let
$\varepsilon=-\partial_\tau I$ and $p_r=\partial_r I$.
In the Painlev\'e metric, the leading massless
Hamilton--Jacobi equation becomes
\begin{equation}
f p_r^2-2\sqrt{1-f}(\varepsilon-e\Phi)p_r
-(\varepsilon-e\Phi)^2=0.
\label{eq:charge_transfer_HJ}
\end{equation}
Its outgoing solution is
\begin{equation}
p_r^{\rm out}
=\frac{\varepsilon-e\Phi}{1-\sqrt{1-f}}
=\frac{\varepsilon-e\Phi}{\dot r},
\label{eq:charge_transfer_outgoing_momentum}
\end{equation}
where $\dot r$ is given by
Eq.~\eqref{eq:gen_outgoing_null_geodesic}. Mass and transverse
kinetic terms do not change the leading simple pole when
$\varepsilon-e\Phi_+>0$.

In the reduced KPW action, Hamilton's equation converts the
radial momentum integration into an integration over this energy
transfer. Conservation along ${\cal C}$ gives
$\dd\Mcal'=-\dd\Ecal'$ and
$\dd\Qcal'=-\dd\Qcal_{\rm em}$, so the singular increment is
\begin{equation}
\dd p_r\big|_{\rm pole}
=\left.
\frac{\dd\Ecal'-\Phi(r;\Mcal',\Qcal')\dd\Qcal_{\rm em}}
{\dot r(r;\Mcal',\Qcal')}
\right|_{\rm pole}.
\label{eq:charge_transfer_momentum_increment}
\end{equation}
Here $\dd p_r$ denotes the momentum increment of the reduced
action after subtraction of the electromagnetic canonical term,
not the total variation of
Eq.~\eqref{eq:charge_transfer_outgoing_momentum} through changing
backgrounds. Hamilton's equation is applied at fixed $r$ in each
instantaneous state. The profile $\Qcal_{\rm em}(\Ecal')$ fixes
the charge transferred between successive black hole states.

Inserting $p_r=\int\dd p_r$ into the reduced radial action
$I=\int p_r\dd r$ and parametrizing the transfers by $\Ecal'$ gives
\begin{align}
I\big|_{\rm pole}
={}&\int_0^{\Ecal}\dd\Ecal'
\int_{\gamma(\Ecal')}\dd r  
\frac{1-\Phi(r;\Mcal',\Qcal')
\dfrac{\dd\Qcal_{\rm em}}{\dd\Ecal'}}
{\dot r(r;\Mcal',\Qcal')}
\bigg|_{\rm pole}.
\label{eq:charge_transfer_radial_action}
\end{align}
Here $\gamma(\Ecal')$ crosses the instantaneous outer zero
$R_+(\Ecal')$ and follows its displacement along ${\cal C}$.
The positive-energy contour prescription is imposed before
exchanging the integrations. Terms regular at the horizon do not
contribute to this imaginary part.

To evaluate the running residue, center each radial contour
at its pole by writing $r=R_+(\Ecal')+y$.
Each radial integral is taken at fixed $\Ecal'$, so
$\dd r=\dd y$. The local expansions are
\begin{equation}
\begin{aligned}
\dot r=2\pi T_{\text H}(\Ecal')y+\Order(y^2),
\qquad
\Phi(r)=\Phi_+(\Ecal')+\Order(y).
\end{aligned}
\label{eq:charge_transfer_running_pole}
\end{equation}
The radial variation of $\Phi$ is regular after division by
$\dot r$ and leaves the residue unchanged. Here the positive
shell transfer is paired with the local outgoing contour, oriented
from inside to outside. With this orientation, the prescription
$y-i0$ gives $\operatorname{Im}\int\dd y/(y-i0)=\pi$.
Thus
\begin{align}
\operatorname{Im}I
&=\frac12\int_0^{\Ecal}
\frac{1-\Phi_+(\Ecal')\,
\dfrac{\dd\Qcal_{\rm em}}{\dd\Ecal'}}
{T_{\text H}(\Ecal')}\dd\Ecal'
 =-\frac12\int_{\cal C}
\frac{\dd\Mcal'-\Phi_+\dd\Qcal'}{T_{\text H}}.
\label{eq:charge_transfer_residue}
\end{align}
This reproduces Eq.~\eqref{eq:gen_charged_KPW_residue}.
The stationary first law identifies the last integrand with
$\dd\Scal$, giving the finite entropy difference.

A change of gauge reference must be accompanied by the
corresponding change of canonical energy, preserving
$\varepsilon-e\Phi$. The conserved mass,
charge and potential for the first-order
$\mathcal{H}(\mathcal{P})$ realization are identified by the
fixed-coupling Hamiltonian derivation in Appendix~\ref{appx:termo}.

\section{Puiseux--Lagrange series for degenerate horizons
\label{appx:Puiseux}}

At a multiple horizon, the inverse in
Appendix~\ref{appx:Lagrange} is naturally expanded in fractional
powers of the emitted energy for neutral emission, or of the mass
difference from the critical curve for charged emission. We derive
the coefficients by
reducing the problem to an ordinary inverse with a nonzero linear
term. This gives the particular Puiseux construction needed here. More
general treatments may be found in
Refs.~\cite{Puiseux:1850,Sambale:2022fps}.

\subsection{A horizon of arbitrary multiplicity}

Let $r_d$ be a zero of multiplicity $m\geq2$ at the reference mass
$\Mcal$, so that
\begin{align}
\partial_r^j f(r_d;\Mcal,\lambda^A)
&=
0,
\qquad(0\leq j<m),
\nonumber\\
\partial_r^m f(r_d;\Mcal,\lambda^A)
&\ne
0.
\label{eq:PL_multiplicity}
\end{align}
We assume analyticity and
$\partial_{\Mcal}f(r_d;\Mcal,\lambda^A)\ne0$, so the local mass
function in Eq.~\eqref{eq:gen_local_mass} remains well defined.
Repeated differentiation of
$f\bigl[r,\mgot(r),\lambda^A\bigr]=0$ then gives
\begin{align}
\partial_r^j\mgot(r_d)={}&0,
\qquad \qquad\,\,\,(1\leq j<m),
\nonumber \\
\partial_r^m\mgot(r_d)
=
{}&-\frac{\partial_r^m f}{\partial_{\Mcal}f},
\label{eq:PL_mass_multiplicity}
\end{align}
with the partial derivatives evaluated at the degenerate horizon.

Taking $r_d$ as the reference radius in
Eq.~\eqref{eq:gen_h_definition}, we write
\begin{equation}
h(y)
:=
\mgot(r_d)-\mgot(r_d+y),
\qquad
\Ecal=h(y).
\label{eq:PL_local_mass_difference}
\end{equation}
where $y$ is the displaced radius relative to $r_d$. The ratio
$h(y)/y^m$ is analytic and nonzero at the origin. Choose one of its
analytic $m$-th roots and introduce the local variable
\begin{equation}
u
=
y\left(\frac{h(y)}{y^m}\right)^{1/m},
\qquad
u^m=\Ecal.
\label{eq:PL_uniformizing_coordinate}
\end{equation}
Since $d u/d y\ne0$ at $y=0$, the inversion formula in
Appendix~\ref{appx:Lagrange} applies to $y(u)$. The $m$ values
$u=\zeta_j\Ecal^{1/m}$, with
$\zeta_j=\exp(2\pi\mathrm{i}j/m)$ and $j=0,\ldots,m-1$, give
\begin{align}
\Delta r_j(\Ecal)
&=
\sum_{n=1}^{\infty}a_n^{(j)}\Ecal^{n/m},
\nonumber\\
a_n^{(j)}
&=
\frac{\zeta_j^n}{n!}
\left[
\partial_y^{\,n-1}
\left(\frac{y^m}{h(y)}\right)^{n/m}
\right]_{y=0}.
\label{eq:PL_general_coefficients}
\end{align}
The fractional power in this expression is the $n$th power of
the reciprocal of the chosen analytic $m$-th root of $h(y)/y^m$. Changing that choice only permutes the
$m$ solutions. Their series converge for sufficiently small
$|\Ecal|^{1/m}$, or can be read as formal Puiseux series.

For the affine mass dependence of
Eq.~\eqref{eq:gen_affine_mass}, one substitutes
$h(y)=-f_0(r_d+y)/c$ in
Eq.~\eqref{eq:PL_general_coefficients}. In particular,
\begin{equation}
\bigl(a_1^{(j)}\bigr)^m
=
-\frac{c\,m!}{\partial_r^m f_0(r_d)}.
\label{eq:PL_leading_balance}
\end{equation}
For real analytic $f_0$ and sufficiently small
$\Ecal>0$, a zero of even multiplicity therefore produces two
nearby real zeros when $\partial_r^m f_0(r_d)<0$, and none when
$\partial_r^m f_0(r_d)>0$. An odd multiplicity produces one real
zero for either sign. The remaining zeros are complex. This is a
local count: horizons away from $r_d$ are unaffected by this
classification.

In this specialization, the slopes at the displaced zeros satisfy
\begin{equation}
\partial_r f\bigl[r_d+\Delta r_j(\Ecal);
\Mcal-\Ecal,\lambda^A\bigr]
=-\frac{mc}{a_1^{(j)}}\Ecal^{(m-1)/m}
+\Order(\Ecal).
\label{eq:PL_displaced_slopes}
\end{equation}
Thus each of these zeros is simple for sufficiently small nonzero
energy. When a real zero represents an event horizon with positive
slope, its temperature inherits the power
$\Ecal^{(m-1)/m}$.

\subsection{Application to double and triple zeros}

For $m=2$ and $\partial_r^2f_0(r_d)<0$, choose the positive
square root at $y=0$ and label the two real solutions by
$\varepsilon=\pm1$. The coefficients through order $\Ecal$ are
those in Eq.~\eqref{eq:degenerate_first_terms}; the next one is
\begin{align}
a_3^{(\varepsilon)}
=\frac{c\,a_1^{(\varepsilon)}}
{36[\partial_r^2f_0(r_d)]^3}
\left\{
3\partial_r^2f_0(r_d)\partial_r^4f_0(r_d)
-5[\partial_r^3f_0(r_d)]^2
\right\}.
\label{eq:PL_double_coefficients}
\end{align}
The identity $a_n^{(-)}=(-1)^n a_n^{(+)}$ holds at every order.

For $m=3$, choose
\begin{equation}
a_1^{(0)}
=
\sqrt[3]{-\frac{6c}{\partial_r^3f_0(r_d)}},
\label{eq:PL_real_cubic_coefficient}
\end{equation}
where the cube root is real. With
$\zeta_j=\exp(2\pi\mathrm{i}j/3)$, the first coefficients are
\begin{align}
a_1^{(j)}
&=
\zeta_j a_1^{(0)},
\nonumber\\
a_2^{(j)}
&=
-\frac{\zeta_j^2[a_1^{(0)}]^2\partial_r^4f_0(r_d)}
{12\partial_r^3f_0(r_d)},
\nonumber\\
a_3^{(j)}
&=
[a_1^{(0)}]^3
\left\{
\frac{[\partial_r^4f_0(r_d)]^2}
{48[\partial_r^3f_0(r_d)]^2}
-\frac{\partial_r^5f_0(r_d)}
{60\partial_r^3f_0(r_d)}
\right\}.
\label{eq:PL_triple_coefficients}
\end{align}
The solution $j=0$ is real, while $j=1,2$ form a complex-conjugate
pair. Thus a stationary inflection does not split into three real
horizons under the uniform mass shift. The real zero moves toward
smaller radii if $\partial_r^3f_0(r_d)>0$, and toward larger radii
if $\partial_r^3f_0(r_d)<0$. Its slope vanishes as
$\Ecal^{2/3}$.

\subsection{Charged emission about the critical curve}

For a double zero, the charged inverse is centered at
$r_{\rm d}(\Qcal')$, as in
Sec.~\ref{sec:degenerate_charged_unfolding}. Since
$\partial_r^2\mgot\neq0$, the equation
 $\partial_r\mgot\bigl[r_{\rm d}(\Qcal'),\Qcal'\bigr]=0$ defines this radius
analytically for nearby charges. Differentiating this equation
and the critical mass
$\Mcal_{\rm d}(\Qcal')=\mgot\bigl[r_{\rm d}(\Qcal'),\Qcal'\bigr]$
gives
\begin{align}
\frac{d r_{\rm d}}{d\Qcal'}
&=-\frac{\partial_r\partial_\Qcal\mgot}{\partial_r^2\mgot},
\nonumber\\
\frac{d\Mcal_{\rm d}}{d\Qcal'}
&=\partial_\Qcal\mgot,
\nonumber\\
\frac{d^2\Mcal_{\rm d}}{d\Qcal'^2}
&=\partial_\Qcal^2\mgot
-\frac{(\partial_r\partial_\Qcal\mgot)^2}{\partial_r^2\mgot}.
\label{eq:degenerate_critical_curve_derivatives}
\end{align}
All derivatives of $\mgot$ are evaluated at
$(r_{\rm d}(\Qcal'),\Qcal')$.

At the endpoint charge $\Qcal'(\Ecal)$, define
\begin{align}
h_{\rm d}(y;\Qcal'(\Ecal))
&:=\Mcal_{\rm d}(\Qcal'(\Ecal))
-\mgot\bigl[r_{\rm d}(\Qcal'(\Ecal))+y,\Qcal'(\Ecal)\bigr],
\nonumber\\
\Delta\Mcal_{\rm d}
&=h_{\rm d}(y;\Qcal'(\Ecal)).
\label{eq:degenerate_charged_h_definition}
\end{align}
Here $y$ is the dummy radial variable of the inverse construction;
at a root it equals the displacement $\Delta r_{\rm d}$ measured
from the critical radius at the final charge. The second equality
is then the final horizon condition, with
$\Delta\Mcal_{\rm d}$ defined in
Eq.~\eqref{eq:degenerate_charged_mass_offset}. Since
$h_{\rm d}(y;\Qcal'(\Ecal))=-(\partial_r^2\mgot)y^2/2+\Order(y^3)$,
the construction in Eq.~\eqref{eq:PL_uniformizing_coordinate}
applies with $m=2$ and $\Delta\Mcal_{\rm d}$ in place of
$\Ecal$.

For  $\partial_r^2\mgot<0$ and $\Delta\Mcal_{\rm d}>0$, choose
the positive square root at $y=0$. Accordingly, the two real branches are
\begin{equation}
r_{{\rm d},f}^{(\epsilon)}
=r_{\rm d}(\Qcal'(\Ecal))
+\sum_{n=1}^{\infty}a_n^{(\epsilon)}(\Qcal'(\Ecal))
(\Delta\Mcal_{\rm d})^{n/2},
\label{eq:degenerate_charged_puiseux_series}
\end{equation}
where $\epsilon=\pm1$ and
\begin{equation}
a_n^{(\epsilon)}(\Qcal'(\Ecal))
=\frac{\epsilon^n}{n!}
\left[
\partial_y^{\,n-1}
\left(\frac{y^2}{h_{\rm d}(y;\Qcal'(\Ecal))}\right)^{n/2}
\right]_{y=0}.
\label{eq:degenerate_charged_puiseux_coefficients}
\end{equation}
The first coefficients follow from
Eqs.~\eqref{eq:degenerate_first_terms} and
\eqref{eq:PL_double_coefficients} with the replacement
\[
\frac{\partial_r^j f_0(r_d)}{c}
\longrightarrow
\partial_r^j\mgot\bigl[r_{\rm d}(\Qcal'(\Ecal)),\Qcal'(\Ecal)\bigr].
\]
At a local minimum, $\partial_r^2\mgot>0$, the real pair instead
requires $\Delta\Mcal_{\rm d}<0$. Replacing
$h_{\rm d}$ by $-h_{\rm d}$ and
$\Delta\Mcal_{\rm d}$ by $-\Delta\Mcal_{\rm d}$ in the
inverse construction gives its real coefficients. In either case,
the first terms take the form of
Eq.~\eqref{eq:degenerate_charged_first_terms}. The neutral
formulas are recovered when $\Qcal'(\Ecal)=\Qcal$ and
$\Delta\Mcal_{\rm d}=\Ecal$; in the affine specialization,
$\partial_r^j\mgot=\partial_r^j f_0/c$ for $j$ radial
derivatives. No affine mass dependence is needed for the charged
construction itself.

The expansion along ${\cal C}$ follows by inserting its charge
profile into the critical mass. For an analytic profile near the
initial state,
\begin{equation}
\Delta\Mcal_{\rm d}(\Ecal)
=\Ecal+\sum_{k=1}^{\infty}
\frac{(-1)^k}{k!}
\frac{d^k\Mcal_{\rm d}}{d\Qcal^k}
\big[\Qcal_{\rm em}(\Ecal)\big]^k,
\label{eq:PL_charged_path_mass_difference}
\end{equation}
where the critical derivatives are evaluated at the initial charge.
With $\eta=(\partial_\Ecal\Qcal_{\rm em})(0)$, this gives
\begin{align}
\Delta\Mcal_{\rm d}(\Ecal)
={}&\left(1-\eta\frac{d\Mcal_{\rm d}}{d\Qcal}\right)\Ecal
 +\frac12\left[
\eta^2\frac{d^2\Mcal_{\rm d}}{d\Qcal^2}
-\frac{d\Mcal_{\rm d}}{d\Qcal}(\partial_\Ecal^2\Qcal_{\rm em})(0)
\right]\Ecal^2
\nonumber\\
&+\Order(\Ecal^3).
\label{eq:PL_charged_path_expansion}
\end{align}
Thus the quadratic term in a tangent channel depends both on the
curvature of the critical curve and on that of the charge profile.
Together with the common displacement
$r_{\rm d}\bigl[\Qcal-\Qcal_{\rm em}(\Ecal)\bigr]$, this gives the
path-dependent splitting described in
Sec.~\ref{sec:degenerate_charged_unfolding}.

These formulas apply while the critical zero remains double.
If $\partial_r^2\mgot$ also vanishes, its continuation requires a higher-multiplicity analysis, since charge variation need not
preserve that multiplicity.

\subsection{Approach to the outer coalescence}
\label{appx:Puiseux_outer_endpoint}

The next radial term in
Eq.~\eqref{eq:degenerate_outer_radius} follows from the double-zero
inverse at the limiting minimum $r_{\rm e}$,
\begin{align}
R_+(\Ecal)
={}&
r_{\rm e}+
\sqrt{\frac{2c}{\partial_r^2 f_c(r_{\rm e})}}
(\Ecal_c-\Ecal)^{1/2}
 -
\frac{c\,\partial_r^3 f_c(r_{\rm e})}{3[\partial_r^2 f_c(r_{\rm e})]^2}
(\Ecal_c-\Ecal)
+\Order\bigl((\Ecal_c-\Ecal)^{3/2}\bigr).
\label{eq:PL_outer_endpoint_radius}
\end{align}

For charged emission, the expansion in
Eq.~\eqref{eq:degenerate_charged_first_terms} is evaluated at the
outer minimum $r_{\rm e}(\Qcal')$, with
$\partial_r^2\mgot>0$ also at the endpoint and the mass difference
$\Delta\Mcal_{\rm e}<0$ of
Eq.~\eqref{eq:degenerate_charged_outer_offset}.
In the affine family with a regular positive coefficient $c$,
$T_{\text H}=c\,\partial_r\mgot/(4\pi)$ at the simple event
horizon, and hence
\begin{equation}
T_{\text H}(\Ecal)
=\frac{c}{4\pi}
\sqrt{-2\partial_r^2\mgot\Delta\Mcal_{\rm e}}
+\Order(|\Delta\Mcal_{\rm e}|).
\label{eq:PL_charged_outer_endpoint_temperature}
\end{equation}
Here $\Qcal'$ denotes
$\Qcal'(\Ecal)=\Qcal-\Qcal_{\rm em}(\Ecal)$, and
$\partial_r^2\mgot$ is evaluated at
$(r_{\rm e}(\Qcal'),\Qcal')$. If $c$ depends on the charge, it
is evaluated at the same $\Qcal'$.

For an analytic channel with
$\Delta\Mcal_{\rm e}\sim-K(\Ecal_c-\Ecal)^n$, where
$K>0$ and $n\geq1$ is an integer, the radial correction is of
order $(\Ecal_c-\Ecal)^{n/2}$, while
$r_{\rm e}\bigl[\Qcal'(\Ecal)\bigr]$ varies analytically near the endpoint.
If the entropy is analytic in the radius and charge there, its
derivative therefore receives a regular contribution from the
motion along the critical curve and a term of order
$(\Ecal_c-\Ecal)^{n/2-1}$. The first law gives
\begin{align}
\frac{1-\Phi_+(\Ecal)\,(\partial_\Ecal\Qcal_{\rm em})(\Ecal)}
{T_{\text H}(\Ecal)}
&=-\frac{d}{d\Ecal}
\Scal\bigl[R_+(\Ecal),\Qcal'(\Ecal)\bigr]
\nonumber\\
&=\Order(1)
+\Order\!\left((\Ecal_c-\Ecal)^{n/2-1}\right).
\label{eq:PL_charged_endpoint_entropy_derivative}
\end{align}
Although $1/T_{\text H}$ alone is not integrable for $n\geq2$,
the full charged integrand remains integrable for every $n\geq1$.
Tangency can thus change the endpoint power without making the
tunneling action divergent.

\section{Details on the nonlinearly charged BTZ black hole
\label{appx:BTZ_NLED}}

This Appendix collects the supporting matter, horizon-adapted
parametrizations, and fixed-coupling thermodynamics of the static
nonlinearly charged BTZ family used in
Sec.~\ref{sec:BTZ_laboratory}. The parametrizations extend the
stationary-point analysis of Ref.~\cite{Mendez-Zavaleta:2026rgg} by
resolving the one, two, and three-horizon configurations explicitly.

\subsection{Plebanski formulation of the supporting matter}
\label{appx:BTZNLED_Plebanski}

Consider Einstein gravity coupled to nonlinear electrodynamics
in Plebanski's $\mathcal{H}(\mathcal{P})$ formulation
\cite{Plebanski1968}. The electromagnetic fields are the gauge
potential $A_\mu$ and an independent antisymmetric tensor
$P^{\mu\nu}$, with invariant
$\mathcal{P}:= P_{\mu\nu}P^{\mu\nu}/4$. The first-order action is
\begin{align}
I[g_{\mu \nu},A_{\mu},P^{\mu \nu}]
={}&
\int d^3x\,\sqrt{-g}
\left[
\frac{1}{2\kappa}
\left(R+\frac{2}{\ell^2}\right)
\right.
\nonumber\\
&\left.
\hspace{2.5em}
+\frac12 P^{\mu\nu}F_{\mu\nu}
-\mathcal{H}(\mathcal{P})
\right],
\label{eq:BTZNLED_Plebanski_action}
\end{align}
with $F_{\mu\nu}=2\partial_{[\mu}A_{\nu]}$. It is worth pointing out that this ($\mathcal P, \mathcal H$) formulation is especially suitable for constructing  {non-rotating} and rotating black hole configurations (see Refs. \cite{Ayon-Beato:1998hmi,Ayon-Beato:1999qin,Garcia-Diaz:2021bao,DiazGarcia:2022jpc,Bravo-Gaete:2025fwx,Bravo-Gaete:2026hoe}).
The theory supporting the black hole family of
Ref.~\cite{Mendez-Zavaleta:2026rgg} is specified by
\begin{equation}
\mathcal{H}(\mathcal{P})
=
\frac{\beta\sqrt{-4\mathcal{P}}}{|\beta\gamma|}
\left[
1-
\ln\left(
\frac{\sqrt{-4\mathcal{P}}}{|\beta\gamma|}
\right)
\right],
\qquad
\mathcal{P}<0.
\label{eq:BTZNLED_Plebanski_structural}
\end{equation}
The couplings $(\kappa,\ell,\beta,\gamma)$ are held fixed, with
$\beta\gamma\ne0$. The additive constant in $\mathcal{H}$ is set to
zero, keeping the cosmological scale $\ell$ fixed as in the metric.

Variation with respect
to $A_\mu$ and $P^{\mu\nu}$ gives, respectively,
\begin{equation}
\nabla_\mu P^{\mu\nu}=0,
\qquad
F_{\mu\nu}
=
\mathcal{H}_{\mathcal{P}}P_{\mu\nu},
\label{eq:BTZNLED_Plebanski_equations}
\end{equation}
where $\mathcal{H}_{\mathcal{P}}:=\partial\mathcal{H}/\partial\mathcal{P}$.
The metric variation yields standard Einstein equations
$G_{\mu\nu}-\ell^{-2}g_{\mu\nu}=\kappa T_{\mu\nu}$, with
\begin{equation}
T_{\mu\nu}
=
-\mathcal{H}_{\mathcal{P}}
P_{\mu\alpha}P_\nu{}^\alpha
+
\left(2\mathcal{P}\mathcal{H}_{\mathcal{P}}
-\mathcal{H}\right)g_{\mu\nu}.
\label{eq:BTZNLED_Plebanski_stress_tensor}
\end{equation}
Here the constitutive relation has been used after varying the
action.

For the static metric \eqref{eq:static_3D_metric} and a purely
electric potential $A=A_t(r)\dd t$, the first equation in
Eq.~\eqref{eq:BTZNLED_Plebanski_equations} integrates to
\begin{equation}
rP^{rt}=q,
\qquad
P_{rt}=-\frac{q}{r},
\qquad
\mathcal{P}(r)=-\frac{q^2}{2r^2}.
\label{eq:BTZNLED_Plebanski_electric}
\end{equation}
In the sector $\beta\gamma/q>0$, the constitutive relation gives
\begin{equation}
F_{rt}
=
\partial_r A_t
=
\frac{\sqrt{2}}{\gamma}
\ln\left(
\frac{\beta\gamma r}{\sqrt{2}q}
\right),
\label{eq:BTZNLED_Plebanski_field_strength}
\end{equation}
Integrating this expression
recovers the potential in Eq.~\eqref{eq:BTZNLED_gauge_potential},
with its additive gauge constant kept as specified there.

In this sector, the stress tensor obeys
$T^t{}_t=T^r{}_r=-\mathcal{H}$ and
$T^\varphi{}_\varphi=2\mathcal{P}\mathcal{H}_{\mathcal{P}}-\mathcal{H}$.
The radial Einstein equation therefore becomes
\begin{align}
\frac{\partial_r f(r)}{2r}-\frac{1}{\ell^2}
=
-\kappa\mathcal{H}\bigl[\mathcal{P}(r)\bigr]
 =
-\frac{\sqrt{2}\kappa q}{\gamma r}
\left[
1+\ln\left(
\frac{\beta\gamma r}{\sqrt{2}q}
\right)
\right].
\label{eq:BTZNLED_Plebanski_radial_Einstein}
\end{align}
Its integration gives Eq.~\eqref{eq:BTZNLED_blackening}, with $M$
as the remaining integration constant. The angular Einstein equation
is satisfied by the same solution.

It is worth noticing that the structural function and constitutive relation are single-valued
and smooth on the electric domain $\mathcal{P}<0$, including
$\mathcal{P}=-\beta^2\gamma^2/4$, where
$\mathcal{H}_{\mathcal{P}}=0$ and the electric field vanishes.
Since $\mathcal{P}(r)$ is monotonic for nonzero $q$, this formulation
provides the supporting matter throughout $r>0$. Moreover, neither
$M$ nor $q$ enters the structural function: both integration constants
may vary within the same theory while $(\kappa,\ell,\beta,\gamma)$
remain fixed.

\subsection{Horizon-adapted parametrizations}

For fixed $(q,\beta,\gamma,\ell)$, the positive roots of
Eq.~\eqref{eq:BTZNLED_blackening} determine the Killing horizons of the
geometry. It is nevertheless useful to eliminate combinations of the
original parameters in favor of the horizon radii themselves. This
produces representations adapted to a prescribed root configuration
and makes the corresponding degeneracy conditions manifest.
For fixed couplings, the horizon radii remain subject to the
parameter relations inherited from Eq.~\eqref{eq:BTZNLED_blackening}.

Consider first a geometry with a positive horizon at $r=r_+$. Using
$f(r_+)=0$ to eliminate the constant logarithmic contribution, the
blackening function can be written as
\begin{align}
f(r)
={}&
\frac{r^2}{\ell^2}
-
M
-
\frac{2\sqrt{2}\kappa q r}{\gamma}
\ln\left(
\frac{r}{r_+}
\right)
 -
\frac{r}{r_+}
\left(
\frac{r_+^2}{\ell^2}-M
\right).
\label{eq:f_1hor}
\end{align}
This form satisfies $f(r_+)=0$ identically. In the sector
$q/\gamma<0$, Eq.~\eqref{eq:BTZNLED_mass_derivatives} shows that the
mass--radius curve is strictly convex. Together with
$f(0^+)=-M<0$ and $f(r)\rightarrow+\infty$ as
$r\rightarrow\infty$, this ensures that a positive mass geometry has
exactly one simple positive horizon.

A form adapted to two prescribed positive roots $0<r_-<r_+$
follows by imposing $f(r_-)=f(r_+)=0$.
Eliminating the logarithmic coefficient gives
\begin{align}
f(r)
={}&
-M
+
\frac{r^2}{\ell^2}
-
\frac{r}{r_+}
\left(
\frac{r_+^2}{\ell^2}-M
\right)
\nonumber\\
&-
\frac{r}{r_-}
\left[
\frac{r_-^2}{\ell^2}
-
M
-
\frac{r_-}{r_+}
\left(
\frac{r_+^2}{\ell^2}-M
\right)
\right]
\frac{\ln(r/r_+)}{\ln(r_-/r_+)}.
\label{eq:f_2hor}
\end{align}
The conditions $f(r_-)=f(r_+)=0$ alone do not exclude an additional
positive root between them. The two-horizon configuration selected in
Sec.~\ref{sec:BTZ_laboratory} is obtained when the inner zero
is degenerate,
\begin{equation}
f(r_-)
=
\partial_r f(r_-)
=
0,
\qquad
\partial_r^2 f(r_-)<0,
\label{eq:BTZNLED_inner_degeneracy}
\end{equation}
while $r_+$ remains a simple event horizon.

 To express this restriction more efficiently, we define
\begin{equation}
\alpha
:=
\frac{r_-}{r_+},
\qquad
\mu
:=
\frac{M\ell^2}{r_+^2},
\qquad
0<\alpha<1.
\label{eq:BTZNLED_alpha_mu}
\end{equation}
The condition $\partial_r f(r_-)=0$ fixes the dimensionless
mass,
\begin{equation}
\mu(\alpha)
=
\frac{
\alpha\left(
1-\alpha+\alpha\ln\alpha
\right)
}{
\alpha-1-\ln\alpha
},
\label{eq:mu2-alpha}
\end{equation}
with $M=r_+^2\mu(\alpha)/\ell^2$.
Thus, the initial degenerate configuration may be parametrized by
$(r_+,\alpha)$, with $r_-=\alpha r_+$. This is the origin of the
dependence on $r_+/r_-$ displayed by the inverse coefficients in
Eq.~\eqref{eq:BTZNLED_two_horizon_coefficients}.

Equations~\eqref{eq:f_2hor} and \eqref{eq:mu2-alpha} parametrize
the initial degenerate member of the family. The subsequent horizon
positions follow from the original blackening function at the
running mass and charge, as in
Eq.~\eqref{eq:BTZNLED_final_horizon_equation} for neutral emission.

\subsection{The three-horizon parametrization}

Before either pair of roots coalesces, the geometry admits three
distinct positive horizons,
\begin{equation}
0<r_-<r_0<r_+.
\label{eq:BTZNLED_appendix_root_ordering}
\end{equation}
Starting from Eq.~\eqref{eq:f_2hor}, the remaining condition
$f(r_0)=0$ fixes the mass in terms of the three radii. Introducing
\begin{equation}
\widetilde r_-
:=
\frac{r_-}{r_+},
\qquad
\widetilde r_0
:=
\frac{r_0}{r_+},
\qquad
0<\widetilde r_-<\widetilde r_0<1,
\label{eq:BTZNLED_reduced_radii}
\end{equation}
one obtains
\begin{align}
M
=
\widetilde M
:={}&
\frac{
r_+^2\widetilde r_0\widetilde r_-
}{\ell^2}
\frac{
(1-\widetilde r_-)\ln\widetilde r_0
-
(1-\widetilde r_0)\ln\widetilde r_-
}{
\widetilde r_-(1-\widetilde r_0)\ln\widetilde r_-
-
\widetilde r_0(1-\widetilde r_-)\ln\widetilde r_0
}.
\label{eq:M_3hor}
\end{align}
This expression is positive throughout the ordered domain
\eqref{eq:BTZNLED_reduced_radii}. Dividing both its numerator and
denominator by $(1-\widetilde r_-)(1-\widetilde r_0)>0$ leaves
differences of $\ln x/(1-x)$ and $x\ln x/(1-x)$, respectively.
For $0<x<1$, the first function is strictly increasing and the
second strictly decreasing; the ordering
$\widetilde r_-<\widetilde r_0$ therefore makes both differences
positive.

The ordering \eqref{eq:BTZNLED_appendix_root_ordering}, together with
$f(0^+)=-M<0$ and $f(r)\rightarrow+\infty$ at large radius, yields the
alternating sign structure stated in
Sec.~\ref{subsec:BTZNLED_fine-structure}. Accordingly, $r_+$ is the event
horizon, $r_0$ is the Cauchy horizon associated with the exterior
region, and $r_-$ is the innermost positive Killing horizon.

\subsection{Extremal limits of the three-horizon configuration}

The two boundaries of the three-horizon sector follow directly from
Eq.~\eqref{eq:M_3hor}. When the middle root approaches the
innermost one, $\widetilde r_0\rightarrow\widetilde r_-=\alpha$,
\begin{equation}
\lim_{\widetilde r_0\rightarrow\widetilde r_-}\widetilde M
=\frac{r_+^2}{\ell^2}\mu(\alpha)>0.
\label{eq:M-inner-coalescence}
\end{equation}
This reproduces Eq.~\eqref{eq:mu2-alpha} and corresponds to the upper
critical configuration \eqref{eq:BTZNLED_upper_boundary}, where the
two inner branches merge while the event horizon remains simple.

The opposite boundary is reached when the middle root approaches the
event horizon, $\widetilde r_0\rightarrow1$. Equation~\eqref{eq:M_3hor}
then gives
\begin{equation}
\lim_{\widetilde r_0\rightarrow1}
\widetilde M
=
\frac{r_+^2\widetilde r_-}{\ell^2}
\frac{
\widetilde r_--1-\ln\widetilde r_-
}{
1-\widetilde r_-+\widetilde r_-\ln\widetilde r_-
}
>
0.
\label{eq:M-outer-coalescence}
\end{equation}
This limit describes the lower critical configuration
\eqref{eq:BTZNLED_lower_boundary}. The middle and event horizons merge
into a double outer root, while the innermost horizon remains simple.
Unlike the inner coalescence, this boundary has
$\partial_r f(r_+)=0$ and hence vanishing Hawking temperature. It is the
configuration approached by the neutral emission trajectory in
Sec.~\ref{sec:BTZ_laboratory}. The limiting behavior through simple
outer horizons is described in
Sec.~\ref{sec:degenerate_outer_coalescence}.

\subsection{Revisiting the black hole thermodynamics}
\label{appx:termo}

We derive the first law by computing the Hamiltonian charge associated
with the Killing vector $\xi=\partial_t$ by means of the covariant phase space
formalism~\cite{Wald:1993nt,Iyer:1994ys}. Variations are taken within the nonlinearly charged BTZ family at fixed $(\kappa,\ell,\beta,\gamma)$, so only
the integration constants $(M,q)$ vary. The variation of the first-order action
\eqref{eq:BTZNLED_Plebanski_action} gives the surface term
\begin{equation}
\mathcal{J}^\mu
=
\sqrt{-g}
\left[
2\mathcal P^{\mu(\alpha\beta)\gamma}
\nabla_\gamma\delta g_{\alpha\beta}
+P^{\mu\nu}\delta A_\nu
\right],
\label{eq:surface}
\end{equation}
with
\begin{equation*}
\mathcal P^{\mu\nu\rho\sigma}
:=
 \frac{1}{2\kappa} \left(\frac{\partial R}{\partial R_{\mu\nu\rho\sigma}}\right)
=
\frac{1}{4\kappa}
\left(
g^{\mu\rho}g^{\nu\sigma}
-g^{\mu\sigma}g^{\nu\rho}
\right).
\end{equation*}
Since $\mathcal P^{\mu\nu\rho\sigma}$ is covariantly constant, its derivatives do not contribute to the surface term.

The corresponding symplectic potential is the two-form
\begin{equation*}
\Theta_{(2)}
=
\frac12\epsilon_{\mu\nu\rho}
\frac{\mathcal{J}^\mu}{\sqrt{-g}}\,
d x^\nu\wedge d x^\rho.
\end{equation*}
On shell, the Noether current associated with $\xi$ is locally
exact. Using the orientation $\epsilon_{tr\varphi}=r$, the charge
one-form is
\begin{equation}
Q_{(1)}
=
\frac12\epsilon_{\alpha\mu\nu}
Q^{\mu\nu}d x^\alpha,
\label{eq:noether}
\end{equation}
with
\begin{equation}
Q^{\mu\nu}
=
-2\mathcal P^{\mu\nu\rho\sigma}\nabla_\rho\xi_\sigma
+P^{\mu\nu}\xi^\rho A_\rho.
\label{eq:noether-off-shell}
\end{equation}

For variations between stationary solutions, the one-form
$\delta Q_{(1)}-i_\xi\Theta_{(2)}$ is closed, where $i_\xi$ denotes
contraction with $\xi$. The Hamiltonian variation
\begin{equation}
\delta\mathcal{H}
=
\int_{\Sigma^{(1)}}
\left(
\delta Q_{(1)}-i_\xi\Theta_{(2)}
\right)
\label{var_diff}
\end{equation}
is therefore independent of the  circle
$\Sigma^{(1)}$ on which it is evaluated in the exterior.

For the static ansatz, the electromagnetic first integral reads
 $r P^{rt}=q$. The pullbacks of the charge and symplectic
potential to a circle of constant $t$ and $r$ are then
\begin{align*}
Q_{(1)}
&=
\left(
\frac{r\partial_r f}{2\kappa}-qA_t
\right)d\varphi,
\\
i_\xi\Theta_{(2)}
&=
\left[
\frac{r\,\partial_r\delta f+\delta f}{2\kappa}
-q\,\delta A_t
\right]d\varphi.
\end{align*}
Here, variations act on $(M,q)$ at fixed $r$. Integrating over
$\varphi\sim\varphi+2\pi$ gives
\begin{equation*}
\delta\mathcal{H}
=
-\frac{\pi}{\kappa}\delta f(r)
-2\pi A_t(r)\delta q.
\end{equation*}
From  Eqs.~\eqref{eq:BTZNLED_blackening}-
\eqref{eq:BTZNLED_gauge_potential}, the metric variation satisfies
\begin{equation*}
\delta f(r)
=
-\delta M-2\kappa A_t(r)\delta q.
\end{equation*}
Thus the growing terms in the metric and gauge sectors cancel,
leaving the finite, integrable variation
$\delta\mathcal{H}=\pi\delta M/\kappa$. With the mass normalization
adopted in Eq.~\eqref{eq:BTZNLED_physical_quantities},
\begin{equation}
\delta\mathcal{H}_\infty
=
\frac{\pi}{\kappa}\delta M
=
\delta\Mcal.
\label{eq:deltaHinf}
\end{equation}
The gauge is kept as specified in
Eq.~\eqref{eq:BTZNLED_gauge_potential} throughout this construction.

At the event horizon, varying $f(r_+)=0$ along the stationary family
gives
\begin{equation*}
\delta f(r_+)
=
-\partial_r f(r_+)\delta r_+.
\end{equation*}
Evaluating the same Hamiltonian variation on the limiting horizon
circle therefore yields
\begin{align}
\delta\mathcal{H}_+
&=
\frac{\pi}{\kappa}\partial_r f(r_+)\delta r_+
-2\pi A_t(r_+)\delta q
\nonumber\\
&=
T_{\text H}\delta\Scal
+\Phi_e\delta\Qcal,
\label{eq:deltaHrh}
\end{align}
where the entropy, charge, temperature, and potential are those in
Eqs.~\eqref{eq:BTZNLED_physical_quantities}--
\eqref{eq:BTZNLED_electric_potential}. The equality
$\delta\mathcal{H}_\infty=\delta\mathcal{H}_+$ establishes the first
law \eqref{eq:BTZNLED_first_law}.

Finally, the fixed-coupling homogeneity of the mass function,
Eq.~\eqref{eq:BTZNLED_homogeneity}, gives through Euler's theorem
\begin{align}
2\Mcal
&=
r_+\frac{\partial\mgot(r_+,q)}{\partial r_+}
+q\frac{\partial\mgot(r_+,q)}{\partial q}
\nonumber\\
&=
T_{\text H}\Scal+\Phi_e\Qcal.
\label{eq:smarr}
\end{align}
The second equality follows from the first law and the linear
dependence of $\Scal$ and $\Qcal$ on $r_+$ and $q$, respectively.
This recovers the fixed-coupling Smarr relation stated in
Eq.~\eqref{eq:BTZNLED_Smarr}. A related derivation from Euler
homogeneity for Maxwell-charged AdS$_3$ black holes, with boundary
conditions preserving the scaling symmetry, is given in
Ref.~\cite{Erices:2017nta}. Here the identity follows directly from
the NLED mass function and the Hamiltonian normalization above.

\section[Tunneling coefficients for the two-horizon nonlinearly charged BTZ black hole]{Tunneling coefficients for the two-horizon\texorpdfstring{\\}{} nonlinearly charged BTZ black hole
\label{appx:BTZNLED_coefficients}}

In this appendix we derive the functions $\mathcal{A}_n$ entering
Eq.~\eqref{eq:BTZNLED_two_horizon_coefficients}. The initial inner
degeneracy constrains the derivatives of $f$ at the event horizon
through Eq.~\eqref{eq:f_2hor}, allowing the coefficients to be
written in terms of $r_+/r_-$.

Writing $\rho:=r_+/r_->1$, the initial conditions
$f_0(r_-)=\partial_r f_0(r_-)=f_0(r_+)=0$
allow the parameters entering the logarithmic term to be expressed in
terms of $(r_+,\rho)$. In particular, the two dimensionless
combinations relevant to the expansion at the event horizon are
\begin{align}
\nu(\rho)
&:=
\frac{2\sqrt{2}\kappa q\ell^2}{\gamma r_+}
=
\frac{
(\rho-1)^2
}{
\rho\left(
1-\rho+\rho\ln\rho
\right)
},
\label{eq:BTZNLED_nu_rho}
\\
\sigma(\rho)
&:=
\frac{\ell^2}{r_+}\,
\partial_r f_0(r_+)
=
\frac{
(\rho-1)
\left[
(\rho+1)\ln\rho-2(\rho-1)
\right]
}{
\rho\left(
1-\rho+\rho\ln\rho
\right)
}.
\label{eq:BTZNLED_sigma_rho}
\end{align}
Thus, $\sigma$ is the dimensionless slope of the initial event horizon,
whereas $\nu$ measures the strength of the logarithmic contribution.
For the two-horizon configuration considered here, $\sigma>0$.

A considerable simplification follows from the specific radial
dependence of the blackening function
\eqref{eq:BTZNLED_blackening}. Beyond second order, only the term
proportional to $r\ln r$ contributes, and its derivatives satisfy
\begin{equation}
\partial_r^n f_0(r)
=
(-1)^{n+1}
(n-2)!
\frac{2\sqrt{2}\kappa q}
{\gamma r^{\,n-1}},
\qquad
n\geq3.
\label{eq:BTZNLED_derivative_tower}
\end{equation}
At the initial event horizon,
$\partial_r^2 f_0(r_+)=(2-\nu)/\ell^2$.
With the dimensionless displacement
$x:=\Delta r_+/r_+$ and emitted energy
$u:=\kappa\ell^2\Ecal/(\pi r_+^2)$,
Eq.~\eqref{eq:BTZNLED_final_horizon_equation} becomes
\begin{equation}
0 = u + \sigma x + x^2 + \nu
\left[ x-(1+x)\ln(1+x) \right].
\label{eq:BTZNLED_resummed_horizon_equation}
\end{equation}
This exact equation retains the full radial dependence of the
blackening function and resums the derivative tower above.

Since $\sigma\neq0$, the solution connected to the initial event
horizon admits a Maclaurin expansion about $u=0$,
\begin{equation}
x(u)
=
\sum_{n=1}^{\infty}
\mathcal{A}_n(\rho)\,u^n.
\label{eq:BTZNLED_reduced_inverse_series}
\end{equation}
Applying the Lagrange--B\"urmann formula directly to
Eq.~\eqref{eq:BTZNLED_resummed_horizon_equation} gives the closed
expression
\begin{equation}
\mathcal{A}_n(\rho)
=\frac{(-1)^n}{n!}
\Bigg[\frac{d^{\,n-1}}{d x^{\,n-1}}
\left\{\sigma+x+\nu\left[1-\frac{1+x}{x}\ln(1+x)\right]\right\}^{-n}
\Bigg]_{x=0}.
\label{eq:BTZNLED_closed_coefficients}
\end{equation}
The expression in braces is understood by its regular
extension at $x=0$, where its value is $\sigma$.
Equation~\eqref{eq:BTZNLED_closed_coefficients} determines
every order solely in terms of the initial ratio $\rho=r_+/r_-$.

The first coefficients are
\begin{align}
\mathcal{A}_1
={}&
-\frac{1}{\sigma},
\nonumber\\
\mathcal{A}_2
={}&
-\frac{
1-\nu/2
}{
\sigma^3
},
\nonumber\\
\mathcal{A}_3
={}&
\frac{
\sigma\nu/6
-
2\left(
1-\nu/2
\right)^2
}{
\sigma^5
},
\label{eq:BTZNLED_first_coefficients}
\\
\mathcal{A}_4
={}&
\frac{1}{\sigma^7}
\left[
-5\left(
1-\frac{\nu}{2}
\right)^3
+
\frac{5}{6}\sigma\nu
\left(
1-\frac{\nu}{2}
\right)
+\frac{1}{12}\sigma^2\nu\right],
\nonumber
\end{align}
with $\sigma=\sigma(\rho)$ and $\nu=\nu(\rho)$ as given in
Eqs.~\eqref{eq:BTZNLED_sigma_rho} and
\eqref{eq:BTZNLED_nu_rho}.

Restoring dimensions in
Eq.~\eqref{eq:BTZNLED_reduced_inverse_series} reproduces
Eq.~\eqref{eq:BTZNLED_two_horizon_coefficients}. In particular,
\[
a_1=-\frac{\kappa}{\pi\partial_r f_0(r_+)}
=-\frac{\kappa}{4\pi^2T_{\text H}},
\]
so the leading contribution to the tunneling exponent is
$-\Ecal/T_{\text H}$, as required.

The inverse can be continued locally while this outer zero
remains simple. The Maclaurin series in $u$ about $u=0$,
however, reconstructs that branch only within its own convergence
domain. At the later merger of the middle and event horizons, the
outer slope vanishes and the local inverse takes the Puiseux form
developed in Appendix~\ref{appx:Puiseux}.

\bibliography{refs}

@article{PainleveGullstrand,
    author  = {Painlev{\'e}, Paul},
    title   = {La m{\'e}canique classique et la th{\'e}orie
               de la relativit{\'e}},
    journal = {C. R. Acad. Sci. (Paris)},
    volume  = {173},
    pages   = {677--680},
    year    = {1921},
    report  = {A.~Gullstrand,
               ``Allgemeine L{\"o}sung des statischen
               Eink{\"o}rperproblems in der Einsteinschen
               Gravitationstheorie'',
               Ark. Mat. Astron. Fys. \textbf{16}, No.~8,
               1--15 (1922)},
    note    = {A.~Gullstrand,
               ``Allgemeine L{\"o}sung des statischen
               Eink{\"o}rperproblems in der Einsteinschen
               Gravitationstheorie'',
               Ark. Mat. Astron. Fys. \textbf{16}, No.~8,
               1--15 (1922)}
}

@article{Parikh:1999mf,
    author = "Parikh, Maulik K. and Wilczek, Frank",
    title = "{Hawking radiation as tunneling}",
    eprint = "hep-th/9907001",
    archivePrefix = "arXiv",
    reportNumber = "PUPT-1775, SPIN-1998-12, IASSNS-HEP-98-22",
    doi = "10.1103/PhysRevLett.85.5042",
    journal = "Phys. Rev. Lett.",
    volume = "85",
    pages = "5042--5045",
    year = "2000"
}

@article{Massar:1999wg,
    author = "Massar, S. and Parentani, R.",
    title = "{How the change in horizon area drives black hole evaporation}",
    eprint = "gr-qc/9903027",
    archivePrefix = "arXiv",
    doi = "10.1016/S0550-3213(00)00067-5",
    journal = "Nucl. Phys. B",
    volume = "575",
    pages = "333--356",
    year = "2000"
}

@article{Sarkar:2007sx,
    author = "Sarkar, Sudipta and Kothawala, Dawood",
    title = "{Hawking radiation as tunneling for spherically symmetric black holes: A Generalized treatment}",
    eprint = "0709.4448",
    archivePrefix = "arXiv",
    primaryClass = "gr-qc",
    doi = "10.1016/j.physletb.2007.11.056",
    journal = "Phys. Lett. B",
    volume = "659",
    pages = "683--687",
    year = "2008"
}

@article{Ryskin:2014wua,
    author = "Ryskin, Gregory",
    title = "{Boltzmann factor and Hawking radiation}",
    eprint = "1810.07520",
    archivePrefix = "arXiv",
    primaryClass = "physics.gen-ph",
    doi = "10.1016/j.physletb.2014.05.085",
    journal = "Phys. Lett. B",
    volume = "734",
    pages = "394--395",
    year = "2014"
}

@article{Zhang:2005xt,
    author = "Zhang, Jing-Yi and Zhao, Zheng",
    title = "{Hawking radiation of charged particles via tunneling from the Reissner-Nordstroem black hole}",
    doi = "10.1088/1126-6708/2005/10/055",
    journal = "{JHEP}",
    volume = "10",
    pages = "055",
    year = "2005"
}

@article{Israel:2010gk,
    author = "Israel, Werner and Yun, Zinkoo",
    title = "{Band-aid for information loss from black holes}",
    eprint = "1009.0879",
    archivePrefix = "arXiv",
    primaryClass = "hep-th",
    doi = "10.1103/PhysRevD.82.124036",
    journal = "Phys. Rev. D",
    volume = "82",
    pages = "124036",
    year = "2010"
}

@article{Vanzo:2011wq,
    author = "Vanzo, L. and Acquaviva, G. and Di Criscienzo, R.",
    title = "{Tunnelling Methods and Hawking's radiation: achievements and prospects}",
    eprint = "1106.4153",
    archivePrefix = "arXiv",
    primaryClass = "gr-qc",
    doi = "10.1088/0264-9381/28/18/183001",
    journal = "Class. Quant. Grav.",
    volume = "28",
    pages = "183001",
    year = "2011"
}

@article{Angheben:2005rm,
    author = "Angheben, Marco and Nadalini, Mario and Vanzo, Luciano and Zerbini, Sergio",
    title = "{Hawking radiation as tunneling for extremal and rotating black holes}",
    eprint = "hep-th/0503081",
    archivePrefix = "arXiv",
    doi = "10.1088/1126-6708/2005/05/014",
    journal = "{JHEP}",
    volume = "05",
    pages = "014",
    year = "2005"
}

@article{Gessel:2016Lagrange,
    author        = "Gessel, Ira M.",
    title         = "{Lagrange inversion}",
    eprint        = "1609.05988",
    archivePrefix = "arXiv",
    primaryClass  = "math.CO",
    doi           = "10.1016/j.jcta.2016.06.018",
    journal       = "J. Comb. Theory A",
    volume        = "144",
    pages         = "212--249",
    year          = "2016"
}

@article{Sambale:2022fps,
    author = "Sambale, Benjamin",
    title = "{An Invitation to Formal Power Series}",
    eprint = "2205.00879",
    archivePrefix = "arXiv",
    primaryClass = "math.CO",
    doi = "10.1365/s13291-022-00256-6",
    journal = "Jahresber. Dtsch. Math.-Ver.",
    volume = "125",
    number = "1",
    pages = "3--69",
    year = "2023"
}

@article{Puiseux:1850,
    author  = "Puiseux, Victor",
    title   = "{Recherches sur les fonctions alg\'ebriques}",
    journal = "J. Math. Pures Appl.",
    volume  = "15",
    pages   = "365--480",
    year    = "1850",
    url     = "https://www.numdam.org/item/JMPA_1850_1_15__365_0/"
}

@article{veberivc2012lambert, title={Lambert W function for applications in physics}, author={Veberi{\v{c}}, Darko}, journal={Computer Physics Communications}, volume={183}, number={12}, pages={2622--2628}, year={2012}, publisher={Elsevier} }

@article{Wald:1993nt,
    author = "Wald, Robert M.",
    title = "{Black hole entropy is the Noether charge}",
    eprint = "gr-qc/9307038",
    archivePrefix = "arXiv",
    reportNumber = "EFI-93-42",
    doi = "10.1103/PhysRevD.48.R3427",
    journal = "Phys. Rev. D",
    volume = "48",
    number = "8",
    pages = "R3427--R3431",
    year = "1993"
}

@article{Iyer:1994ys,
    author = "Iyer, Vivek and Wald, Robert M.",
    title = "{Some properties of Noether charge and a proposal for dynamical black hole entropy}",
    eprint = "gr-qc/9403028",
    archivePrefix = "arXiv",
    doi = "10.1103/PhysRevD.50.846",
    journal = "Phys. Rev. D",
    volume = "50",
    pages = "846--864",
    year = "1994"
}

@article{Aman:2003ug,
    author = "Aman, Jan E. and Bengtsson, Ingemar and Pidokrajt, Narit",
    title = "{Geometry of black hole thermodynamics}",
    eprint = "gr-qc/0304015",
    archivePrefix = "arXiv",
    doi = "10.1023/A:1026058111582",
    journal = "Gen. Rel. Grav.",
    volume = "35",
    pages = "1733",
    year = "2003"
}

@book{Plebanski1968,
  author    = {Pleb{\'a}nski, Jerzy},
  title     = {Lectures on Non-Linear Electrodynamics},
  year      = {1968},
  publisher = {Nordita},
  address   = {Copenhagen}
}

@article{Erices:2017nta,
    author = "Erices, Cristi{\'a}n and Fuentealba, Oscar and Riquelme, Miguel",
    title = "{Electrically charged black hole on AdS$_3$ : Scale invariance and the Smarr formula}",
    eprint = "1710.05962",
    archivePrefix = "arXiv",
    primaryClass = "hep-th",
    doi = "10.1103/PhysRevD.97.024037",
    journal = "Phys. Rev. D",
    volume = "97",
    number = "2",
    pages = "024037",
    year = "2018"
}

@article{Mendez-Zavaleta:2026rgg,
    author = "M{\'e}ndez-Zavaleta, Julio A. and Rojas, Efra{\'\i}n and Su{\'a}rez-Garibay, Jos{\'e} Joaqu{\'\i}n",
    title = "{Classical emergence of the quantum-backreacted BTZ black hole from exponential electrodynamics}",
    eprint = "2601.18967",
    archivePrefix = "arXiv",
    primaryClass = "gr-qc",
    doi = "10.1103/wq8b-rpkf",
    journal = "Phys. Rev. D",
    volume = "113",
    number = "8",
    pages = "084064",
    year = "2026"
}

@article{Chernicoff:2024dll,
    author = "Chernicoff, Mariano and Giribet, Gaston and Moreno, Javier and Oliva, Julio and Rojas, Raul and Torres, Cielo Ramirez de Arellano",
    title = "{Quantum backreactions in (A)dS3 massive gravity and logarithmic asymptotic behavior}",
    eprint = "2404.10127",
    archivePrefix = "arXiv",
    primaryClass = "hep-th",
    doi = "10.1103/PhysRevD.110.044021",
    journal = "Phys. Rev. D",
    volume = "110",
    number = "4",
    pages = "044021",
    year = "2024"
}

@article{Martinez:1999qi,
    author = "Martinez, Cristian and Teitelboim, Claudio and Zanelli, Jorge",
    title = "{Charged rotating black hole in three space-time dimensions}",
    eprint = "hep-th/9912259",
    archivePrefix = "arXiv",
    reportNumber = "CECS-PHY-99-22",
    doi = "10.1103/PhysRevD.61.104013",
    journal = "Phys. Rev. D",
    volume = "61",
    pages = "104013",
    year = "2000"
}

@article{Martinez:1996uv,
    author = "Martinez, Cristian and Zanelli, Jorge",
    title = "{Back reaction of a conformal field on a three-dimensional black hole}",
    eprint = "gr-qc/9610050",
    archivePrefix = "arXiv",
    doi = "10.1103/PhysRevD.55.3642",
    journal = "Phys. Rev. D",
    volume = "55",
    pages = "3642--3646",
    year = "1997"
}

@article{Page:2015rxa,
    author = "Page, Don N.",
    title = "{Finite upper bound for the Hawking decay time of an arbitrarily large black hole in anti{\textendash}de Sitter spacetime}",
    eprint = "1507.02682",
    archivePrefix = "arXiv",
    primaryClass = "hep-th",
    reportNumber = "ALBERTA-THY-13-15",
    doi = "10.1103/PhysRevD.97.024004",
    journal = "Phys. Rev. D",
    volume = "97",
    number = "2",
    pages = "024004",
    year = "2018"
}

@article{Barcelo:2010pj,
    author = "Barcelo, Carlos and Liberati, Stefano and Sonego, Sebastiano and Visser, Matt",
    title = "{Minimal conditions for the existence of a Hawking-like flux}",
    eprint = "1011.5593",
    archivePrefix = "arXiv",
    primaryClass = "gr-qc",
    doi = "10.1103/PhysRevD.83.041501",
    journal = "Phys. Rev. D",
    volume = "83",
    pages = "041501",
    year = "2011"
}

@article{Carballo-Rubio:2018pmi,
    author = "Carballo-Rubio, Ra{\'u}l and Di Filippo, Francesco and Liberati, Stefano and Pacilio, Costantino and Visser, Matt",
    title = "{On the viability of regular black holes}",
    eprint = "1805.02675",
    archivePrefix = "arXiv",
    primaryClass = "gr-qc",
    doi = "10.1007/JHEP07(2018)023",
    journal = "{JHEP}",
    volume = "07",
    pages = "023",
    year = "2018"
}

@article{Preskill:1991tb,
    author = "Preskill, John and Schwarz, Patricia and Shapere, Alfred D. and Trivedi, Sandip and Wilczek, Frank",
    title = "{Limitations on the statistical description of black holes}",
    reportNumber = "IASSNS-HEP-91-34, CALT-68-1730",
    doi = "10.1142/S0217732391002773",
    journal = "Mod. Phys. Lett. A",
    volume = "6",
    pages = "2353--2362",
    year = "1991"
}

@misc{Madler:2026fve,
    author = {M{\"a}dler, Thomas and Gannouji, Radouane and Liebling, Steven L. and Gallo, Emanuel},
    title = "{Near-Extremal Black-Hole Formation Observed from Null Infinity}",
    eprint = "2610.23486",
    archivePrefix = "arXiv",
    primaryClass = "gr-qc",
    month = "10",
    year = "2026"
}

@article{Poisson:1989zz,
    author = "Poisson, Eric and Israel, W.",
    title = "{Inner-horizon instability and mass inflation in black holes}",
    doi = "10.1103/PhysRevLett.63.1663",
    journal = "Phys. Rev. Lett.",
    volume = "63",
    pages = "1663--1666",
    year = "1989"
}

@article{Singha:2022bvr,
    author = "Singha, Chiranjeeb and Chakraborty, Sumanta and Dadhich, Naresh",
    title = "{Strong cosmic censorship conjecture for a charged BTZ black hole}",
    eprint = "2203.07708",
    archivePrefix = "arXiv",
    primaryClass = "gr-qc",
    doi = "10.1007/JHEP06(2022)028",
    journal = "{JHEP}",
    volume = "06",
    pages = "028",
    year = "2022"
}

@article{McMaken:2023uue,
    author = "McMaken, Tyler",
    title = "{Semiclassical instability of inner-extremal regular black holes}",
    eprint = "2303.03562",
    archivePrefix = "arXiv",
    primaryClass = "gr-qc",
    doi = "10.1103/PhysRevD.107.125023",
    journal = "Phys. Rev. D",
    volume = "107",
    number = "12",
    pages = "125023",
    year = "2023"
}

@article{Carballo-Rubio:2026gwg,
    author = "Carballo-Rubio, Ra{\'u}l and Di Filippo, Francesco and Liberati, Stefano and Visser, Matt",
    title = "{Semiclassical regularity of compact trapped regions: From dynamical horizons to inner extremality}",
    eprint = "2607.03916",
    archivePrefix = "arXiv",
    primaryClass = "gr-qc",
    doi = "10.1103/6jvv-wyyy",
    journal = "Phys. Rev. D",
    volume = "114",
    number = "10",
    pages = "104008",
    year = "2026"
}

@article{Luo:2026epp,
    author = "Luo, Shu and Pando Zayas, Leopoldo A.",
    title = "{Quantum-corrected evaporation and absorption cross-section of near-extremal rotating black holes}",
    eprint = "2601.06720",
    archivePrefix = "arXiv",
    primaryClass = "hep-th",
    reportNumber = "LITP-25-14",
    doi = "10.1007/JHEP05(2026)194",
    journal = "JHEP",
    volume = "05",
    pages = "194",
    year = "2026"
}

@article{Wald:1974hkz,
    author = "Wald, Robert",
    title = "{Gedanken experiments to destroy a black hole}",
    doi = "10.1016/0003-4916(74)90125-0",
    journal = "Annals Phys.",
    volume = "82",
    number = "2",
    pages = "548--556",
    year = "1974"
}

@article{Sorce:2017dst,
    author = "Sorce, Jonathan and Wald, Robert M.",
    title = "{Gedanken experiments to destroy a black hole. II. Kerr-Newman black holes cannot be overcharged or overspun}",
    eprint = "1707.05862",
    archivePrefix = "arXiv",
    primaryClass = "gr-qc",
    doi = "10.1103/PhysRevD.96.104014",
    journal = "Phys. Rev. D",
    volume = "96",
    number = "10",
    pages = "104014",
    year = "2017"
}

@article{Gonzalez:2021vwp,
    author = "Gonz{\'a}lez, P. A. and Rinc{\'o}n, {\'A}ngel and Saavedra, Joel and V{\'a}squez, Yerko",
    title = "{Superradiant instability and charged scalar quasinormal modes for (2+1)-dimensional Coulomb-like AdS black holes from nonlinear electrodynamics}",
    eprint = "2107.08611",
    archivePrefix = "arXiv",
    primaryClass = "gr-qc",
    doi = "10.1103/PhysRevD.104.084047",
    journal = "Phys. Rev. D",
    volume = "104",
    number = "8",
    pages = "084047",
    year = "2021"
}

@article{Konewko:2023gbu,
    author = "Konewko, Sebastian and Winstanley, Elizabeth",
    title = "{Charge superradiance on charged BTZ black holes}",
    eprint = "2301.01169",
    archivePrefix = "arXiv",
    primaryClass = "hep-th",
    doi = "10.1140/epjc/s10052-024-12910-7",
    journal = "Eur. Phys. J. C",
    volume = "84",
    number = "6",
    pages = "594",
    year = "2024"
}

@article{Bardeen:1973gs,
    author = "Bardeen, James M. and Carter, B. and Hawking, S. W.",
    title = "{The Four laws of black hole mechanics}",
    doi = "10.1007/BF01645742",
    journal = "Commun. Math. Phys.",
    volume = "31",
    pages = "161--170",
    year = "1973"
}

@article{Hawking:1974rv,
    author = "Hawking, S. W.",
    title = "{Black hole explosions}",
    doi = "10.1038/248030a0",
    journal = "Nature",
    volume = "248",
    pages = "30--31",
    year = "1974"
}

@article{Bekenstein:1973ur,
    author = "Bekenstein, Jacob D.",
    title = "{Black holes and entropy}",
    doi = "10.1103/PhysRevD.7.2333",
    journal = "Phys. Rev. D",
    volume = "7",
    pages = "2333--2346",
    year = "1973"
}

@article{Hawking:1975vcx,
    author = "Hawking, S. W.",
    editor = "Gibbons, G. W. and Hawking, S. W.",
    title = "{Particle Creation by Black Holes}",
    doi = "10.1007/BF02345020",
    journal = "Commun. Math. Phys.",
    volume = "43",
    pages = "199--220",
    year = "1975",
    report = "[Erratum: Commun.Math.Phys. 46, 206 (1976)]",
    note = "[Erratum: Commun.Math.Phys. 46, 206 (1976)]"
}

@article{Drori:2018ivu,
    author = "Drori, Jonathan and Rosenberg, Yuval and Bermudez, David and Silberberg, Yaron and Leonhardt, Ulf",
    title = "{Observation of Stimulated Hawking Radiation in an Optical Analogue}",
    eprint = "1808.09244",
    archivePrefix = "arXiv",
    primaryClass = "gr-qc",
    doi = "10.1103/PhysRevLett.122.010404",
    journal = "Phys. Rev. Lett.",
    volume = "122",
    number = "1",
    pages = "010404",
    year = "2019"
}

@article{Hawking:1976ra,
    author = "Hawking, S. W.",
    title = "{Breakdown of Predictability in Gravitational Collapse}",
    doi = "10.1103/PhysRevD.14.2460",
    journal = "Phys. Rev. D",
    volume = "14",
    pages = "2460--2473",
    year = "1976"
}

@article{Kraus:1994by,
    author = "Kraus, Per and Wilczek, Frank",
    title = "{Selfinteraction correction to black hole radiance}",
    eprint = "gr-qc/9408003",
    archivePrefix = "arXiv",
    reportNumber = "PUPT-1490, IASSNS-HEP-94-61",
    doi = "10.1016/0550-3213(94)00411-7",
    journal = "Nucl. Phys. B",
    volume = "433",
    pages = "403--420",
    year = "1995"
}

@article{Kraus:1994fh,
    author = "Kraus, Per and Wilczek, Frank",
    title = "{Some applications of a simple stationary line element for the Schwarzschild geometry}",
    eprint = "gr-qc/9406042",
    archivePrefix = "arXiv",
    reportNumber = "PUPT-1474, IASSNS-94-46",
    doi = "10.1142/S0217732394003567",
    journal = "Mod. Phys. Lett. A",
    volume = "9",
    pages = "3713--3719",
    year = "1994"
}

@article{Keski-Vakkuri:1996lbi,
    author = "Keski-Vakkuri, Esko and Kraus, Per",
    title = "{Tunneling in a time dependent setting}",
    eprint = "hep-th/9604151",
    archivePrefix = "arXiv",
    reportNumber = "CALT-68-2050",
    doi = "10.1103/PhysRevD.54.7407",
    journal = "Phys. Rev. D",
    volume = "54",
    pages = "7407--7420",
    year = "1996"
}

@article{Parikh:2004ih,
    author = "Parikh, Maulik K.",
    title = "{A Secret tunnel through the horizon}",
    eprint = "hep-th/0405160",
    archivePrefix = "arXiv",
    reportNumber = "CU-TP-1114",
    doi = "10.1142/S0218271804006498",
    journal = "Int. J. Mod. Phys. D",
    volume = "13",
    pages = "2351--2354",
    year = "2004"
}

@article{Wu:2006pz,
    author = "Wu, Shuang-Qing and Jiang, Qing-Quan",
    title = "{Remarks on Hawking radiation as tunneling from the BTZ black holes}",
    eprint = "hep-th/0602033",
    archivePrefix = "arXiv",
    doi = "10.1088/1126-6708/2006/03/079",
    journal = "JHEP",
    volume = "03",
    pages = "079",
    year = "2006"
}

@article{Unruh:1980cg,
    author = "Unruh, W. G.",
    title = "{Experimental black hole evaporation}",
    doi = "10.1103/PhysRevLett.46.1351",
    journal = "Phys. Rev. Lett.",
    volume = "46",
    pages = "1351--1353",
    year = "1981"
}

@inproceedings{Bernal:2026lme,
    author = "Bernal, Isaac and Cort{\'e}s-Ortiz, Miguel A. and Bermudez, David",
    title = "{Analogue Hawking radiation in nonlinear quantum optics}",
    eprint = "2603.03525",
    archivePrefix = "arXiv",
    primaryClass = "gr-qc",
    month = "3",
    year = "2026"
}

@article{Banados:1992wn,
    author = "Banados, Maximo and Teitelboim, Claudio and Zanelli, Jorge",
    title = "{The Black hole in three-dimensional space-time}",
    eprint = "hep-th/9204099",
    archivePrefix = "arXiv",
    reportNumber = "PRINT-92-0151 (CHILE), IASSNS-HEP-92-29",
    doi = "10.1103/PhysRevLett.69.1849",
    journal = "Phys. Rev. Lett.",
    volume = "69",
    pages = "1849--1851",
    year = "1992"
}

@article{Medved:2001ca,
    author = "Medved, A. J. M.",
    title = "{Radiation via tunneling in the charged BTZ black hole}",
    eprint = "hep-th/0110289",
    archivePrefix = "arXiv",
    doi = "10.1088/0264-9381/19/3/313",
    journal = "Class. Quant. Grav.",
    volume = "19",
    pages = "589--598",
    year = "2002"
}

@article{Hemming:2000as,
    author = "Hemming, Samuli and Keski-Vakkuri, Esko",
    title = "{Hawking radiation from AdS black holes}",
    eprint = "gr-qc/0005115",
    archivePrefix = "arXiv",
    reportNumber = "HIP-2000-25-TH",
    doi = "10.1103/PhysRevD.64.044006",
    journal = "Phys. Rev. D",
    volume = "64",
    pages = "044006",
    year = "2001"
}

@article{Page:1976df,
    author = "Page, Don N.",
    title = "{Particle Emission Rates from a Black Hole: Massless Particles from an Uncharged, Nonrotating Hole}",
    doi = "10.1103/PhysRevD.13.198",
    journal = "Phys. Rev. D",
    volume = "13",
    pages = "198--206",
    year = "1976"
}

@article{Bekenstein:1974ax,
    author = "Bekenstein, Jacob D.",
    title = "{Generalized second law of thermodynamics in black hole physics}",
    doi = "10.1103/PhysRevD.9.3292",
    journal = "Phys. Rev. D",
    volume = "9",
    pages = "3292--3300",
    year = "1974"
}

@article{Hiscock:1990ex,
    author = "Hiscock, William A. and Weems, Lance D.",
    title = "{Evolution of Charged Evaporating Black Holes}",
    reportNumber = "Print-90-0082 (MONTANA STATE)",
    doi = "10.1103/PhysRevD.41.1142",
    journal = "Phys. Rev. D",
    volume = "41",
    pages = "1142",
    year = "1990"
}

@article{Kraus:1994fj,
    author = "Kraus, Per and Wilczek, Frank",
    title = "{Effect of selfinteraction on charged black hole radiance}",
    eprint = "hep-th/9411219",
    archivePrefix = "arXiv",
    reportNumber = "PUPT-1511, IASSNS-94-101",
    doi = "10.1016/0550-3213(94)00588-6",
    journal = "Nucl. Phys. B",
    volume = "437",
    pages = "231--242",
    year = "1995"
}

@article{Brown:2024ajk,
    author = "Brown, Adam R. and Iliesiu, Luca V. and Penington, Geoff and Usatyuk, Mykhaylo",
    title = "{The evaporation of charged black holes}",
    eprint = "2411.03447",
    archivePrefix = "arXiv",
    primaryClass = "hep-th",
    doi = "10.1007/JHEP01(2026)109",
    journal = "JHEP",
    volume = "01",
    pages = "109",
    year = "2026"
}

@article{Ayon-Beato:1998hmi,
    author = "Ayon-Beato, Eloy and Garcia, Alberto",
    title = "{Regular black hole in general relativity coupled to nonlinear electrodynamics}",
    eprint = "gr-qc/9911046",
    archivePrefix = "arXiv",
    doi = "10.1103/PhysRevLett.80.5056",
    journal = "Phys. Rev. Lett.",
    volume = "80",
    pages = "5056--5059",
    year = "1998"
}

@article{Ayon-Beato:1999qin,
    author = "Ayon-Beato, Eloy and Garcia, Alberto",
    title = "{Nonsingular charged black hole solution for nonlinear source}",
    eprint = "gr-qc/9911084",
    archivePrefix = "arXiv",
    doi = "10.1023/A:1026640911319",
    journal = "Gen. Rel. Grav.",
    volume = "31",
    pages = "629--633",
    year = "1999"
}

@article{Garcia-Diaz:2021bao,
    author = "Garcia-Diaz, Alberto A.",
    title = "{Stationary Rotating Black Hole Exact Solution within Einstein--Nonlinear Electrodynamics}",
    eprint = "2112.06302",
    archivePrefix = "arXiv",
    primaryClass = "gr-qc",
    month = "12",
    year = "2021"
}

@article{DiazGarcia:2022jpc,
    author = "D{\'\i}az Garc{\'\i}a, Alberto A.",
    title = "{AdS{\textendash}dS stationary rotating black hole exact solution within Einstein-nonlinear electrodynamics}",
    eprint = "2201.10682",
    archivePrefix = "arXiv",
    primaryClass = "gr-qc",
    doi = "10.1016/j.aop.2022.168880",
    journal = "Annals Phys.",
    volume = "441",
    pages = "168880",
    year = "2022"
}

@article{Bravo-Gaete:2025fwx,
    author = "Bravo-Gaete, Mois{\'e}s and Guajardo, Luis and Higuita-Borja, Daniel F. and M{\'e}ndez-Zavaleta, Julio A.",
    title = "{Transport coefficients of charged Gauss-Bonnet black holes with arbitrary topology}",
    eprint = "2509.03838",
    archivePrefix = "arXiv",
    primaryClass = "hep-th",
    doi = "10.1103/gs9f-dx3h",
    journal = "Phys. Rev. D",
    volume = "113",
    number = "2",
    pages = "024039",
    year = "2026"
}

@article{Bravo-Gaete:2026hoe,
    author = "Bravo-Gaete, Mois{\'e}s and Ahmed, Moaathe Belhaj and Mann, Robert B. and Quijada, Constanza",
    title = "{Multicritical points of gravitational solitons and a black hole in four dimensions}",
    eprint = "2605.24783",
    archivePrefix = "arXiv",
    primaryClass = "hep-th",
    doi = "10.1103/jnk2-zc4k",
    journal = "Phys. Rev. D",
    volume = "114",
    number = "4",
    pages = "044037",
    year = "2026"
}

@article{Deshpande:2024vbn,
    author = "Deshpande, Rhucha and Lunin, Oleg",
    title = "{Rotating Einstein-Maxwell black holes in odd dimensions}",
    eprint = "2411.01795",
    archivePrefix = "arXiv",
    primaryClass = "hep-th",
    doi = "10.1007/JHEP06(2025)066",
    journal = "JHEP",
    volume = "06",
    pages = "066",
    year = "2025"
}

@article{Hale:2024zvu,
    author = "Hale, Tom{\'a}{\v{s}} and Hull, Brayden R. and Kubiz{\v{n}}{\'a}k, David and Mann, Robert B. and Men{\v{s}}{\'\i}kov{\'a}, Jana",
    title = "{New interpretation of the original charged BTZ black hole spacetime}",
    eprint = "2412.04329",
    archivePrefix = "arXiv",
    primaryClass = "gr-qc",
    doi = "10.1088/1361-6382/adc9f2",
    journal = "Class. Quant. Grav.",
    volume = "42",
    number = "9",
    pages = "09LT01",
    year = "2025"
}

@article{Bravo-Gaete:2025lgs,
    author = "Bravo-Gaete, Moises and Cisterna, Adolfo and Hassaine, Mokhtar and Kubiznak, David",
    title = "{Cardy entropy of charged and rotating asymptotically AdS and Lifshitz solutions with a generalized Chern{\textendash}Simons term}",
    eprint = "2506.04854",
    archivePrefix = "arXiv",
    primaryClass = "hep-th",
    doi = "10.1016/j.physletb.2025.139721",
    journal = "Phys. Lett. B",
    volume = "868",
    pages = "139721",
    year = "2025"
}

@article{Bravo-Gaete:2026bdq,
    author = "Bravo-Gaete, Mois{\'e}s and Herrera-Mendoza, Jhony A. and Oliva, Julio and Zhang, Xiangdong",
    title = "{Novel five-dimensional rotating Lifshitz black holes with electric and axionic charges}",
    eprint = "2601.18048",
    archivePrefix = "arXiv",
    primaryClass = "hep-th",
    doi = "10.1140/epjc/s10052-026-15742-9",
    journal = "Eur. Phys. J. C",
    volume = "86",
    number = "5",
    pages = "502",
    year = "2026"
}
\bibliographystyle{JHEP}

\end{document}